\documentclass{amsart}
\usepackage[utf8]{inputenc}
\usepackage{appendix}
\usepackage{amsmath,amssymb,euscript}
\usepackage{bbm}
\usepackage{array}
\usepackage{enumitem}
\usepackage{comment}
\usepackage[acronym]{glossaries}

\makeglossaries

\newglossaryentry{rho}
{
name={\ensuremath{\rho}},
sort=spectral radius,
description={spectral radius of a matrix}
}

\newglossaryentry{snorm}
{
name={\ensuremath{\|\cdot \|}},
sort=euclidian norm,
description={euclidian norm if applied to a vector, spectral norm if applied to a matrix}
}

\newglossaryentry{hadamard}
{
name={\ensuremath{\circ}},
sort=hadamard product,
description={Hadamard product of matrices $(A\circ B)_{ij} = A_{ij} B_{ij}$}
}

\newglossaryentry{diagmatrix}
{
name={\ensuremath{\mathrm{diag}(\cdot)}},
sort=diagonal matrix,
description={$\mathrm{diag}(\mathbf{x})$ is a diagonal matrix with diagonal entries the components of vector $\mathbf{x}$}
}

\newglossaryentry{convlaw}
{
name={\ensuremath{\xrightarrow[]{\mathcal L}}},
sort=convergence in law,
description={convergence in distribution}
}

\newglossaryentry{convas}
{
name={\ensuremath{\xrightarrow[]{a.s.}}},
sort=convergence almost sure,
description={almost sure convergence}
}

\newglossaryentry{Mgaussian}
{
name={\ensuremath{{\mathcal N}_M(\mathbf{a}, C})},
sort=multivariate gaussian,
description={Multivariate Gaussian random variable of dimension $M$, with mean $\mathbf{a}$ and covariance matrix $C$}
}

\newglossaryentry{bigO}
{
name={\ensuremath{{\mathcal O}}},
sort=big O,
description={Standard big O notation, usually associated to an index $n\to \infty, \varepsilon\to 0$, etc. For example, $a_n={\mathcal O}(b_n)$ means that there exists $K>0$ such that $|a_n|\le K|b_n|$ as $n\to \infty$}
}

\newacronym{LV}{LV}{Lotka-Volterra}
\newacronym{RMT}{RMT}{Random Matrix Theory}
\newacronym{LCP}{LCP}{Linear Complementarity Problem}
\newacronym{ER}{ER}{Erdös-Rényi}
\newacronym{SBM}{SBM}{Stochastic Block Model}
\newacronym{IBM}{IBM}{Individual-based model}

\newtheorem{theo}{Theorem}[section]
\newtheorem{heur}{Heuristics}[section]
\newtheorem{lemma}{Lemma}[section]
\newtheorem{prop}{Proposition}[section]
\newtheorem{remark}{Remark}[section]

\usepackage{graphicx}
\usepackage{caption}
\usepackage{subcaption}
\usepackage{epic}
\usepackage{epsfig}
\usepackage{pstricks}
\usepackage{psfrag}
 \usepackage{curves}
 \usepackage{picture}
\usepackage{mathrsfs}
\usepackage{rotating}
\usepackage{color}
\usepackage[linkcolor=blue,colorlinks=true]{hyperref}
\usepackage{ulem}
\newcommand{\beq}{\begin{equation}}
\newcommand{\eeq}{\end{equation}}
\def\beqa#1\eeqa{\begin{align}#1\end{align}}
\newcommand{\R}{\mathbb{R}}
\newcommand{\N}{\mathbb{N}}

\newcommand{\mean}[1]{\overline{ #1}}

\newcommand{\eye}{\mathbb I}
\newcommand{\dd}{\mathrm d}

\newcommand{\ind}{{\bf 1}}

\DeclareMathOperator{\var}{Var}
\DeclareMathOperator{\Cov}{Cov}

\definecolor{lightblue}{rgb}{.80,.85,1}
\definecolor{malachite}{rgb}{.04,.85,0.32}
\definecolor{darkgray}{rgb}{.2,.25,0.5}

\newcommand{\walid}[1]{\color{orange}{(WH) #1}\color{black}}

\newcommand{\bs}{\boldsymbol}
\newcommand{\x}{\bs{x}}
\newcommand{\Sur}{{\mathcal S}}

\newcommand{\mhat}{\hat{m}}

\newcommand{\phiLV}{\phi^{LV}_i}

\newcommand{\NN}{\mathbb N}

\newcommand{\CC}{\mathbb{C}}
\newcommand{\RR}{\mathbb{R}}
\newcommand{\PP}{\mathbb P}
\newcommand{\EE}{\mathbb E}

\newcommand{\Var}{\mbox{Var}}

\title[A review on Lotka-Volterra models]{Complex systems in Ecology: a guided tour with large Lotka-Volterra models and random matrices}
\author[Akjouj et al.]{Imane Akjouj, Matthieu Barbier, Maxime Clenet, Walid Hachem,\\ Myl\`ene Ma\"ida, Fran\c{c}ois Massol, Jamal Najim, Viet Chi Tran}
\date{\today}

\begin{document}

\maketitle

\setcounter{tocdepth}{1}

\begin{abstract}
   Ecosystems represent archetypal complex dynamical systems, often modelled by coupled differential equations of the form
$$
\frac{\dd x_i}{\dd t} = x_i  \varphi_i(x_1,\cdots, x_N)\ ,
$$
where $N$ represents the number of species and $x_i$, the abundance of species $i$. Among these families of coupled diffential equations, Lotka-Volterra (LV) equations 
$$
\frac{\dd x_i}{\dd t} = x_i ( r_i - x_i +(\Gamma \mathbf{x})_i)\ ,
$$
play a privileged role, as the LV model represents an acceptable trade-off between complexity and tractability. Here, $r_i$ represents the intrinsic growth of species $i$ and $\Gamma$ stands for the interaction matrix: $\Gamma_{ij}$ represents the effect of species $j$ over species $i$. For large $N$, estimating matrix $\Gamma$ is often an overwhelming task and an alternative is to draw $\Gamma$ at random, parametrizing its statistical distribution by a limited number of model features. Dealing with large random matrices, we naturally rely on Random Matrix Theory (RMT).

The aim of this review article is to present an overview of the work at the junction of theoretical ecology and large random matrix theory. It is intended to an interdisciplinary audience spanning theoretical ecology, complex systems, statistical physics and mathematical biology. 
\end{abstract}

\tableofcontents

\noindent Keywords:

\noindent \textbf{Acknowledgments:}\\
{\footnotesize All authors are supported by the CNRS 80 prime project KARATE. I.A., M.C., W.H., M.M., J.N. are supported by GdR MEGA.
I.A. and M.M. are funded by Labex CEMPI (ANR-11-LABX-0007-01). 
M.C., W.H., J.N. and V.C.T. are supported by Labex B\'ezout (ANR-10-LABX-0058). 
F.M. and V.C.T. are funded by ANR EcoNet (ANR-18-CE02-0010).
I.A. and V.C.T. acknowledges support of the ``Chaire Mod\'elisation Math\'ematique et Biodiversit\'e (MMB)'' of Veolia Environnement-Ecole Polytechnique-Museum National d'Histoire Naturelle-Fondation X. V.C.T. is partly funded by the European Union (ERC-AdG SINGER-101054787). M.M. and V.C.T. thank for its hospitality the CRM Montr\'eal (IRL CNRS 3457) where part of this work was completed.}\\

\section{Introduction: Complex networks, randomness and large dimension}

\subsection{Goals of this study}
Ecosystems can be seen as archetypal complex dynamical systems, as they  usually consist of a large number of interacting components with heterogeneous properties. In the present article, these components are species or sub-populations that evolve according to their own demographic dynamics and interact through various mechanisms (such as competition, predation or facilitation). These dynamics generally take the form:
\begin{equation}
\frac{\dd x_i}{\dd t} = x_i \phi_i(x_1,\cdots, x_N)
    \label{eq:premiere}
\end{equation}
where $x_i(t)$ represents the (dimensionless) abundance or density of species $i$ population, the function $\phi_i$ encapsulates all the sources of growth and mortality, and $N \gg 1$ is the number of interacting species. Notice that 
$$
\phi_i= \frac 1{x_i}\frac{\dd x_i}{\dd t}
$$
represents the growth rate of abundance $x_i$, i.e. its per capita rate of abundance change, hence $\phi_i$ is often referred to as the {\bf \textit{net growth rate}}.

Equation \eqref{eq:premiere} incorporates two essential properties of biological dynamics: first, a species can be extinct (equilibrium at $x_i=0$), and second, a small population $x_i  \to 0$ displays an exponential growth or decay with rate $\phi_i\mid_{x_i=0}$ referred to as the {\bf \textit{invasion growth rate}} (e.g. \cite{Marrow1992,champagnatferrieremeleard}, see also Appendix \ref{section:AD}).


To analyze and model such complex dynamical systems, a set of formal tools, including network theory, dynamical systems and \gls{RMT}, have proven successful across a variety of scientific disciplines. This review is meant to offer a constructive viewpoint on the connection between these mathematical tools, especially RMT, and ecological systems, addressed to an interdisciplinary audience spanning theoretical ecology, complex systems, statistical physics and mathematical biology. We propose to focus on formal problems inspired by complex ecosystems, with the ultimate aim of answering ecological questions regarding the conditions of species coexistence, community diversity and ecosystem stability. Although the standing of mathematical models is quite different in ecology from that in physics or computer science, partly due to the lack of fundamental laws or canonical theories, some dynamical models are present in both ecological and mathematical studies, especially the well-known \gls{LV} model~\cite{lotka,volterra}. 

Through the prism of the Lotka-Volterra model with random interactions, our aim here is to give a diverse overview of concepts and questions that have proven fruitful in that line of work.  In particular, we wish to point out results that we conjecture may also hold qualitatively, or even quantitatively, beyond this particular model, in the hope that such ``generic'' behaviors might be shown in the future to capture some aspects of real ecosystems.

    \subsection{Historical context}
    \label{sub:history}

 The first eminent proponent of applying results from complex dynamical systems to ecosystems was Robert May~\cite{may1972will}. Inspired by the success of RMT for modelling unknown interactions in complex physical systems, such as large atomic nuclei~\cite{wigner}, he proposed the following argument:
 \begin{description}
     \item[Obs] Empirical observation: ecosystems with a large number of species appear to exist and persist for long times.
     \item[Hyp 1]  If we assume that species abundances are poised at some dynamical equilibrium $\bs{x}^*=(x_1^*,\cdots, x_N^*)$ with all $x_i^* > 0$
     \beq\label{ode-gale}
     \frac{\dd x_i}{\dd t}=x_i^* \phi_i(\bs{x}^*) = 0\eeq
     and
     \item[Hyp 2] If we assume that their interactions are \textit{sufficiently complex} to be modelled as random, i.e. more precisely, that the Jacobian matrix governing the linearized dynamics around the equilibrium
     \beq J_{ij} = \dfrac{\partial (x_i \phi_i) }{ \partial x_j} (\bs{x}^*) \label{eq:jacobien}\eeq
     is modelled as 
     $$
     \left\{
     \begin{array}{cclc}
     J_{ii}&=& -1 + M_{ii} C_{ii}&\\
     J_{ij} &=& M_{ij} C_{ij}& (i\neq j)
     \end{array}
     \right.\qquad i.e. \qquad J= -I +C\circ M 
     $$ where the random variables $M_{ij}$ are independent identically distributed (i.i.d.) centered random variables of variance $V$, $C_{ij}$ are i.i.d. with Bernoulli distribution with parameter $C,$ called the connectance (such that a species has $N \times C$ links on average), and $\circ$ denote the Hadamard product of matrices (pointwise multiplication of entries),\\
     
     \item[Res] Then the equilibrium is stable for $NCV < 1 $ and unstable for $NCV >1$. \\
 \end{description}
 
 
 
May presented his conclusion as a paradox: earlier ecologists had imagined the observed complexity of natural ecosystems as a positive feature favoring their persistence (i.e. the more connected and/or strongly interacting species were, the more likely they would coexist), which could be contradicted by the loss of stability predicted in May's simplistic model. May's  result, however, did not spell out where in the above argument lies the ``paradox'', i.e. which part of the argument should be reevaluated in the light of empirical evidence, or instead, should lead us to reevaluate earlier assumptions about ecosystems. 

A formal presentation of May's result is provided in Appendix \ref{ann:may_model} (see in particular Proposition \ref{prop:may-connectance}). 
This result  had a lasting influence on the theoretical ecology literature, yet it never received clear support from ecological data \cite{tangpawarallesina2014,jacquet2016,neutel2007,James2015}. 
Theoretical and empirical work on this topic over the following decades has been diverse in scope and focus, but can be interpreted as questioning each of the four points summarized above\footnote{May's own interpretation was probably focused on the third (Nature had to employ ``devious strategies'', in his words, to allow high-diversity ecosystems to exist in a stable equilibrium).}.  In response to \textbf{Obs}, ecosystems might be less diverse and complex than they appear~\cite{hubbell2001unified} (e.g. few species may coexist at any location, or species may exist in large number $N$ but with low interaction variance  $V$ ). Contra \textbf{Hyp 1}, real ecosystem dynamics might be far from equilibrium, e.g. dominated by transient excursions~\cite{hastings2018transient}. Contra  \textbf{Hyp 2}, ecological interactions might be structured so that the resulting dynamics deviate importantly from predictions from full randomness~\cite{allesina2012stability} (suggesting a greater role of non-random structure in ecology). Finally, even admitting \textbf{Res}, the loss of stability when $NCV>1$ does not imply the extinction of species: the ecosystem may still persist in a steady out-of-equilibrium state~\cite{roy2020complex}, and thus this mathematical statement may not be interpreted ecologically as a limit on a realistic number of species.

The discussion that May's result generated \cite{LehmanTilman2000,kokkoris2002} also highlights the issues arising from misunderstanding mathematical notions of equilibrium existence, equilibrium stability, and system persistence, as well as the gap between measurable quantities in ecology and those relevant to the analysis of ODEs \cite{arnoldi2016unifying,loreaudemaz2013,wangloreau2014}.
Nonetheless, the ``May-Wigner transition'' to instability was an influential result because, while it was proven for a particular case (a random linearized dynamical system), there were many reasons -- both mathematical and empirical, from successes of Random Matrix Theory in other fields -- to conjecture that it could have far broader generalizations. 


\begin{figure}
\begin{center}
\includegraphics*[width=\linewidth]{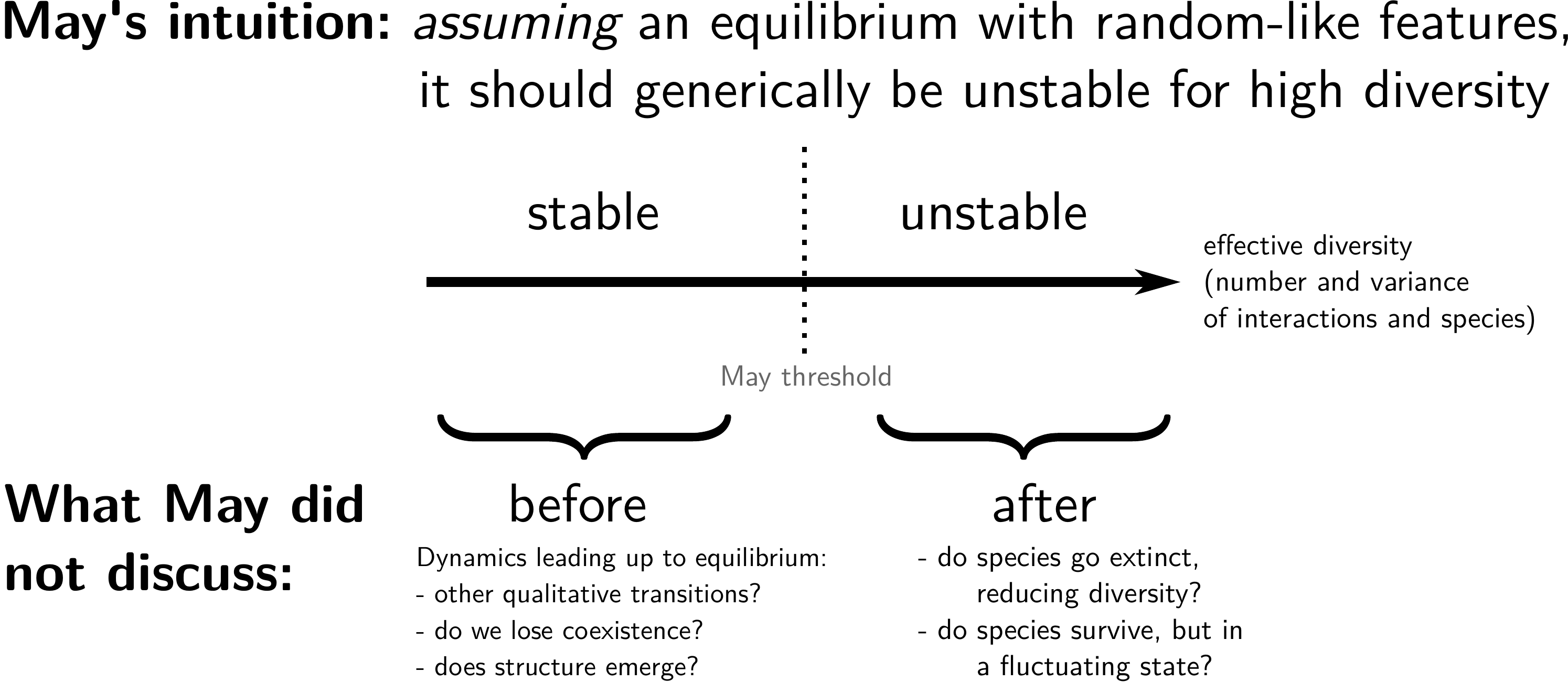}
\end{center}
\caption{Schematic cartoon of early arguments from randomness in complex ecological systems. Robert May (1972) used a mathematical result to anchor a broader qualitative intuition: assuming a system with random features at dynamical equilibrium, there is an axis of effective diversity (equal to the product $NCV$ in his model), combining system size (species number) and statistics (notably variance) of interaction, along which a qualitative transition from stability to instability is expected. Yet how likely is such an equilibrium to arise in the first place, and are there other qualitative transitions along this axis (or other axes that are as broadly relevant)?} 
\label{fig:May_cartoon}
\end{figure}

\subsection{Qualitative questions}
\label{sub:intro_questions}
May's original work did not actually state what happens in arbitrary nonlinear ecological dynamics, besides the fact that it may exhibit a transition to instability (cf. Fig.~\ref{fig:May_cartoon}). For instance, can some species go extinct without a loss of stability? \cite{GardnerAshby1970,roberts74,gilpin75}.

The aim of this review is to demonstrate that such questions, which cannot be addressed in May's setting, can be and have been fruitfully studied in subsequent decades using the LV model introduced in Section \ref{sec:model-LV}, a model that is simple, can be derived as an approximation of individual-based models (see Appendix \ref{annexe:micro1}), and is rich enough to demonstrate various dynamical phenomena.

Another important line of inquiry that has grown in response to May's work on stability  has focused on \textit{imposed structure}, searching for various deviations from full randomness that might possibly allow to restore feasibility, stability, or both. These structural features can be motivated by empirical networks, such as reproducing in broad strokes the hierarchical or group structure of a foodweb~\cite{allesina2015stability}, the size-dependence of trophic interactions~\cite{Otto2007}, or accounting for the spatial structure of ecosystems~\cite{gravel2016,mougi2016}. They can also arise from theoretical mechanisms, such as interactions deriving from underlying ecological traits and niches~\cite{eklof2013}, or their evolution~\cite{allhoff2015}. We will discuss results and questions arising when we allow a combination of deterministic structure and randomness in Lotka-Volterra systems.

\subsection{Outline of the article} In Section \ref{sec:model-LV}, we introduce the LV system of coupled differential equations. We discuss and motivate the large random interactions model, and introduce two such models, with independent interactions (i.i.d. model), and cross-diagonal correlated interactions (elliptic model). 

In Section~\ref{sec:unique}, we present results for the two reference models (i.i.d. and elliptic) and address the question of the existence of a unique equilibrium $\bs{x}^*=(x_i^*)$ for a LV system with random interactions and its stability, that is the conditions for which $\bs{x}(t)\to \bs{x}^*$. We also focus on the feasibility of the equilibrium, that is the conditions for which no species vanishes at the equilibrium: $x_i^*>0$ for all $i.$
If feasibility is not reached, we describe the composition of the equilibrium in terms of surviving species $x_i^*>0$ and vanishing species $x_i^*=0$, together with various properties of the equilibrium.
Beyond i.i.d. and elliptic models, there is a need to consider more realistic structures of interaction matrices which take into account important features of real foodwebs. In Section~\ref{sec:struct}, we focus on sparse models and kernel matrices. We present a quick survey of the related literature in theoretical ecology and recall associated basic RMT results.

In Section~\ref{sec:otherextensions}, we extend Lotka-Volterra models by adding extra randomness yielding to stochastic differential equations (SDEs).

Finally, we summarize in Section~\ref{sec:discussion} what has been covered in this review and what has not. We also present some directions of investigation and open problems of interest.

Several appendices complete the main exposition. In Appendix \ref{ann:may_model}, we present May's model. In Appendix \ref{section:LV-2d}, we present the deterministic LV model of size 2, for pedagogical purpose. Appendix \ref{annexe:micro1}
establishes the connexion between individual-based models and LV model, while Appendix \ref{annexe:micro2} shows how individual-based models studied at a different asymptotics yield community models with noise. In Appendix \ref{app:structuredmodapp}, we provide mathematical details related to Section \ref{sec:struct}, in particular a certain type of deterministic networks and kernel matrices. More precisely, concerning the latter, we emphasize the connexion between general kernel matrices and Marchenko-Pastur distribution associated to large random covariance matrices.

A glossary gathering the main mathematical notations and a list of acronyms is provided at the end of the article.

\section{The Lotka-Volterra model}\label{sec:model-LV}

\subsection{The Lotka-Volterra system of differential equations}

We are interested in many-species ecological dynamics and our main object of study will be the following LV system of differential equations:
\begin{equation}
\label{eq:LV} 
    \frac{\dd x_i}{\dd t}=x_i\left(r_i-x_i+ (\Gamma \boldsymbol{x})_i \right)\,,
\end{equation}
where $i\in [N]:=\{1,\cdots, N\}$ and $\boldsymbol{x}=(x_1,\cdots, x_N).$ The parameter $N$ represents the number of species, supposed large, $x_i=x_i(t)$ is a dimensionless quantity evolving in time $t$, in relation with the abundance of species $i$, $r_i$ represents the intrinsic growth of species $i$ and $\Gamma=(\Gamma_{ij})$ is a $N\times N$ matrix reflecting the interaction effect of species $j$ on the growth of species $i$:
$$
(\Gamma \boldsymbol{x})_i= \sum_{j} \Gamma_{ij} x_j\, .
$$
Notice that if we introduce the following function
\begin{equation}
\label{def:phiLV}
\phiLV(\boldsymbol{x}) := r_i-x_i+ (\Gamma \boldsymbol{x})_i \,,
\end{equation}
Eq. \eqref{eq:LV} follows the generic form introduced in \eqref{eq:premiere}:
$$
\frac{\dd x_i}{\dd t}=x_i\phiLV(\boldsymbol{x})\, ,
$$
with $\phiLV$ a function of all the $x_i$'s representing the net growth rate of species $i$.

As mentioned above, we detail in Appendix \ref{annexe:micro1} how this LV model naturally emerges from an individual-based model 
when we consider  a certain asymptotics of the birth, death and interaction rates. 
Let us also formalise in Remark \ref{rem:dimensionless} below how one can obtain the dimensionless, normalized system of equations \eqref{eq:LV}, where each species has a comparable importance, from a similar system dealing with the actual abundances of species.

. 

\begin{remark}[Obtaining the dimensionless LV model \eqref{eq:LV}]\label{rem:dimensionless}
In ecology, a natural formalization of the LV model involving the actual abundances or densities $X_i$ of the modelled species reads as:

\begin{equation}
\label{dimensional_LV}
    \frac{\dd X_i}{\dd t} = X_i \left[r_i - D_i X_i +  \left(M\boldsymbol{X} \right)_i \right],
\end{equation}
where $r_i$ is species $i$ intrinsic growth rate, $D_i$ is species $i$ density-dependent term and $\boldsymbol{M}$ is the matrix of interaction coefficients. In order to obtain the dimensionless version of the Lotka-Volterra model presented in Equation~\eqref{eq:LV}, the following changes of variables are needed:
\begin{equation}
x_i := D_i X_i\,,\quad 
\Gamma_{ij} := M_{ij}/D_j.
\end{equation}
Plugging these new variables into Eq.~\eqref{dimensional_LV}, one naturally obtains Eq.~\eqref{eq:LV}.
\end{remark}

In a large ecosystem consisting of $N$ species ($N\gg 1$), the precise knowledge of the interaction matrix $\Gamma=(\Gamma_{ij})$ among these species is often out of reach. An interesting alternative is to model the $N\times N$ matrix $\Gamma$ with random entries and to rely on RMT. The statistical properties of the entries may then reflect a partial knowledge on the ecological interaction network.

\subsection{Two random interactions models}
\label{sec:models-typo-hypo}

In this section, we precisely describe two models for the matrix $\Gamma$ with random interactions. The first is the simplest theoretical baseline of i.i.d. entries with zero mean, while the second model is a natural extension that allows to represent more types of ecological interactions~\cite{tangallesina2014}.
Other (more involved) models will be discussed in Section~\ref{sec:struct}.

\subsubsection*{Independent and Identically Distributed entries.} In this model, the entries $\Gamma_{ij}$ are i.i.d., with a $N$-dependent common distribution. The entries may write
\begin{equation}\label{eq:iid}
(i)\quad \Gamma_{ij}= \frac{A_{ij}}{\sqrt{N}}\qquad \text{or}\qquad(ii)\quad  \Gamma_{ij}=\frac{A_{ij}}{\alpha_N\sqrt{N}}
\end{equation}
where the $A_{ij}$'s are i.i.d. centered $(\mathbb{E}\, A_{ij}=0$) with variance $\mathbb{E}\, A_{ij}^2=1$ and a distribution independent from $N$.

In the case $(i)$, the $N^{-1/2}$-normalization casts the interaction matrix $\Gamma=\frac{A}{\sqrt{N}}$ into the framework of RMT, where the limiting properties of the spectrum and eigenvectors of matrix $\Gamma$ are well described. 

\begin{figure}
\centering
\begin{subfigure}[b]{.32\textwidth}
    \includegraphics[width=\textwidth]{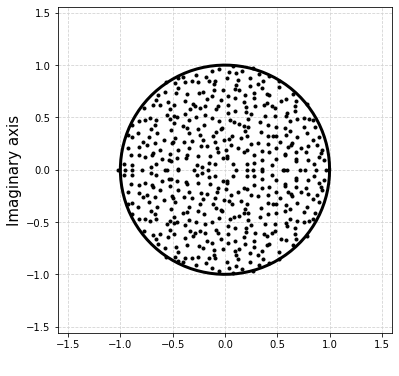}
    \caption{$\xi=0$}
  \end{subfigure}%
  \hfill   
  \begin{subfigure}[b]{0.32\textwidth}
    \includegraphics[width=\textwidth]{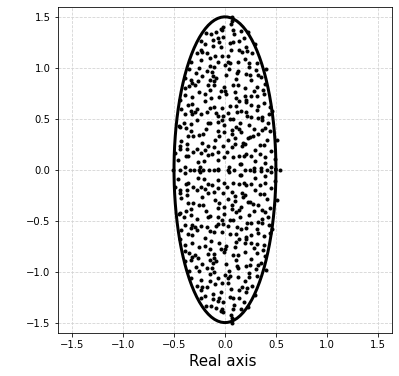}
    \caption{$\xi=-0.5$}
  \end{subfigure}%
\hfill   
  \begin{subfigure}[b]{0.32\textwidth}
    \includegraphics[width=\textwidth]{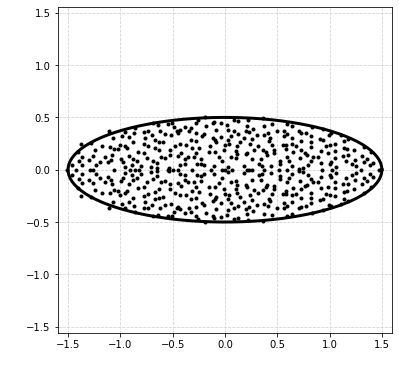}
    \caption{$\xi=0.5$}
  \end{subfigure}%
\caption{Spectrum of $500\times 500$ centered random matrices with (a) $N^{-1/2}$-normalized i.i.d. standard Gaussian entries (left), (b) elliptic distribution $\xi=-0.5$, (c) elliptic distribution $\xi=0.5$. The points represent the eigenvalues. The solid curve represents the boundary of the support of (a) the circular law (uniform law on the disc), (b) and (c) the elliptic distribution with parameter $\xi$.}
\label{fig:circular}
\end{figure}

The circular law (cf. \cite{bordenave2012around}) asserts that the spectrum of $\Gamma$ converges toward the uniform distribution on the disk of radius $1$, see for instance Figure \ref{fig:circular}-(a). Moreover, if we denote by  \gls{rho}$(\Gamma)$  the spectral radius  of $\Gamma$:
\[ 
\rho(\Gamma) :=\max\{ |\lambda(\Gamma)|\,,\ \lambda(\Gamma)\in \mathbb{C}\ \textrm{eigenvalue of}\ \Gamma\},\
\]
then its asymptotic behaviour is well-known: 
$$
\rho(\Gamma)\xrightarrow[N\to\infty]{a.s} 1\,,
$$
where \gls{convas} stands for the almost sure convergence. As a consequence, matrix $\Gamma$ has a non-vanishing macroscopic effect on the dynamical system \eqref{eq:LV} even for large $N$.


In the case $(ii)$, there is an extra normalization term $\alpha_N$ which may or may not depend on $N$. If it does not depend on $N$, then it is simply a parameter that allows to tune the variance of the entries since $\text{var}(\Gamma_{ij})=\frac{1}{\alpha^2 N}$. If $\alpha_N$ grows to infinity as $N\to \infty$, it has the effect of asymptotically squeezing to zero the contribution of the interaction matrix $\Gamma$ as
$$
\rho(\Gamma) =\frac {\rho(A/\sqrt{N})}{\alpha_N} \xrightarrow[N\to\infty]{} 0\, .
$$
Despite this feature, we will see later that $\alpha^*_N=\sqrt{2\log(N)}$ is the threshold to reach feasibility, see Section \ref{subsec:feasibility}.\\

\subsubsection*{The elliptic model.} Two assumptions of the i.i.d. model are commonly relaxed to describe a wider range of ecological scenarios. First, while the i.i.d. model assumes that interactions have zero mean, ecological networks often contain interactions with a prescribed sign. Second, the i.i.d. model enforces the reciprocal interactions $\Gamma_{ij}$ and $\Gamma_{ji}$ to be uncorrelated. However, a large literature, for random as well as for deterministic interactions, deals with symmetric matrices $\Gamma_{ij}=\Gamma_{ji}$ 
, which can for instance arise in ecology in the case of competition~\cite{mac1969species}, or skew-symmetric matrices $\Gamma_{ij}=-\Gamma_{ji}$ ~\cite{redheffer1985volterra}, which were originally proposed for predator-prey interactions~\cite{volterra}. The random symmetric case is well-known in RMT and is referred to as the Wigner model, see for instance  ~\cite[Chap. 2]{bai2010spectral}.

These various cases can be unified into the elliptic model~\cite{girko1986elliptic,Sommers1988,o2014low}, which displays a richer statistical structure than the i.i.d. model. The entries of matrix $\Gamma$ write
\begin{equation}\label{def:ellipticalmodel}
\Gamma_{ij} = \frac{A_{ij}}{\sqrt{N}} + \frac{\mu}{N}\, ,
\end{equation}
where $\mu$ is a deterministic quantity, $A_{ij}$ is centered with variance equal to 1. In particular, $\frac {\mu}N$ stands for $\Gamma_{ij}$'s expectation. The second feature of the model is the existence of a correlation between the random variables $A_{ij}$ and $A_{ji}$ ($i\neq j$):
\begin{equation}\label{def:ellipticalmodel-corr}
\text{corr}(A_{ij},A_{ji})=\xi\in [-1,1]\, 
\end{equation}
while $A_{kk}$ and $\{A_{ij}, A_{ji}\}$ are independent for $k,i,j$ and $i< j$.

\begin{figure}
    \centering
    \includegraphics[width=0.6\linewidth]{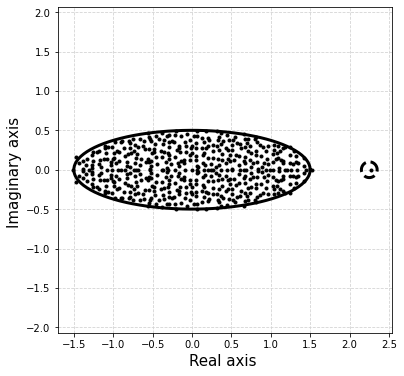}
    \caption{Spectrum of a deformed elliptic matrix of size $500\times500$ with parameter $\xi = 0.5$ and $\mu=2$. The solid line represents the ellipse $\mathcal{E}_{\xi}$ which is the boundary of the support of the limiting spectral distribution for an elliptic model. There is an extra outlier in the small dashed circles centered at $2+\frac{\rho}{2}$ as witnessed.}
    \label{fig:elliptic+outlier}
\end{figure}

Under the elliptic model and in the case where $\mu=0$ and $|\xi| \neq 1$, the spectrum of $\Gamma$ converges toward the uniform law on the ellipse 
\[{\mathcal E}_{\xi}=\left\{ z=x+\mathbf{i}y,\ \frac {x^2}{(1+\xi)^2} + \frac{y^2}{(1-\xi)^2} \le 1\right\}\, , 
\] see Figure \ref{fig:circular}, (b) and (c) (hence the name). If $\mu >1$ and does not belong to $\mathcal E_\xi,$ then it has been shown in \cite{orourkerenfrew2014} that we witness an extra outlier located near $\mu+ \frac{\xi}{\mu}$ (i.e. a single random eigenvalue of $\Gamma$ will converge to $\mu +\frac{\xi}{\mu}$ as $N\to \infty$),  see Figure \ref{fig:elliptic+outlier}.

We may also consider this model with an extra normalization, as in \eqref{eq:iid}-$(ii)$:
\begin{equation}\label{eq:elliptic-extra-norm}
\Gamma_{ij} =\frac{A_{ij}}{\alpha_N\sqrt{N}} +\frac {\mu}{N}\, .
\end{equation}

Among the key issues we will address in the sequel is the study of the various regimes of the dynamical system \eqref{eq:LV} with random interactions: existence of an equilibrium, study of its stability, etc.



\begin{remark}
Using random interaction models means that we hope to predict essential aspects of an ecosystem only on the basis of statistical features of the network, rather than its detailed structure. In May's framework (see Sec.~\ref{sub:history}), the main parameter turned out to be the complexity $NCV,$ combining the number $N$ of species, the connectance $C$ and the variance $V$ of the interactions. 
The system's equilibrium is predicted to go from stable to unstable as its complexity parameter $N C V$ crosses a threshold. 

Our LV models are parametrized by the variance $\frac{1}{\alpha^2}$ of the interactions, their bias $\mu$ and their correlations $\xi$ etc. One of our goal in the sequel will be to investigate how these parameters can be combined to characterize the behavior of the corresponding dynamical system.  
\end{remark}

\section{Equilibrium, coexistence and stability}
\label{sec:unique}

Due to the form of the dynamical system \eqref{eq:LV} with $\phiLV$ defined in \eqref{def:phiLV}:
 $$
 \frac{\dd x_i}{\dd t} = x_i \phiLV(\bs{x})\,,\quad i\in [N]\,,
 $$
 standard ODE results yield that if $\bs{x}(t=0)=\bs{x}_0>0$ (resp. $\bs{x}(t=0)=\bs{x}_0\ge0$) componentwise, then $\bs{x}(t)>0$ (resp. $\bs{x}(t)\ge 0$) for all $t>0$. We are thus interested in equilibrium points $\bs{x}^*=(x_i^*)_{i\in [N]}$ satisfying 
 $$
 x_i^* \phiLV(\bs{x}^*)= x_i^*(r_i -x_i^* +(\Gamma \bs{x}^*)_i) = 0\, ,\quad i\in [N]\quad \textrm{and}\quad x^*_i\ge 0\,.
 $$

Let us now present the main properties of the equilibria that have been investigated in the literature.



\subsection{Terminology and main questions} 


Consider the general system \eqref{eq:premiere}. If needed, we will use $\phiLV$ instead of $\phi_i$. 
 
 \subsubsection{Various notions of stability}
 
 The most common notion of stability in dynamical systems is the so-called {\bf \textit{Lyapounov stability}}: The equilibrium $\bs{x}^*$ is Lyapounov stable if for any neighborhood $U$ of $\bs{x}^*,$ there exists a neighborhood $W$ of $\bs{x}^*$
    such that $$\bs{x}(0) \in W \quad \Longrightarrow\quad \bs{x}(t) \in U\qquad \textrm{for all}\ t\ge 0
    \,.$$ 
From now on, we will simply refer to {\bf\textit{stability}} instead of Lyapounov stability.

The equilibrium    $\bs{x}^*$ is said to be {\bf\textit{asymptotically stable}}, if and only if it is stable  and the neighborhood $W$ can be chosen so that $$\bs{x}(0) \in W \quad \Longrightarrow \quad   \bs{x}(t) \xrightarrow[t \rightarrow \infty]{}  \bs{x}^*\,.$$

 

One can also get interested in {\bf\textit{global stability}}, in the sense that $\bs{x}^*$ is stable and
$$
\forall\ \bs{x}(0)\in (0,\infty)^N\,,\qquad \bs{x}(t) \xrightarrow[t\to \infty]{} \bs{x}^*\, .
$$
The following theorem provides a necessary condition for stability to hold (notice that the result below holds for general functions $\phi_i$).
\begin{theo}[see Takeuchi \cite{takeuchi1996global}, Theorem 3.2.5]
\label{th:fitness-stability}
Consider the system 
\begin{equation}\label{eq:generic-system}
    \frac{\dd x_i}{\dd t} = x_i \phi_i(x_1,\cdots, x_N),\quad  i\in [N]\ ,
\end{equation}
with all $\phi_i$ continuous, If a non-negative equilibrium point $\bs{x}^*$ of \eqref{eq:generic-system} is stable, then 
$$
\phi_i(x_1^*,\cdots, x_N^*) \le 0,\quad i\in [N]\, .
$$
\end{theo}

We will explain below how to relate the properties of the interaction matrix $\Gamma$ and the stability of the equilibrium.

\subsubsection{Invasion, extinction, feasability and uninvadability}

An important feature of Lotka-Volterra dynamics  \eqref{eq:LV} is the fact that species can be extinct: $x_i=0$ entails $\dd x_i/\dd t=0$ and is always a possible equilibrium value for species $i$. On the other hand, if at some time $x_i(t)> 0$, then positivity is maintained at all later times \cite{HofbauerSigmund1988}. However, the solution can tend to zero asymptotically (see also Section~\ref{sec:otherextensions} for finite populations).
  
The fact that, for all species $i$, $x_i=0$ is always associated to some possible equilibria has motivated the notion of {\bf\textit{species invasion}} where one takes a system where $x_i(0)=0$, and perturbs it at time $t$ by setting $x_i$ to a nonzero value. In particular, we often ask whether a species can invade \textit{from rare}, i.e. what is the asymptotic behavior when $x_i(t)$ is set to a small positive value? This can be answered by considering the net growth rate $\phiLV$, defined for species $i$ in Eq. \eqref{def:phiLV} and representing its per capita rate of abundance change.
{The constant 
\begin{equation}\label{def:fitness}
\phiLV\mid_{x_i=0}\ =\ \phiLV(x_1,\cdots, x_{i-1},0,x_{i+1},\cdots, x_N)=r_i-\sum_{j\not= i} \Gamma_{ij}x_j\end{equation}defines the rate of exponential growth or decay of a small population $x_i \approx 0$ in an environment where the other species start with sizes given by $\bs{x}$.}  This quantity is called the {\bf \textit{invasion growth rate}} in the community ecology literature, or invasion fitness (or just fitness) in evolution. The net growth rate $\phiLV$ also has a probabilistic individual-based interpretation since it is related to the survival probability of a birth and death process that approximates locally $x_i(t)$ when $x_i$ is small and considering that the other sizes also do not vary much (see Appendix \ref{section:AD}). 
The species is said to be deterministically (or asymptotically) extinct if $x_i \to 0$, and permanent otherwise~\cite{hofbauer1998evolutionary,law1996permanence,metzgeritzmeszenajacobsheerwaarden}.

Ecologists have investigated \textit{invasion sequences}, e.g. whether different dynamics and attractors are reached depending on the order in which species are introduced~\cite{robinson1987does}.
The same notions arise in an evolutionary context, where mutant types can be modelled as initially absent species.
 Champagnat and Méléard \cite{champagnatmeleard2011} construct the \textit{polymorphic evolution sequence} that alternates phases where the dynamics is described by the LV system \eqref{eq:LV} and phases of invasion of new arriving species, generalizing the trait substitution sequence process introduced by Metz et al. \cite{metzgeritzmeszenajacobsheerwaarden}. Merging coevolution and invasion sequences in simulation studies has also been a long-standing endeavour among ecologists \cite{Rummel1985,TaperCase1992,Calcagno2017,Romanuk2019,Case1990}.

Beyond deterministic extinctions and invasions, we explain in Sec. \ref{sec:otherextensions} and Appendix \ref{annexe:micro1} how the expression for fitness arises in a probabilistic individual-based description and connects to survival probability in such a context.

The notion of {\bf\textit{feasibility}} will refer to an equilibrium $\bs{x}^*=(x_i^*)$ of \eqref{eq:premiere} where all the species coexist:
$$
x^*_i>0 \quad \textrm{for}\quad i\in [N]\, .
$$
Such an equilibrium will be called {\bf\textit{feasible}}.

The notion of {\bf\textit{uninvadability}} will refer to an equilibrium $\bs{x}^*=(x_i^*)$ of \eqref{eq:premiere} satisfying:
$$
x_i^*\phi_i(\bs{x}^*) =0\quad \textrm{for} \quad i\in [N]
$$
where either
\begin{itemize}
    \item[(a)] $\phi_i(\bs{x}^*) =0$ and $x^*_i > 0$, in which case the species $x_i^*$ is said to {\bf \textit{survive}} or
    \item[(b)] $x^*_i=0$ and $\phi_i(\bs{x}^*) \leq 0$, in which case the species is said to {\bf \textit{vanish}} (or to be extinct).
\end{itemize} 
Such an equilibrium will be called {\bf\textit{uninvadable}}. Notice that it is stated in Theorem \ref{th:fitness-stability} that an equilibrium, to be stable, must be uninvadable.

 

\subsubsection{Main questions we want to address}

This raises a number of fundamental questions, in particular:
\begin{enumerate}
    \item Given parameter $\bs{r}$ and a random model for matrix $\Gamma$, is there a unique equilibrium? Is it locally stable? globally stable ? What is the proportion of surviving species and what are their statistical features?
    \item Given parameter $\bs{r}$ and a random model for matrix $\Gamma$, what are the conditions to get a feasible equilibrium? Is it unique? stable?
    \item Given an interaction matrix $\Gamma$, can we characterize the domain of growth rates $\bs{r}$ that allow feasibility?  
\end{enumerate}
We focus hereafter on questions 1 and 2. We will not develop answers to question 3, which assumes that $\bs{r}$ is likely to vary with environmental conditions while $\Gamma$ is biologically fixed, since the main object of this study is LV systems when $N$ is large and the interaction matrix $\Gamma$ contains a random component. However, extensive works on question 3 can be found elsewhere, in the context of structural stability of LV equilibria \cite{bastolla2005,saavedra2017structural}.

There are several levels of answers to these questions, depending of the expected level of rigor. As we will see, we can provide fairly precise mathematical answers for a narrow range of assumptions (tight assumptions on the interactions $\Gamma$). Theoretical physics tools and computer experiments will substantially relax these assumptions and widen our understanding.

\subsection{Linear Complementarity Problem: an important concept to study equilibrium dynamics}

In this section, we provide the definition of the \gls{LCP}, which is part of the theory of mathematical programming (see \cite{murty1997,cottle2009linear} for standard references). LCP has already been used in ecological contexts in \cite[Chapter 3]{takeuchi1996global}.

Given a $N\times N$ matrix $M$ and a $N\times 1$ vector $q$, we say that the $LCP(M,q)$ admits a solution $(z,w)$ where $z$ and $w$ are $N\times 1$ vectors if there exist two such vectors satisfying the following set of constraints
$$
\left\{ \begin{array}{lcl}
     w&=&M z +q\ \ge\  0\,,  \\
     z&\ge&0\,,\\
     w^Tz&=&0\quad \Leftrightarrow \quad w_i z_i=0\quad \textrm{for all}\  i\in [N]\, .
\end{array}\right.
$$
In this case, we simply write $z\in LCP(M,q)$ since $w=Mz+q$ can be inferred from $z$. 

Consider the LV dynamics \eqref{eq:LV}. An uninvadable equilibrium $\bs{x}^*=(x_i^*)$ (if it exists) will satisfy
\begin{equation}\label{eq:LCP}
\left\{ \begin{array}{lcl}
     x^*_i \phiLV(\bs{x}^*)&=& 0 \, ,\\
     x^*_i&\ge&0\quad \,,\\
     \phiLV(\bs{x}^*)&\le&0
\end{array}\right. \quad \textrm{for all}\  i\in [N]\, .
\end{equation}
Taking into account the explicit form of $\phiLV(\bs{x}^*)=r_i - x_i^* +(\Gamma \bs{x}^*)_i$, this exactly means that $\bs{x}^*\in LCP (I-\Gamma, -\bs{r})$.


\subsection{Criteria for existence and uniqueness of a globally stable equilibrium.}

As mentioned above, an equilibrium point $\bs{x}^*=(x_i^*)$, if it exists, should satisfy
\begin{equation}\label{eq:equilibrium-general}
x_i^*(r_i - x_i^* +(\Gamma \bs{x}^*)_i)=0,\quad i\in [N]\, .
\end{equation}
Hence, either $x_i^*=0$, and the species $i$ vanishes at equilibrium or $r_i - x_i^* +(\Gamma \bs{x}^*)_i=0$. We a priori do not know beforehand which species vanish and which ones remain. Moreover, uniqueness of the equilibrium may not be guaranteed.

A systematic way to find all the solutions is to arbitrarily fix the vanishing species $x_i^*=0$ for $i\in A$, $A$ being any subset of $[N]$ ($2^N$ possibilities), then to solve the remaining set of equations
\begin{equation}\label{lineareq}
r_j - x_j^* +(\Gamma \bs{x}^*)_j=0,\quad j\in A^c := [N]\setminus A\, .
\end{equation}
If the obtained $x_j^*$ are positive, then $(0,i\in A;x_j^*, j\in A^c)$ is a possible solution.

Adding an uninvadability condition may considerably reduce the number of solutions, which can then be analyzed in the LCP framework.




The following result due to Takeuchi and Adachi \cite{takeuchi1996global} provides a sufficient condition for the existence of a unique equilibrium and the global stability of the LV system. It is the cornerstone to establish single equilibrium/stability conditions for interaction matrices with random entries  and is based on the explicit construction of a Lyapunov function (notice that Goh \cite{Goh1977AmNat} provides similar stability results in the case of the existence of a feasible equilibrium).  

\begin{theo}[Takeuchi and Adachi 1980, see \cite{takeuchi1996global} Th. 3.2.1]
\label{theo:TAKEUCHI-stability}
Consider the system \eqref{eq:LV}:
$$
\frac{\dd x_i}{\dd t} = x_i \left( r_i - x_i + (\Gamma\bs{x})_i\right) = 
x_i \left( r_i +[(-I+\Gamma)\bs{x}]_i\right)
$$
and assume that there exists a diagonal matrix $\Delta$ with positive diagonal elements such that 
matrix $\Delta(I-\Gamma)+ (I-\Gamma)^*\Delta$ is positive definite. Then there exists a unique equilibrium point $\bs{x}^*$ solution of the $LCP(I-\Gamma,-\bs{r})$ and this equilibrium is globally stable in the sense that it is stable and
$$
\forall\ \bs{x}(0)\in (0,\infty)^N,\quad \bs{x}(t) \xrightarrow[t\to \infty]{} \bs{x}^*\, .
$$
\end{theo}
Combining Theorem \ref{theo:TAKEUCHI-stability} and standard results in RMT on the limit of the spectral radius of a Wigner matrix, we prove the following result:

\begin{theo}[Unique equilibrium and stability under uninvadability condition] \label{th:unique-LCP} Let $\alpha_N=\alpha$ be fixed. If one of the following conditions is satisfied:
\begin{enumerate}
\item \label{ass:iid} Matrix $\Gamma$ is given by model \eqref{eq:iid}-(ii) and $\alpha>\sqrt{2}\, ,$
\item Matrix $\Gamma$ is given by model \eqref{eq:elliptic-extra-norm} and the parameters $(\alpha, \mu, \xi)\in \mathbb{R}^+\times \mathbb{R}\times [-1,1]$ satisfy
$$
\alpha>\sqrt{2(1+\xi)} \quad \textrm{and}\quad \mu<\frac 12 +\frac 12\sqrt{1-\frac{2(1+\xi)}{\alpha^2}}\, .
$$
\end{enumerate}
Then almost surely, there exists $N$ large enough such that the system \eqref{eq:LCP} admits a unique solution $\bs{x}^*=(x_i^*)$, $x_i^*\ge 0$. Moreover, this equilibrium is globally stable in the sense that it is stable and 
$$
\forall\ \bs{x}(0)\in (0,\infty)^N,\quad \bs{x}(t) \xrightarrow[t\to \infty]{} \bs{x}^*\, .
$$
\end{theo}
We provide the proof under Assumption \eqref{ass:iid} hereafter. For the general proof, see \cite[Prop. 2.6]{clenet2022equilibrium}. 
\begin{proof}[Elements of proof] Let Assumption (\ref{ass:iid}) hold. We consider the condition of Theorem \ref{theo:TAKEUCHI-stability} with $\Delta=I$ and compute
\begin{equation}\label{eq:lyapunov-criterion}
I-\Gamma +I-\Gamma^* =2I -(\Gamma+\Gamma^*) = 2I - \frac{\sqrt{2}}{\alpha}\left( \frac{A_{ij} + A_{ji}}{\sqrt{2N}}\right)\, . 
\end{equation}
Notice that matrix $W=\left( \frac{A_{ij}+A_{ji}}{\sqrt{2N}}\right)$ is a symmetric matrix with independent centered entries on and above the diagonal with variance $\textrm{var}\left( \frac{A_{ij}+A_{ji}}{\sqrt{2}}\right)=1+\delta_{ij}$, where $\delta_{ij}$ is the Kronecker symbol with value 1 if $i=j$ and zero else. In RMT, Matrix $W$ is referred to as a Wigner matrix, and its properties are well-studied. In particular, it is known that its largest eigenvalue $\lambda_{\max} (W)$ behaves as follows:
$$
\lambda_{\max} (W) \xrightarrow[N\to\infty]{a.s.} 2\,,
$$
see for instance \cite[Theorem 5.1]{bai2010spectral}. Going back to \eqref{eq:lyapunov-criterion}, we get:
$$
I-\Gamma +I-\Gamma^* = 2I - \frac{\sqrt{2}}{\alpha} W$$
and the smallest eigenvalue of this matrix is 
$$
\lambda_{\min}\left(2I - \frac{\sqrt{2}}{\alpha} W\right) = 2 - \frac{\sqrt{2}}{\alpha} \lambda_{\max}(W)\xrightarrow[N\to\infty]{a.s.} 2\left( 1-\frac{\sqrt{2}}{\alpha}\right)\, . 
$$
Taking $\alpha>\sqrt{2}$ yields the desired result.

\end{proof}

It is worth noticing that
the two conditions are sufficient to establish the theorem but simulations indicate that these conditions  are not tight, i.e. one could observe a unique equilibrium for $\alpha$ smaller than the bounds above.

\begin{remark}
In a similar vein, Champagnat et al. \cite{champagnat2010convergence} provide a variant to Takeuchi and Adachi's result (Theorem \ref{theo:TAKEUCHI-stability}) with a similar RMT interpretation. 

Consider Model \eqref{eq:LV} and assume first that matrix $\Gamma$ is such that there exists a positive diagonal matrix $\Delta$ such that $\Gamma\Delta=\Delta \Gamma$. Assume moreover that $\Delta(I- \Gamma)$ is positive definite. Then there exists a unique stable equilibrium to \eqref{eq:LV}. Notice that this set of assumptions is (a.s. eventually) satisfied in a RMT context with $\Delta=I$, $\Gamma=\frac{W}{\alpha}$ where $W$ is a Wigner matrix, and $\alpha>2$. 
\end{remark}





\subsection{Unique feasible equilibrium.} \label{subsec:feasibility}
Once the conditions are met so that the LV system of coupled equations admits a unique equilibrium, a natural question arises: is this equilibrium feasible, in the sense that $\bs{x}^*>0$ componentwise? A negative answer has been brought by Dougoud et al. \cite{dougoud2018feasibility} in the case where $\alpha$ is fixed. As a first conclusion, getting feasibility requires $\alpha=\alpha_N \nearrow \infty$. This implies a qualitative change in the nature of the interactions since the random part of the interaction matrix would have a (macroscopic) vanishing effect:
$$
\rho(\Gamma) = {\mathcal O}\left( \frac 1{\alpha_N}\right)\to 0 \quad (\textrm{i.i.d. model})\ ,\quad 
\rho(\Gamma) = |\,\mu\,|\, +\, {\mathcal O}\left( \frac 1{\alpha_N}\right)\quad (\textrm{elliptical model})\ ,
$$
where \gls{bigO} stands for the standard big O notation.
In Bizeul and Najim \cite{bizeul2021positive}, the feasibility threshold $\alpha_N \sim \sqrt{2\log(N)}$ is established. We first present the argument of Dougoud et al. \cite{dougoud2018feasibility}.

\subsubsection*{No feasibility if $\alpha$ is fixed.} Assume that the equilibrium point $\bs{x}^*$ is feasible, then $x_i^*>0$ for $i\in [N]$ and the equations \eqref{eq:equilibrium-general} are equivalent to the linear system
$$
(I-\Gamma)\bs{x}^*=\bs{r}.
$$
Let $\Gamma$ be given by the i.i.d. model \eqref{eq:iid}. It is well-known in RMT that the spectral radius $\rho(A/\sqrt{N})$ almost surely converges to 1 (see e.g. \cite{geman1986spectral}). As a consequence, for every $\alpha>1$, $\rho(\Gamma) <1$ 
eventually (i.e. almost surely for large $N$) and matrix $(I-\Gamma)$ is almost surely invertible for large $N$. Hence, the following algebraic representation of the equilibrium:
\begin{equation}\label{eq:equilibrium-algebraic}
\bs{x}^*=\left( I - \frac{A}{\alpha\sqrt{N}}\right)^{-1} \bs{r}\, .
\end{equation}
In the simpler case where $\bs{r}=\bs{1}_N$, the $N\times 1$ vector of ones, Geman and Hwang \cite{geman1982chaos} have proved that asymptotically, for every finite $M$,
$$
(x_1^*,\cdots, x_M^*)\xrightarrow[N\to\infty]{\mathcal L} {\mathcal N}_M(\bs{1}_M, \sigma_{\alpha}^2 I_M)\, ,
$$
where $\sigma_{\alpha}^2 = \frac{1}{4\alpha^2 - 1}$ depends on $\alpha$, \gls{Mgaussian} is the multivariate normal distribution with mean $\mathbf{a}$ and covariance matrix $C$,
and \gls{convlaw} stands for the convergence in distribution. As a consequence, Dougoud et al. \cite{dougoud2018feasibility} argued that under this interaction regime (fixed $\alpha$) observing a feasible equilibrium was unlikely. In fact, the theoretical result by Geman and Hwang \cite{geman1982chaos} asserts that each component $x_i^*$ of the equilibrium asymptotically behaves as an independent Gaussian random variable centered at 1, with a variance independent from $N$, hence the heuristics
$$
\mathbb{P}(\min_{i\in [N]} x^*_i>0) \simeq \prod_{i\in [N]} \mathbb{P}(x^*_i>0) \xrightarrow[N\to\infty]{} 0\, .
$$
Otherwise stated, the initial assumption that $\bs{x}^*>0$ is very unlikely to happen and is asymptotically a large deviation.
This a priori analysis motivates the study of a feasible equilibrium under the regime $\alpha=\alpha_N\xrightarrow[N\to\infty]{} \infty$.

\subsubsection*{Feasibility when $\alpha_N$ grows to infinity.} In the case where $\bs{r}=\bs{1}$, there is a sharp phase transition around the threshold value $\alpha^*_N\sim\sqrt{2\log(N)}$ for both models \eqref{eq:iid}-(ii) and \eqref{eq:elliptic-extra-norm}. Below the threshold, there is no feasibility with very high probability while above the threshold, feasibility occurs with probability growing to 1.

\begin{theo} \label{theo:faisibility-iid-model} Let $\Gamma$ be either given by model \eqref{eq:iid}-(ii) or \eqref{eq:elliptic-extra-norm}; in the latter case assume moreover that $\mu<1$. Assume $\alpha_N \xrightarrow[N\to\infty]{} \infty$ and denote by $\alpha^*_N=\sqrt{2\log(N)}$. Let $\bs{r}=\bs{1}$. 

Then $\bs{x}^*$ given by \eqref{eq:equilibrium-algebraic} is well-defined and 
\begin{enumerate}
    \item If $\exists\, \varepsilon>0$ such that $\alpha_N\le (1-\varepsilon) \alpha^*_N$ then $\mathbb{P}(\min_{i\in [N]} x^*_i>0)\xrightarrow[N\to\infty]{} 0$ ,
    \item If $\exists\, \varepsilon>0$ such that $\alpha_N\ge (1+\varepsilon) \alpha^*_N$ then $\mathbb{P}(\min_{i\in [N]} x^*_i>0)\xrightarrow[N\to\infty]{} 1$ .
\end{enumerate}
\end{theo}
Although the full proofs of the theorem are involved, a simple heuristics captures the phase transition in the i.i.d. case and is presented hereafter.
\subsubsection*{Remarks}
\begin{itemize}
\item These results are established in \cite{bizeul2021positive} and \cite{clenet2022equilibrium} for the i.i.d. case and the elliptic case respectively.
\item In the case where vector $\bs{r}$ is still positive componentwise but different from $\bs{1}$, there is not a sharp threshold at $\alpha^*_N=\sqrt{2\log(N)}$ but rather a \textit{transition buffer} $[\alpha_{\min,N}^*,\alpha_{\max,N}^*]$ from non-feasibility ($\alpha_N< \alpha_{\min,N}^*$) to feasibility ($\alpha_N> \alpha_{\max,N}^*$). Details can be found in \cite[Section 4.2]{bizeul2021positive}.
\end{itemize}

\begin{proof}[A heuristics to understand the phase transition in Theorem \ref{theo:faisibility-iid-model}]
Assume that $\Gamma$ is given by model \eqref{eq:iid}-(ii). In the representation \eqref{eq:equilibrium-algebraic} of the equilibrium, expand the inverse matrix as a Neumann series and only consider the first order expansion:
$$
\bs{x}^* =\left( I - \frac{A}{\alpha_N \sqrt{N}}\right)^{-1} \bs{1}
= \bs{1} +\frac{A}{\alpha_N \sqrt{N}} \bs{1} + \cdots
$$
Every component $x_k^*$ of $\bs{x}^*$ writes 
$$
x_k^* = 1+\frac{Z_k}{\alpha_N} +\cdots \quad \text{where}\quad 
Z_k=\frac{\sum_{j\in [N]} A_{kj}} {\sqrt{N}}\, .
$$
Notice that the $Z_k$'s are i.i.d. ${\mathcal N}(0,1)$. Going one step further in the approximation and taking the minimum yields
$$
\min_{k\in [N]} x_k^* \simeq 1+\frac{\min_{k\in [N]} Z_k}{\alpha_N}\, .
$$
Now standard results from extreme value theory yield $\min_{k\in [N]} Z_k \sim -\sqrt{2\log(N)}$, from which we deduce
$$
\min_{k\in [N]} x_k^* \simeq 1-\frac{\sqrt{2\log(N)}}{\alpha_N}\, .
$$
The relative position of $\alpha_N$ with respect to $\alpha_N^*= \sqrt{2\log(N)}$ yields the desired result.
\end{proof}

\subsection{Unique equilibrium with vanishing species} \label{sec:unique-equilibrium+vanishing}
In the case of a unique equilibrium with species vanishing {when $t\rightarrow +\infty$}, it is interesting to understand some properties of the survivors such as the individual distribution of the abundance of a given species, the number of vanishing species, etc. Various techniques (such as the replica method from theoretical physics) yield quantitative heuristics validated by simulations.
A full mathematical analysis remains currently out of reach.

\subsubsection{Number of vanishing species.} \label{sec:vanishing} We mentioned earlier that should the parameter $\alpha$ (related to the strength of the interaction) be constant or less that $\sqrt{2\log(N)}$, the equilibrium $\boldsymbol{x}^*$ will feature vanishing components $x^*_i=0$ representing disappearing species.  

In this section, we address the question of estimating the proportion of surviving species $p=p(\alpha)$ as a function of parameter $\alpha$. In \cite{bunin2017ecological}, Bunin provides a heuristics based on the cavity method to address this question, while in \cite{galla2018dynamically}, Galla establishes equations comparable to those of Bunin based on dynamical generating functionals techniques. Both heuristics apply for the elliptical model \eqref{eq:elliptic-extra-norm}. For the i.i.d. model, a simple order statistics argument 
can be found in \cite{clenet2022preprint,clenet2022gretsi}. These a priori different methods yield the same equations from which one can extract $p(\alpha)$.

Given the random equilibrium $\bs{x}^*$, we introduce the following quantities:

$$
{\mathcal S}=\{ i\in [N], x_i^*>0\}\,,\quad \hat{p}=\frac {|\Sur|}{N}\,,\quad \mhat^2 =\frac 1{|\Sur|} \sum_{i\in [N]} (x_i^*)^2\,.
$$
Denote by $Z\sim{\mathcal N}(0,1)$ a standard Gaussian random variable and by $\Phi$ the cumulative Gaussian distribution function:
$$
\Phi(x)=\int_{-\infty}^x \frac{e^{-\frac{u^2}2}}{\sqrt{2\pi}}\, du\, .
$$

\begin{heur}[Bunin \cite{bunin2017ecological}, Galla \cite{galla2018dynamically}, Clenet et al. \cite{clenet2022preprint,clenet2022gretsi}]
\label{heuristics:surviving-species} Let $\alpha\in \left(\sqrt{2},\sqrt{2\log(N)}\right)$ and assume that $\Gamma$ follows Model \eqref{eq:iid}. The following system of two equations and two unknowns $(p,m)$

\begin{align}\label{eq:heur1}
& m\, \sqrt{p}\, \Phi^{-1}(1-p) +\alpha = 0\,,\\
\label{eq:heur2}
& 1+\frac{2m \sqrt{p}}{\alpha}\mathbb{E}\left(Z\mid Z>-\frac{\alpha}{m\sqrt{p}}\right)
+ \frac{m^2 p}{\alpha^2} \mathbb{E} \left(Z^2\mid Z> -\frac{\alpha}{m\sqrt{p}}\right) =m^2
\end{align}
admits a unique solution $(p^*,m^*)$ and the following convergence holds
 $$
 \hat{p}\xrightarrow[n\to\infty]{a.s.} p^*\qquad \text{and}\qquad \mhat\xrightarrow[n\to\infty]{a.s.} m^*\, .
$$ 
\end{heur}

\begin{remark}
Notice that the condition $\alpha>\sqrt{2}$ guarantees by Theorem \ref{th:unique-LCP} that a.s. eventually there exists a unique equilibrium. This condition is sufficient but might not be necessary. In the simulations hereafter, we also test the case where $\alpha\in (1,\sqrt{2}]$ and observe that with high probability, there exists a unique equilibrium and a good matching with equations \eqref{eq:heur1}-\eqref{eq:heur2}.  
\end{remark}
\textit{Simulations.} We fix $N=1000$ and draw $L$ independent realizations of matrices $A^{(i)}$. We then compute the corresponding equilibria $\bs{x}^{*(i)}(\alpha)$ and their related quantities $(\hat{p}^{(i)}(\alpha),\mhat^{(i)}(\alpha))$ for a given $\alpha>0$. We finally compare the empirical Monte Carlo averages: 
$$
\hat{p}_{L}(\alpha)=\frac 1L \sum_{i=1}^L \hat{p}^{(i)}(\alpha)\quad \text{and}\quad \mhat_L(\alpha)=\frac 1L \sum_{i=1}^L \mhat^{(i)}(\alpha)
$$
to their theoretical counterparts $p^*(\alpha),m^*(\alpha)$, solutions of \eqref{eq:heur1} and \eqref{eq:heur2}. 

For $\alpha\in (\sqrt{2},\sqrt{2\log(1000)})$, we consider $L=500$ repeated samplings. For $\alpha\in (1,\sqrt{2}]$, we take $L=100$. 

\begin{figure}[htb]
\begin{minipage}[b]{0.98\linewidth}
  \centering
  \centerline{\includegraphics[width=8.5cm]{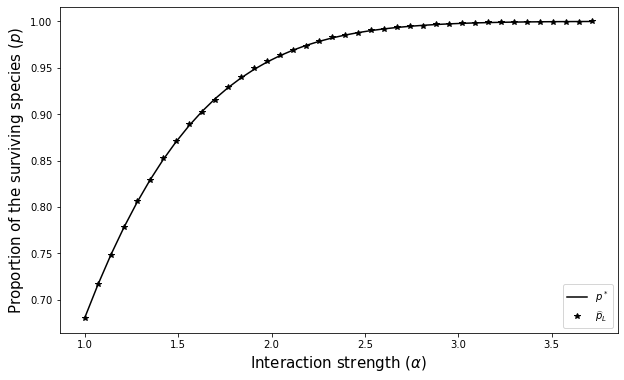}}
    \centerline{\includegraphics[width=8.5cm]{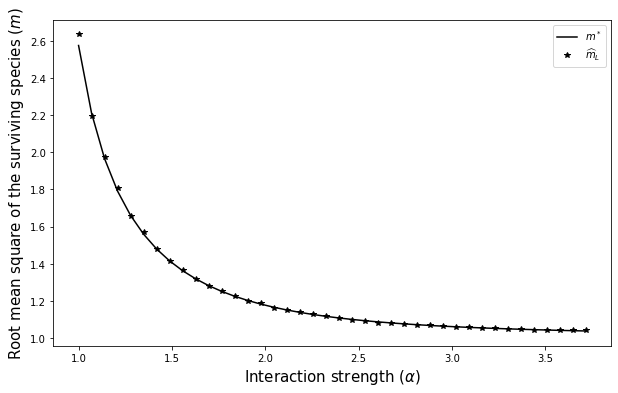}}
   \caption{The plot represents a comparison between the theoretical proportion of surviving species $p^*(\alpha)$ (up) and second moment $m^*(\alpha)$ (down) computed as solutions of \eqref{eq:heur1}-\eqref{eq:heur2}, and their empirical Monte Carlo counterpart $(\hat{p}_L(\alpha),\hat{m}_L(\alpha))$. The parameter $\alpha$ on the $x$-axis ranges from $1$ to $\sqrt{2\log(N)}\simeq3.71$. The quantity $\alpha=\sqrt{2}$ (and above) represents the theoretical lower bound provided by Theorem \ref{th:unique-LCP} that guarantees to have a stable equilibrium; the quantity $\alpha=\sqrt{2\log(N)}$ is the upper-limit above which we have no extinction $(p^*=1)$. Notice that for $\alpha\in (1,\sqrt{2}]$, the simulations show a remarkable matching with the heuristics despite no theoretical guarantee.} 
\label{fig:matching}
\end{minipage}
\end{figure}

As shown in Figure \ref{fig:matching}, the matching is remarkable, even for $\alpha$ below $\sqrt{2}$.

\subsubsection{Single species distribution.} 
The previous heuristics provides an estimation of the proportion of surviving species $p^*(\alpha)$. We go here one step further and describe the distribution of a given abundance $x_i^*$ where index $i$ corresponds to a surviving species. 

\begin{figure}
\centering
    \begin{subfigure}[b]{0.48\textwidth}
    \centering
    \includegraphics[width=\textwidth]{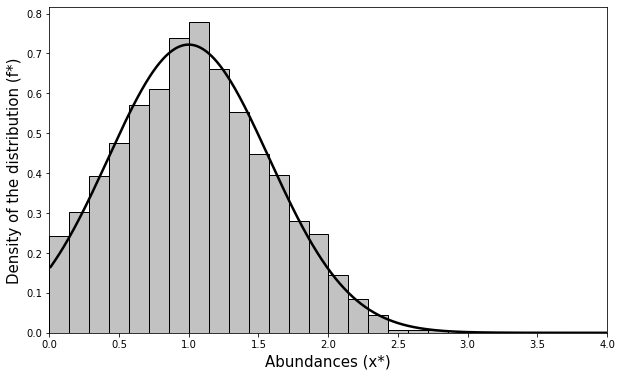}
    \caption{Gaussian entries.}
    \label{subfig:distrib_gaussian}
    \end{subfigure}
    \hfill
    \begin{subfigure}[b]{0.48\textwidth}
    \centering
    \includegraphics[width=\textwidth]{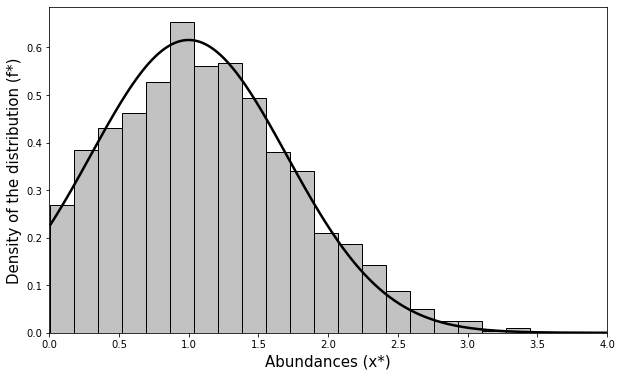}
    \caption{Uniform entries.}
    \label{subfig:distrib_uniform}
    \end{subfigure}
  \caption{Distribution of surviving species. The solid line represents the theoretical distribution $f^*$ as given by Heuristics \ref{heur:truncated-gaussian}. The histogram is built by solving the LCP problem with an interaction matrix of size $n = 2000$. In plot (A), the entries are Gaussian $\mathcal{N}(0,1)$ and the interaction strength is fixed to $\alpha = 2$. In plot (B), the entries are uniform $\mathcal{U}([-\sqrt{3},\sqrt{3}])$ which implies that the entries are centered with variance one, and the parameter is set to $\alpha = \sqrt{3}$.}
  \label{fig:histo}
\end{figure}

\begin{heur}\label{heur:truncated-gaussian}
Let $\alpha\in \left( \sqrt{2}, \sqrt{2\log(N)}\right)$ and let $i\in {\mathcal S}$, i.e. $i$ corresponds to a surviving species.
Let $p^*,m^*$ be the solutions of \eqref{eq:heur1} and \eqref{eq:heur2} and $Z\sim {\mathcal N}(0,1)$ a Gaussian random variable. Then the distribution of $x_i^*$ is a truncated Gaussian:
$$
{\mathcal L}(x_i^*) = {\mathcal L}\left( 1+\frac{m^*\sqrt{p^*}}{\alpha} Z \ \bigg{|}\  Z> - \frac{\alpha}{m^*\sqrt{p^*}}\right)\ .
$$
Otherwise stated, $x_i^*$ admits the following density:
$$
f^*(v) =  \frac{1_{(v>0)}}{\Phi(-\delta)} \, \frac{\delta}{\sqrt{2\pi}} \, \exp \left( - \frac{\delta^2(v-1)^2}{2} \right)
\qquad \textrm{where}\qquad \delta=\frac{\alpha}{m^*\sqrt{p^*}}
$$
and $\Phi$ stands for the cumulative Gaussian distribution.
\end{heur}
The matching between the theoretical density $f^*$ given in Heuristics \ref{heur:truncated-gaussian} and a histogram of a given equilibrium $\bs{x^*}$ is illustrated in Figure \ref{fig:histo}. In particular, the theoretical distribution matches, even  with non-Gaussian entries (see Fig. \ref{subfig:distrib_uniform}).

\subsubsection{Interactions between survivors}

When only a fraction of species survive in the unique LCP equilibrium, one can also ask how the interactions restricted to the survivors are modified. Mathematically, this boils down to consider the submatrix $(\Gamma_{ij})_{i,j \in \mathcal S}.$ Of course, the lines and columns that are selected depend on the initial realisation of the matrix $\Gamma$ and it is not an easy task to predict the new statistical features of the entries. Nevertheless, heuristics for these quantities have been given in~\cite{bunin2016interaction}, using the cavity method. 
These authors have obtained general formulas for the model  \eqref{def:ellipticalmodel} that can be found in~\cite{bunin2016interaction}, but for the sake of simplicity, we present it here in the case $\xi =0.$
\begin{heur}\label{heur:survinteractions}
Let $\Gamma$ be a non-centered Ginibre matrix, that is obeying model \eqref{def:ellipticalmodel} with $\xi = 0.$
Assume that $\alpha\in \left( \sqrt{2}, \sqrt{2\log(N)}\right)$ and, for any $i \in[N]$, $r_i=1.$ Then,  for any  $i, j\in {\mathcal S}$, i.e. $i,j$ correspond to surviving species, the entry $\Gamma_{ij}$ is still Gaussian but with the following bias and correlation :
\beqa  & \mathbb{E}[\Gamma_{ij}|\boldsymbol{x}^*] - \frac{\mu}{N} =- \dfrac{  x^*_i -1+ \mu \mean{\boldsymbol{x}^*}  }{N\mean{(\boldsymbol{x}^*)^2}}x^*_j \nonumber \\
& \mathrm{Corr}{\left[\Gamma_{ij},\Gamma_{ik}|\boldsymbol{x}^*\right]}= - \dfrac{ x^*_j x^*_k}{\mean{N(\boldsymbol{x}^*)^2}}, \label{eq:assembly_pattern} \eeqa
where, for any vector $\bs{x} = (x_i,\,i\in [N])$, , $\bs{x}^2=(x_i^2,i\in [N])$ and $\mean{\bs{x}} := \frac{1}{N} \sum_{i=1}^N x_i.$ 
\end{heur}

In~\cite{barbier2021fingerprints}, a slightly different \textit{maximum likelihood} point of view is adopted. The authors there consider the elliptic model \eqref{def:ellipticalmodel} as a prior distribution and compute the corresponding posterior distribution knowing the equilibrium, formulated as a linear constraint as in \eqref{lineareq}. The formulas~\eqref{eq:assembly_pattern} are recovered with this point of view.  
It is possible to study how these correlations impact spectral properties of the restricted matrix~\cite{barbier2021fingerprints,baron2022non}. Mathematical proofs for all the results in this paragraph remain out of reach.

\subsubsection{An overview of the results and some open questions}
In the next table, we summarize some of the results presented here and list a few open questions.
\begin{table}
\begin{center}
\begin{tabular}{|m{2.5cm}|m{1.5cm}|m{1.5cm}|m{1.5cm}|m{1.5cm}|}
\hline
$\boldsymbol{\alpha}$ & \multicolumn{2}{c|}{ $\boldsymbol{\alpha}$ \bf{fixed}} & \multicolumn{2}{c|}{ $\boldsymbol{\alpha=\alpha_N \nearrow\infty}$ }\\
\hline
\hline
\bf{value}\phantom{\bigg|} & \multicolumn{2}{c|}{$\boldsymbol{\sqrt{2}}$} & \multicolumn{2}{c|}{$\boldsymbol{\sqrt{2\log(n)}}$} \\
\hline
\bf{Equilibrium} & OQ? & \multicolumn{3}{c|}{\tt unique}\\
\hline
\bf{Feasibility} & OQ? & \multicolumn{2}{c|}{\tt no}& {\tt yes}\\
\hline
\bf{Single species distribution} & OQ? & \multicolumn{2}{c|}{\it{truncated gaussian}}& {\tt vanishing gaussian}\\
\hline
\bf{proportion of vanishing species} & OQ? & {\it provided by accurate heuristics} & & {\tt all species are  present}\\
\hline
\end{tabular}
\end{center}
\caption{The table above summarizes the different regimes (depending on $\alpha$) of the LV system \eqref{eq:LV} where $r_i=1$ and $\Gamma$ follows i.i.d. model \eqref{eq:iid}-$(ii)$ with standard Gaussian entries. By "OQ?", we mean open question; in {\tt typewriter font}, the result is proven mathematically; in {\it italics}, the result is established at a physical level of rigor. }
\label{tab:LV-versus-alpha}
\end{table}








   
    



\section{Structured models}
\label{sec:struct}

As illustrated in the previous sections, Large Random Matrices and RMT play a prominent role in the theoretical study of systems of particles in interactions such as ecosystems, foodwebs, etc. In this article, we have studied at large the LV system \eqref{eq:LV} of differential equations:
$$
\frac{\dd x_i}{\dd t} =x_i(r_i - x_i+(\Gamma \bs{x})_i)
$$
where matrix $\Gamma$, supposed random, has either i.i.d. entries or follows the elliptic model, see Section \ref{sec:models-typo-hypo}. Associated to the generic model: 
$$
\frac{\dd x_i}{\dd t} = x_i \varphi_i(\bs{x}),
$$
another line of research focused on a modelisation of the Jacobian of the system near equilibrium:
\begin{equation}\label{eq:jacobian-model}
 J = -I +M
\end{equation}
where $M$ is a random matrix, the question being then to understand the relative localisation of the spectrum of $M$ with respect to -1 to conclude on the stability of the system. This second approach is historically the first one with the influential paper (a detailled presentation of which is provided in Appendix \ref{ann:may_model}) by May \cite{may1972will} being one of its landmarks.  

In order to progress toward a more realistic description of the reality, one is tempted to consider more involved models of random matrices to take into account more properties of the complex systems such as sparsity, existence of underlying structures, randomness beyond independence, etc.  

For instance, the question of the effect of the structure of the ecological network on its feasibility and stability  already appeared in \cite{Pimm79}, following the work of May, where Pimm argues that connectance is not the only parameter that can influence the feasibility and stability of the networks and starts a theoretical study of structured (both deterministic and random) networks. 

In this section, we present a variety of random matrix models beyond the i.i.d. and elliptic ones, emphasizing on their use in theoretical ecology and listing mathematical results and questions of interest.. Often, mathematical results of interest are not available on the shelf for a direct use and massive simulations remain the main approach to exploit the potentialities of such models.

\subsection{An introduction to sparsity for ecological models}\label{subsec:intro-sparse}


Empirically, in an ecosystem with $N$ species, even if the maximal number of interactions is $N^2,$ the real number $L$ of nonzero interactions is often much smaller. For a LV system, this means that the interaction matrix $\Gamma$ has many entries equal to zero. We define the {\bf\textit{connectance}} as $C=L/N^2$.
Before starting the presentation of the models, let us clarify that the notion of sparsity used here is different from the usual one in mathematics, where matrices or networks are said to be \textit{sparse} when the  $L/N^2$ goes to zero with $N$. We will consider a wider range for the connectance (in particular when $C$ can be of order $O(1)$).

Although the interpretation is more doubtful in the jacobian modelization \eqref{eq:jacobian-model}, the connectance already appears in Gardner and Ashby's simulations \cite{GardnerAshby1970} and in May's work \cite{may1972will}. It is an important parameter to capture the sparsity of the models of interest.

In more recent works, Grilli et al. \cite{grilli2017feasibility} work explicitely on the interaction matrix of a LV system and study the stability and feasibility of the equilibrium as a function of various parameters, among which the connectance (see also \cite{marcus2021transition, dunne2002connectance}). Based on empirical evidence, Busiello et al. \cite{busiello2017explorability} suggest that foodwebs can actually be very sparse.

Beyond the connectance, it is possible to take into account the structure of the network by setting some of the entries to zero, thus enforcing an absence of interactions. For this, we may use a matrix $\Delta=(\Delta_{ij})$ where $\Delta_{ij}$ equals 1 if species $j$ has an effect on  species $i$ and 0 otherwise. If one draws the system interactions as a graph, then $\Delta$ can be interpreted as the adjacency matrix of this graph and the interaction matrix $\Gamma$ or the community matrix $M$ can then be represented as proportionnal to $\Delta \circ A$ where \gls{hadamard} represents the Hadamard product of matrices, that is:
$$
(\Delta \circ A)_{ij} = \Delta_{ij} A_{ij}
$$
and $A$ is random either i.i.d. or elliptic. In such a model, $\Delta$ represents the structure of the system and $A$ the (random) intensity of the interactions.

In the following subsections, we consider a number of sparse models, $\Delta$ being deterministic or random. We also refer to \cite{allesina2015stability} for a presentation of many models in connection with RMT.

\subsection{The simplest model for sparsity for ecological networks: Erdös-Rényi  graphs}\label{subsec:ER-graph}
When all species play the same role in the foodweb and
the only parameter of interest  is the average number of interactions for a given species,it is natural to choose $\Delta$ as the adjacency matrix of an \gls{ER} graph of size $N$ : each coefficient of the random matrix $\Delta$ has probability $p$ to be nonzero, equal to $1$, and probability $1-p$ to be put to zero, independently of the others. The average number of edges in the graph is $p N^2,$ hence the connectance $C$ equal to $p$.

\subsubsection*{ER in the mathematical literature}
ER graphs are reference models in the field of random graphs and their geometric properties have been extensively studied (see e.g. \cite{bollobas2001,durrett,vanderhofstad}). 
The spectral properties of their adjacency matrices have also been studied. In the regime when $C= O(1),$ which is called \textit{dense}
by mathematicians, the ER matrix is a rank one deformation of a matrix with centered i.i.d. entries, so that we observe a circular law and one outlier. In fact,
$$
\frac{\Delta}{\sqrt{N}} = \frac 1{\sqrt{N}} (\Delta - \mathbb{E} \Delta) + \frac 1{\sqrt{N}} \mathbb{E} \Delta\quad \textrm{with} \quad \left\| \frac 1{\sqrt{N}} \mathbb{E} \Delta\right\| = \sqrt{N}\, C\, ,
$$
where \gls{snorm} refers to the spectral norm when applied to a matrix. Notice that the precise understanding of the extreme eigenvalues of $\frac{\Delta}{\sqrt{CN}}$ in sparse or very sparse regimes is still an active subject in RMT. A concise overview can be found in the introduction of \cite{AltDucKnow21}.
\subsubsection*{ER in the ecological literature : sparsity increases stability}
As developed in Appendix \ref{ann:may_model},
the case when $M_{ij} = \Delta_{ij} A_{ij},$ with $\Delta$ the adjacency matrix of a dense ER graph and $A$ has i.i.d. centered entries with variance $V$ has been already considered by May. This sparse model is equivalent to the full model, where the entries have variance $CV$ and in this case sparsity increases stability: in fact, the stability condition $NVC<1$ is easily satisfied for small $C$. 

\begin{figure}[h]

 \begin{subfigure}{.3\linewidth}\label{subfig_ER:PP}
    \includegraphics[scale=0.3]{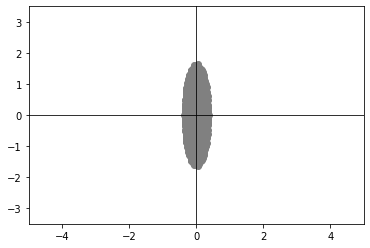}
    \caption{Predator-prey.}
  \end{subfigure}%
  \hspace*{\fill}   
  \begin{subfigure}{0.3\textwidth}\label{subfig_ER:Competitive}
   \includegraphics[scale=0.3]{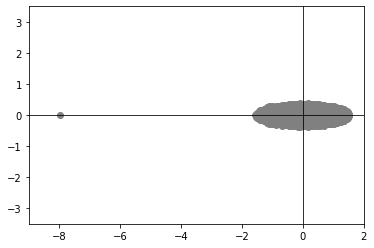}
   \caption{Competitive.}
 \end{subfigure}%
 \hspace*{\fill}   
  \begin{subfigure}{0.3\textwidth}\label{subfig_ER:Mutualistic}
   \includegraphics[scale=0.3]{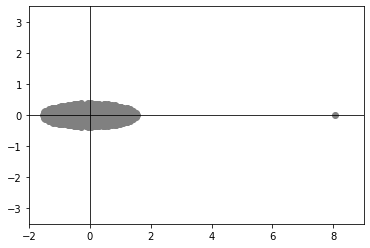}
   \caption{Mutualistic.}
 \end{subfigure}%
  
 \caption{Spectrum of the interaction matrix for $N=1000$ species. $\Delta$ is ER symmetric with $C=0.1$.  For the competitive (resp. mutualistic) model, $A_{ij} = A_{ji}$ with distribution $-|\mathcal{N}(0,1)|$  (resp. $|\mathcal{N}(0,1)|$) variables.  For predator-prey $A_{ij} = - A_{ji},$ distribution $|\mathcal{N}(0,1)|.$ }\label{fig:ER}
\end{figure}

The case when $\Delta$ is the adjacency matrix of an ER graph but the model for the matrix 
$A$ is more involved has been studied in particular in \cite{allesina2012stability}. They use models for $A$ that are of the same flavour as the elliptic model - for example, $(A_{ij}, A_{ji})_{i <j}$ both positive to model  mutualistic systems or with opposite sign to model a prey-predator situation. As illustrated in Figure \ref{fig:ER}, in the mutualistic case, outliers with a large real part may strongly affect the stability. 
In \cite{allesina2012stability}, the authors also establish an explicit stability criterion adapted to each case, generalizing May's criterion and  emphasize  again that sparsity increases stability.


\subsection{Sparsity with a deterministic structure}


An alternative to the ER model is the case where matrix $\Delta$ is deterministic.

Consider a $d$-regular oriented graph with $N$ vertices, that is a graph where each vertex $i$ has exactly $d$ oriented edges exiting from $i$ and $d$ edges coming to $i$. Let $\Delta$ be the adjacency matrix of such a graph, then 
$\Delta$ has $d$ non-null entries per row and per column and $L:= d \times N$ non-null entries overall. Parameter $d$ which may depend on $N$ accounts for the sparsity of the system and in the framework of a LV system, consider the interaction matrix:
$$
\Gamma= \frac 1{\alpha} \frac{\Delta \circ A}{\sqrt{d}}  =\frac 1{\alpha} \left( \frac{\Delta_{ij} A_{ij}}{\sqrt{d}} \right)_{ij}\ ,
$$
where the $A_{ij}$'s are i.i.d. and $\alpha$ is an extra normalization which may depend on $N$. Notice that the normalization is no longer $\sqrt{N}$ but $\sqrt{d}$ accounting for the fact that there are exactly $d$ non-null entries per row. For such a model the connectance $C$ equals:
$$
C=\frac{d}N
$$
and the interest lies in "small" values of $d$.

This model has been studied by Akjouj and Najim in \cite{akjouj2021feasibility} where specific assumptions on $d$ and $\Delta$ are considered, namely either $d$ is proportional to $N$, or $d\gg \log(N)$ and $\Delta$ has a specific block structure, cf. Model (A) in \cite{akjouj2021feasibility} and Appendix \ref{app:deterministic_model}. In this article, it is shown that the same phase transition as in Theorem \ref{theo:faisibility-iid-model} occurs: Feasibility and stability hold iff $\alpha=\alpha_N \gg \sqrt{2\log(N)}$.

\begin{figure}[h!]

 \begin{subfigure}{.46\linewidth}\label{subfig_Reg:A}
    \includegraphics[scale=0.5]{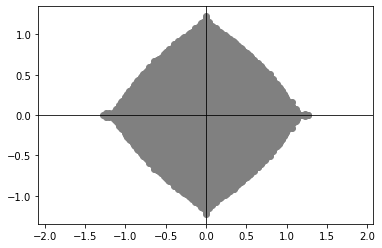}
    \caption{Spectrum of the normalized by $\frac{1}{\sqrt{d}}$ interaction matrix. }
  \end{subfigure}%
  \hspace*{\fill}   
  \begin{subfigure}{.46\textwidth}
   \includegraphics[scale=0.5]{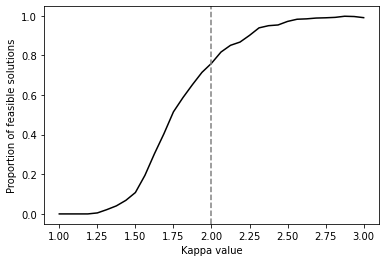}
   \caption{Proportion of feasible equilibrium.}\label{subfig_Reg:B}
 \end{subfigure}%
  
 \caption{Deterministic model with $N=1000$ species and $d=10$. Each species interacts with $d=10$ species. The existing interaction are i.i.d. Gaussian $\mathcal{N}(0,1)$ entries. In figure \ref{subfig_Reg:B}, each point represents the proportion of feasible solutions $\boldsymbol{x}$ over 200 realizations of random matrices $\Gamma_N$ for different values of $\kappa$, with $\alpha_N = \sqrt{\kappa \log(N)}$.}   \label{fig:Reg}
\end{figure}

The spectrum of matrix $\frac{\Delta\circ A}{\sqrt{d}}$ together with the proportion of equilibria near the phase transition thershold are plotted in Figure \ref{fig:Reg}.

\subsection{Introducing modularity through Stochastic Block Model (SBM)}

Beyond the Erdös-Rényi case, when every species equally interacts with any other, it is often more realistic to consider that there exist within the ecosystem \textit{communities} (also called {\bf \textit{modules}}), that is groups of species sharing the same connexion patterns. 
This leads to the celebrated  \gls{SBM}, introduced in  \cite{hollandlaskeyleinhardt}, (see also \cite{abbe,leewilkinson} for reviews). Let $r\in \NN$ be the number of communities. Given
\begin{itemize}
    \item a vector of positive real numbers $(\pi_1,\cdots, \pi_r)$ such that $\sum_{i=1}^r \pi_i=1$,
    \item an $r \times r$ matrix $P,$
\end{itemize}
the corresponding SBM is a random graph whose vertices are partitioned into $r$ communities $C_1, \cdots, C_r,$ where each node belongs to the community $C_i$, $i\in \{1,\cdots, r\}$, with probability $\pi_i$. Then, an edge between a vertex $u \in C_i$ and a vertex $v \in C_j$ exists 
with probability $p_{ij},$ independently of all other edges.

\subsubsection*{SBM in the mathematical literature}
There exists a huge literature on the SBM, initially introduced to analyze social networks, and extensively used in machine learning for modelling complex networks and address the community detection problem. 
The goal there is to design algorithms to cluster the different communities and estimate accurately matrix $P$, see for example \cite{matias2017clustering, baskerville2011bayesian}. 

Again using the Hadamard product $\Delta \circ A$, the spectrum of the adjacency matrix $\Delta$ associated to a SBM can be described, at least in simple cases. Consider for example a SBM with $r=2$ communities of equal size ($\pi_1= \pi_2 = 1/2$) and let 
$$
\begin{pmatrix} p & q \\ q & p \end{pmatrix},
$$ with $p$ and $q$ of order $0(1)$ (dense case). Then $\Delta$ is a rank-two perturbation of a matrix with centered independent entries. Depending on the values of $\frac{p+q}{2}$ and $\frac{p-q}{2},$ there can be up to two outliers in its spectrum. As in the ER case, sparse cases has also been recently considered, see e.g. \cite{BBGK17}.


\subsubsection*{SBM in the ecological literature : modularity increases stability}

In the seventies, May and Pimm already considered rudimental forms of the SBM into the framework of the Jacobian model \eqref{eq:jacobian-model}, to take into account some features of ecosystems such as modularity and {\it \textbf{compartmental models}}. 

\begin{figure}[h!]

 \begin{subfigure}{.46\linewidth}\label{subfig_SBM:A}
    \includegraphics[scale=0.5]{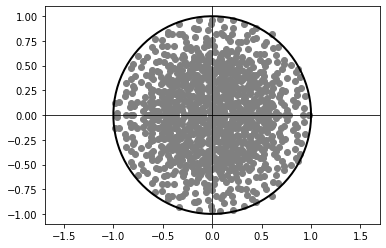}
    \caption{Spectrum of the normalized by $\frac{1}{\sqrt{(c_1+q)N}}$ interaction matrix, where $c_1=0.5>c_2=0.2>q=0.02$. When there exists, the interactions are i.i.d. Gaussian $\mathcal{N}(0,1)$.}
  \end{subfigure}%
  \hspace*{\fill}   
  \begin{subfigure}
  {.46\linewidth}
  \label{subfig_SBM:A2}
    \includegraphics[scale=0.5]{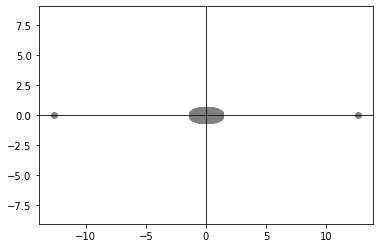}
    \caption{Spectrum of the normalized by $\frac{1}{\sqrt{qN}}$ interaction matrix for a bipartite model, where $q=0.5$ and $c_1=c_2=0$. $\Delta$ is symmetric and $A$ has i.i.d $|\mathcal{N}(0,1)|$ entries.}
  \end{subfigure}%

 \caption{\footnotesize{SBM with two communities of $500$ species, $N=1000$. The probability of interactions inside the first (resp. second) community is $c_1$ (resp. $c_2$) and the probability of interaction with species of the other community is equal to $q$.} }   \label{fig:SBM}
\end{figure}

In \cite{may1972will}, May presents a simple occurrence of SBM. He considers a SBM with $r$ modules and a probability vector $(c_1, \ldots, c_r)$. This SBM corresponds to modules with no interactions,  while within the $i$th block
made of $d_i$ species, the interactions behave like an ER graph with connectance $c_i$ and variance $V_i$.
The stability condition reads : $$\max_{i \in [r]} c_iV_id_i < 1,
$$ hence modularity increases stability.
This phenomenon is illustrated in Figure \eqref{fig:SBM}.

In \cite{Pimm79}, Pimm addresses the following question \textit{"should model systems be organized into compartments of species characterized by strong interactions within compartments, but weak interactions among the compartments ?"} A random version of such a model would correspond to a SBM with a matrix $P$ having large diagonal coefficients and small off-diagonal ones.



 
More recently, the effect  of modularity on the stability of the networks is extensively explored in  \cite{thebaultfontaine} in the framework of a tamed version of LV equations. They evaluate modularity through an index, introduced by \cite{barber2007modularity} and, through simulations, illustrate that persistence, that is the number of surviving species, increases with modularity in trophic networks (see also \cite{stoufferbascompte2011}) but decreases with modularity in mutualistic networks. It would be interesting to investigate whether mathematical results on SBM could help to understand their observations.


The question of stability is for example important for plant-pollinator ecosystems. The latter correspond to bipartite mutualistic networks (see e.g. \cite{thebaultfontaine,billiardlemanreytran}). In \cite{billiardlemanreytran} in particular, the evolution of abundances can be approximated, when the number of species $N$ tends to infinity, by a kinetic integro-differential equation where the dense graphs are replaced by {\bf \textit{graphons}}. The theory of graphons is mathematically well developed  but  beyond the scope of this review (see for example \cite{lovasz}).



\subsection{Nested models : a few generalist and many specialist species.}

In the Erd\"os-Rényi model or in SBM, the network is determined by considering the absence or presence of edges for each pair of vertices independently of the others. Other models of random graphs are defined by specifying the degree distributions. For example in the configuration models (also known as Molloy-Reed-Bollob\'as, see e.g. \cite{bollobas2001,molloyreed,vanderhofstad}), independent random variables distributed with the target degree distribution are associated to each vertex and edges are formed by pairing at random the half-edges. 

By choosing heavy-tailed degree distributions, one can thus create a few vertices with very high degrees (corresponding to generalist species) and a majority of vertices with low degree (corresponding to specialists). Such ecosystems are called {\bf \textit{nested}}. They have also been modelled and studied, at least through simulations. This idea has been implemented in \cite{thebaultfontaine} following \cite{okuyamaholland}.

Nested ecosystems can also be described through random graphs with given \textit{expected degrees}. This model is known as the \textit{Chung-Lu model} : take a deterministic sequence $w=(w_1, \ldots, w_n),$ that will correspond to the expected degrees and draw an edge between vertex $i$ and vertex $j$ with probability $w_iw_j / \sum_{i=1}^n w_i$ independently of all other edges. If we choose all the weights to be equal  to $pn$, we are obviously back to the ER model with connectance $p$ but nested ecosystems can be modeled by choosing a power-law distribution for the weights, that is $w_i= c i^{-\frac{1}{\beta -1}},$ for $i$  greater or equal to some $ i_0.$ In this case, the number of species interacting with $k$ others is proportional to $k^{-\beta}.$.
The spectrum of the adjacency matrix of such a graph has been studied in \cite{ChuLuVu2003} where they point out that a phase transition occurs at $\beta =2.5$ : for $\beta > 2.5,$ the largest eigenvalue behaves like $\sqrt m,$ which is the maximal degree in the graph, whereas for $\beta < 2.5,$ the largest eigenvalue behaves like $\bar d, $ which is the weighted average of the square of the expected degrees. It would be interesting to investigate whether these mathematical results could be effectively used in the study of nested ecosystems.

\subsection{Kernel matrices}

\paragraph{\textit{Definition of the model}.}
Part of the literature on ecological networks considers that the interactions between two species depends on a distance between their respective values of some functional traits.
The examples that we will present below, fit into the mathematical framework of \textbf{kernel matrices}. We have 
\[ \Gamma_{ij} \textrm{ or } M_{ij}= f(g(X_i,X_j)),\]
where $X_i$ is a vector modelling the traits of species $i,$ $g: \RR^p \times \RR^p \longrightarrow \RR$ is a symmetric matrix, denoted as the kernel (corresponding to the measure of the distance), and  $f : \RR  \longrightarrow \RR$ a function called the envelope.
Typical examples are $g(x,y) = x^Ty,$ or $\|x-y\|^2$ and $f(x)= \exp(ex)$ or $(1+x)^a$ etc.

\paragraph{\textit{Kernel matrices in the mathematical literature}.}
Among these models, the first interesting and well studied case is the so-called Wishart case\footnote{That is the empirical covariance matrix of the vectors $X_1, \ldots, X_N.$}, when $g(x,y)= x^T y$ and $f(x)=x.$ 
If the entries of the vectors are i.i.d. centered and normalized, then it is well known (see \cite{marcenkopastur1967math}) that, if the ratio $p/N$ of the number of traits over the number of species  converges to $\tau,$ the spectrum
converges almost surely to the Marcenko-Pastur distribution. We will develop the mathematical theory of kernel matrices in Appendix \ref{appendix:kernel}. The main message is that in the RMT regime, that is when the number of traits is large and of the same order as the number of species, and if $g$ and $f$ are reasonable,
they have almost no influence on the spectrum, meaning that, in the models, "any" kernel matrix could be without harm  replaced by a Wishart matrix or a Gaussian kernel matrix.



\paragraph{\textit{Kernel matrices for ecological networks}.}

Dieckmann and Doebeli consider a simple co-evolutionary model \cite{roughgarden, dieckmanndoebeli, champagnatferrieremeleard}. The birth rate of an individual with trait $x$ is $b(x)=\exp(-x^2/2\sigma_b^2)$, the individual natural death rate is constant $d(x)=d_C$, and the competition between two individuals with traits $x$ and $y$ is $C(x,y)=\eta_c\,\exp(-(x-y)^2/2\sigma^2_c)$, $\sigma_c>0$. This would correspon to a Gaussian kernel (also used in \cite{meszena}).
In \cite{NuJoBa}, they develop \textit{phenotype matching model}, where the interaction is  stronger when the traits of two species are close but also \textit{phenotype difference ( or threshold) model}, in which successful interaction depends on the degree to which the trait of the second species surpasses the trait of the first one\footnote{ One can think of  fruit and beak sizes in a plants-birds interaction network.} (see also  \cite{kisdi, andreazzi}).
Other models involving thresholds can be found in \cite{santamaria} and in \cite{rohrmazza},
the models involving a combination of characteristics of the species taken separately and a measure of the similarity between the traits.

We end this paragraph by mentioning the work of \cite{servan}, which lies in the LV framework, with a kernel matrix $\Gamma$ for which the distance between the traits,  is determined through a distance between species in their phylogenetic tree. He addresses the questions of feasability and stability as we have detailed it for the elliptic case and explicitely uses the link with Wishart matrices mentioned above. It would be interesting to investigate
whether this point of view can be fruitful in other contexts.



\section{Other community models with noise}\label{sec:otherextensions}

 The LV equations \eqref{eq:LV} can arise as limits of \gls{IBM}, see e.g. \cite{champagnatferrieremeleard}, when population sizes are large. We introduce a scaling parameter $K$ often referred to as a \textit{carrying capacity} and we assume that the $N$ species have initially sizes of order $K$. The individuals can give birth to new individuals of the same species or die. 
 More precisely, the natural birth and death rates of an individual of species $i\in[N]$ are $b^K_i$ and $d^K_i$ such that $b^K_i-d^K_i=r^K_i$. The competition pressure (or extra death rate) exerted by an individual of species $j$ on an individual of species $i$ is $\Gamma^K_{ij}$, if the latter is non positive. In case $\Gamma^K_{ij}$ is positive, it can be considered as an extra birth rate due to cooperation between the species $i$ and $j$. 
 Additionally an individual of species $i$ experiences an intra-specific death rate proportional to the size of its species (each individual experiences an extra death term due to the logistic competition and equal to the size of the population $i$ over $K$). The natural IBMs associated with \eqref{eq:LV} have two levels of stochasticity: (i) the interaction matrix $\Gamma^K$ is random, (ii) the occurrence of birth and death events is random. In all this section, we will work conditionally to $\Gamma^K$.

Individual-based models have long been used for simulations in Ecology \cite{deangelis2014individual,deangelis2018individual,ferriere2009stochastic,grimm2006standard}. See also \cite{giorgikaakailemaire,legendreZEN} for software performing IBM simulations. By presenting the fluctuations arising in the convergence of the IBM abundances to LV abundances, we provide a link between the equations considered in this review and these algorithms (refer to Appendix \ref{annexe:micro1} for more mathematical details).

Denote by $Y^K_i(t)$ the size of species $i$ at time $t$ and by $\bs{Y}^K(t)=(Y^K_1(t),\cdots, Y^K_N(t))$ the vector of all the species' sizes: forall $i\in [N]$,
\begin{equation}\lim_{K\rightarrow +\infty}\frac{Y^K_i(0)}{K}=x_i(0),\label{condition_initiale}
\end{equation}
in probability. 

The stochasticity of the birth and death events gives rise to an additional noise process compared to \eqref{eq:LV}. More precisely, $\bs{Y}^K(t)$ now satisfies the following SDE: for all $i\in [N]$,
\begin{eqnarray}
\dd Y^K_i(t)&=&\big(r^K_i-Y^K_i(t)-(\Gamma^K \bs{Y}^K(t))_i\big) Y_i^K(t)\ \dd t + \dd M^K_i(t)
\, ,\label{eq:EDS_Bell}
\end{eqnarray}
where $M^K_i$ is a martingale process, with 
\begin{eqnarray}
\mathbb{E}(M^K_i(t))&=&0\,,\nonumber \\ 
\label{eq:martingale}\ \Var(M^K_i(t))&=&\frac{1}{K}\mathbb{E}\Big(\int_0^t \big(b^K_i+d^K_i+Y^K_i(s)
+(\Gamma^K \bs{Y}^K(s))_i\big)Y^K_i(s)\ \dd s\Big)\,,
\end{eqnarray}
and $\Cov(M^K_i(t),M^K_j(t))=0$ (this equation is the analogue of \eqref{eq:poisson-martingale2} derived for $N=1$ in Appendix \ref{annexe:micro1}). \\

We now detail two different limits that can be derived from the IBM depending on the chosen parameters and rescalings: (i) the Lotka-Volterra ODEs \eqref{eq:LV} with a fluctuation process of Ornstein-Uhlenbeck type when $K\rightarrow +\infty$ without time rescaling; (ii) the Feller-type diffusions when the birth-death dynamics is nearly critical and when time is also rescaled by $K$, that is when considering the process at times $Kt$, $t\in [0,T]$.

\subsection{Large population limit and fluctuation around the ODE \eqref{eq:LV}}

First, we let $K\rightarrow +\infty$ without rescaling time and with a fixed number $N$ of species. 
We consider the rescaled process:
\[\bs{X}^K(t):=\frac{\bs{Y}^K(t)}{K}\,.\]
Here, for all $i,j\in [N]$,
\begin{equation}\label{scaling:edo}
b^K_i=b_i\,,\quad d^K_i=d_i\,,\quad \Gamma^K_{ij}=\frac{\Gamma_{ij}}{K}\,,
\end{equation}
where the quantities $b_i, d_i$ and $\Gamma_{ij}$ do not depend on $K$ and $r_i:=b_i-d_i.$ Notice that the competition term $\Gamma_{ij}^K$ can be understood as the extra death rate exerted by an individual of the species $j$ on an individual of the species $i$. When the population is large and of order $K$, the interaction between the individuals of a given pair are weaker and therefore the competition term is rescaled in $1/K$.\\

An averaging phenomenon appears (similar to what happens for the law of large numbers): from \eqref{eq:martingale}, we can see heuristically that the noise disappear and the evolution equations \eqref{eq:EDS_Bell} can be approximated by Lotka-Volterra ODEs (see e.g. \cite[Theorem 2.1 p.456]{ethierkurtz}, or for more generalizations to measure-valued processes in \cite{champagnatferrieremeleard,fourniermeleard,haegeman2011mathematical,chazottes2019time,bansayemeleard} for rigorous proofs). Indeed, the variance of the martingale part in $X^K_i$ is of order $1/K$ and converges to zero when $K\rightarrow +\infty$.

\begin{prop}\label{prop:LLN}
Assume \eqref{condition_initiale}, \eqref{scaling:edo} and assume that for all $i\in [N]$,
\begin{equation}\label{hyp:moment}\sup_{K\in \N}\EE\left[\left(X^K_i(0)\right)^3\right]<+\infty,\end{equation}
then, when $K\rightarrow +\infty$, the
abundance processes $(\bs{X}^K(t))_{t\geq 0}$ converges uniformly on every compact time interval $[0,T]$ with $T>0$ (as $\R^N$-valued processes) and in probability to population densities $(\bs{x}(t))_{t\geq 0}$, for $i\in [N]$, whose evolution is described by the system of ODEs \eqref{eq:LV}:
\beq \frac{\dd x_i}{\dd t}= x_i\left(r_i - x_i +(\Gamma \bs{x})_i \right),\qquad i\in [N].   \label{eq:LV_dimensions} \eeq
Put formally, this means that 
$$
\forall \varepsilon,T>0\,,\qquad \mathbb{P}\left\{ \sup_{t\le T}\left\|\bs{X}^K(t) -\bs{x}(t)\right\| >\varepsilon 
\right\} \xrightarrow[K\to\infty]{} 0\, .
$$
\end{prop}
We now consider the fluctuation process associated with this convergence:
\begin{equation}
    \bs{\eta}^K(t)=\sqrt{K} \left(\bs{X}^K(t)-\bs{x}(t)\right).
\end{equation}It is a $\R^N$-valued vector process whose $i$th coordinate is $\sqrt{K}(X^K_i(t)-x_i(t))$. Another reformulation is that the stochastic process can be expressed as:
$$
\bs{X}^K(t)=\bs{x}(t)+\frac{\bs{\eta}^K(t)}{\sqrt{K}}\,.
$$
Applying \cite[Theorem 2.3, Chapter 11]{ethierkurtz}, we obtain that:
\begin{prop}\label{prop:TCL}
Under the assumption of Proposition \ref{prop:LLN} and assuming that the following convergence in distribution holds:
\begin{equation}\label{hyp:TCL}
    \lim_{K\rightarrow +\infty}\bs{\eta}^K_0=\bs{\eta}_0 \in \R^N,
\end{equation}
then, when $K\rightarrow +\infty$, the process $(\eta^K(t))_{t\geq 0}$ converges in distribution, and for the topology of uniform convergence on $[0,T]$ for $T>0$, to the solution of the Ornstein-Uhlenbeck SDE:
\begin{multline}\label{eq:fluctuation-OU}
   \dd \bs{\eta}(t)=\Big(\bs{r}-2\bs{x}(t)-\Gamma \bs{x}(t)\Big) \circ \bs{\eta}(t)\, \dd t+ \bs{x}(t)\circ \Big(\Gamma  \bs{\eta}(t)\Big) \, \dd t \\
    + \mathrm{diag}\Big( \left(\bs{b}+\bs{d}+\bs{x}(t)+\Gamma \bs{x}(t) \right)\circ \bs{x}(t)\Big)\, \dd \bs{B}(t)\,,
\end{multline}
with initial condition $\bs{\eta}_0$ defined in \eqref{hyp:TCL}, $\bs{B}$ a standard $N$-dimensional Brownian motion, $\circ$ the Hadamard product and \gls{diagmatrix} the diagonal matrix with diagonal entries the components of the vector in the argument. Equivalently, the componentwise definition of $\bs{\eta}=(\eta_i)$ is given for all $i\in [N]$ by:
\begin{multline*}
    \dd \eta_i(t)=\Big(r_i-2x_i(t)-(\Gamma \bs{x})_i(t)\Big)  \eta_i(t)\, \dd t+ x_i(t) \Big(\Gamma  \bs{\eta}(t)\Big)_i \, \dd t \\
    + \Big(b_i+d_i+x_i(t)+(\Gamma \bs{x})_i(t) \Big)\,  x_i(t)\, \dd B_i(t)\,.
\end{multline*}
\end{prop}


Conditionally on matrix $\Gamma$, the solution $\bs{x}(t)$ of \eqref{eq:LV} is deterministic and hence the stochastic differential equation \eqref{eq:fluctuation-OU} is of Ornstein-Uhlenbeck type, i.e. 
$$
\dd\bs{\eta}(t)=A(t)\bs{\eta}(t)\dd t+\Sigma(t) \dd\bs{B}(t)
$$ 
where
\begin{eqnarray*}
    A(t)&= & \mathrm{diag}\Big(\bs{r}-2\bs{x}(t)-\Gamma \bs{x}(t)\Big)+ \mathrm{diag}\Big(\bs{x}(t)\Big) \, \Gamma,\\ \Sigma(t)& = & \mathrm{diag}\Big( \big(\bs{b}+\bs{d}+\bs{x}(t)+\Gamma \bs{x}(t) \big)\circ \bs{x}(t)\Big)\,.
\end{eqnarray*}
The solution is a centered Gaussian process with covariance function:
$$
\Cov\big(\bs{\eta}(t),\bs{\eta}(s)\big)=\int_0^{t\wedge s} e^{-\int_0^v A(u) \dd u} \Sigma^2(v) e^{-\int_0^v A(u)\dd u}\dd v\, ,
$$
a $N\times N$ matrix-valued function.

This central limit theorem quantifies the convergence rate in Proposition \ref{prop:LLN} and allows for example to compute confidence intervals.

\subsection{Stochastic differential equations corresponding to noisy versions of Lotka-Volterra equations}


\subsubsection{Limiting Feller diffusion for large population with time rescaling}

Another way to exhibit stochastic differential equations is to consider diffusive time rescaling in the almost-critical case (e.g. \cite[Section 4.2]{champagnatferrieremeleard}, when the growth rates are close to zero). 
More precisely, if 
\begin{equation}\label{taux:Feller}
    b^K_i=K \sigma_i +b_i,\quad d^K_i=K\sigma_i + d_i,\quad  \mbox{ and }\quad \Gamma^K_{ij}=\frac{\Gamma_{ij}}{K},
\end{equation}
where $\sigma_i$, $b_i$, $d_i$ and $\Gamma_{ij}$ do not depend on $K$. As usual, we denote by $\bs{b}=(b_i)$, $\bs{d}=(d_i)$, $\bs{r}=(b_i-d_i)$ $N\times 1$ vectors and define the matrix $\Sigma$ as the $N\times N$ diagonal matrix with entries $(\sigma_i)$, i.e. $\Sigma=\mathrm{diag}(\sigma_i)$. The fact that the birth and death rates are of order $K$ corresponds to accelerating the time proportionally to the factor $K$ that also rescaling the population size. 
That the species $i$ is close to criticality appears in the fact that both the birth and the death rate have the same leading term in $K\sigma_i$ with the same coefficient $\sigma_i$. This coefficient can depend on the species.

\begin{prop}\label{prop:Feller}
Assume \eqref{condition_initiale}, \eqref{hyp:moment} and the rates \eqref{taux:Feller}. Then, when $K\rightarrow +\infty$, the process $(\bs{X}^K(t))_{t\geq 0}$ converges uniformly on every compact time interval $[0,T]$ with $T>0$ and in distribution to the solution of the following Feller stochastic differential equation:
\begin{equation}\label{eq:Feller}
    \dd \bs{X}(t)=(\bs{r} -\bs{X}(t) + \Gamma \bs{X}(t)) \circ \bs{X}(t) \, \dd t + \sqrt{2 \Sigma \bs{X}(t)} \circ \dd \bs{B}_t
\end{equation}
where $\bs{B}$ is a $N$-dimensional standard Brownian motion independent of $\Gamma$, and where the function $x\mapsto\sqrt{x}$ is applied 
elementwise to the vector $\Sigma \bs{X}(t)$.
\end{prop}


The random noise appearing in \eqref{eq:Feller} comes from the rapid successions of birth and death events in this accelerated time-scale. Details on the derivation of this equation are given in Appendix \ref{ann:microFeller}.




\subsubsection{Variations around the Feller equations}

In \cite{biroli2018marginally}, Biroli et al. added an immigration factor $\lambda > 0$. The SDEs they consider can be written:
\begin{equation}
\label{sde} 
\dd\bs{X}(t) = \bs{X}(t) \circ \left( \bs 1_N - \bs{X}(t) + \Gamma \bs{X}(t) \right) \dd t
  + \lambda \bs 1_N \dd t 
 + f(\bs{X}(t)) \circ \dd\bs{B}(t), 
\end{equation} 
where $\bs{B}(t)$ is a $N$-dimensional standard Brownian motion independent of $\Gamma$, and where the function $f:
\R_+ \to \R_+$ is applied elementwise to any vector $\bs{x}=(x_i)$, i.e. $f(\bs{x})=(f(x_i))$.

In the framework of the elliptical model for
the interaction matrix $\Gamma$, recall \eqref{def:ellipticalmodel} and \eqref{def:ellipticalmodel-corr}, Biroli et al. considered the extreme case where the correlation $\xi$ between $A_{ij}$ and $A_{ji}$ is $1$, corresponding to a random symmetric matrix $\Gamma$.
Applying the replica method, they unveil the large-$N$ system behavior of $\bs{X}$, recovering the parameter regions for which the system has a single equilibrium or multiple equilibria.

Another generalization was done by Roy et al. in \cite{roy2019numerical}, where $\Gamma$ can follow the general elliptic model. These authors study the large-$N$ limit of the
SDE~\eqref{sde} by making use of a so-called dynamical mean field approach which is based on the dynamical cavity method, as detailed in the classical reference \cite{mez-par-vir-(livre)86}. 

A better understanding of these results from a mathematical perspective, as
well as their generalization to more sophisticated models than the elliptical
model, are interesting and useful research directions which have not been undertaken
so far to the best of our knowledge. One of the first mathematical
formalizations of this class of problems goes back to the work of Ben Arous
and Guionnet \cite{ben-gui-95}, who were interested in the dynamics of
mean-field spin glasses. In that setting, their analogue of our SDE~\eqref{sde}
is a Langevin version of the so-called Sherrington-Kirkpatrick model for the
spin glass dynamics.  In the same line of thought, Faugeras et al. \cite{fau-tou-ces-09,cab-tou-13} used the approach of Ben Arous and Guionnet to study a diffusion version of the so-called Hopfield model for biological neural
networks. Details are given in Appendix \ref{app:mean-field}.

\section{Discussion}\label{sec:discussion}
Our guided tour through large Lotka-Volterra models has highlighted the importance of some features of the interaction matrix to understand the dynamics of complex ecological systems. Among these features, the first two statistical moments of the distribution of interactions, $\mu/N$ and $\sigma^2/(N\alpha^2)$, are of paramount importance, as well as the scaling of the normalisation applied to the interaction matrix ($1/N$ for the deterministic part of the matrix, $1/\sqrt{N}$ or  $1/\alpha_N\sqrt{N}$ for its random part). 

Our review covers some key topics on the analysis of large Lotka-Volterra models constructed using simple random interaction matrices, i.e. conditions leading to a feasible equilibrium, conditions leading to a unique maximal equilibrium and the link between the interaction matrix and equilibrium stability  (Section \ref{sec:unique}). In Section \ref{sec:struct} and Appendix \ref{app:structuredmodapp}, we have mentioned some sophistications of the model that can make it more realistic, in particular the inclusion of structures in the interaction matrices (e.g. block- or trait-based matrices, sparse matrices). Theoretical results obtained on kernel matrices of large size (many traits, many species) (Theorem \ref{theo:kernel}) suggest that the study of LV models obtained from trait-based interaction matrices could be simplified by assuming that many interaction matrix spectra will resemble that of a Wishart matrix. However, this result does not yield any direct conclusion on the feasibility of an equilibrium of the LV model, nor does it help generalize the heuristics \ref{heuristics:surviving-species} to such structured models.

We have not considered models based on fully organized interaction matrices, which are not amenable to an RMT analysis but sometimes allow some direct analysis through classic analytical tools (e.g. Lyapunov functions, see also \cite{gouze1993}). For instance, modelling food webs as strictly organized by trophic levels (i.e. species from trophic level $k$ can only be positively affected by some species of level $k-1$ and negatively affected by some species of level $k+1$), it is possible to express coexistence conditions as mathematical conditions on the covering of the food web by pairs of interacting species \cite{haerter2016,haerter2017}. This finding, which highlights the difficulty of having a feasible equilibrium in level-organized food webs, hints at the potential importance of omnivory (i.e. that interactions are not strictly organized by levels, such that a predator can also feed on the prey of its prey species) to explain the stability of real food webs. An interesting endeavour could be to formalize omnivory in the context of random interaction matrices in order to tease apart the effect of omnivory from that of the food web being acyclic (which would also break the trophic level-based nature of the interaction matrix used by \cite{haerter2016,haerter2017}).


\subsection{Link to empirical data}

There is an enormous literature on statistical models for ecological networks, see e.g. 
\cite{botella_etal,cirtwill,desiqueirasantos_etal,matiasrebafkavillers,mielematias2017,miele_etal_2021,momalrobinambroise,momalrobinambroise2} to name a few papers. The variety of data available and the specificity of each ecosystem explain this vast corpus, which in itself would deserve an entire review -- this is out of the scope of the present work. However, mentioning the relations to data and estimation questions is essential.

Let us state first that a high number of papers propose models that are variations of the Lotka-Volterra systems, for instance integrating functional responses such as in \cite{bansayebilliardchazottes} (and references therein). The inference and empirical testing of models is limited by the type and volume of data that can be obtained through the observation of natural systems or controlled experiments. For most systems, we can only access abundances $x_i$ (often in a single snapshot that may not be at equilibrium, less frequently in time series). For some systems, especially with predator-prey or pollinator-plant interactions, we can also know which interactions $\Gamma_{ij}$ are zero or nonzero, as species may be limited in which partners they can interact with, but quantitative estimates of nonzero interactions are problematic. Therefore, exhaustive information on parameters $r_i$ and $\Gamma_{ij}$ is not accessible in large systems when no additional assumption is made (e.g. \cite{jacquet2016}): the main method for estimating them reliably is to observe the growth rates and/or equilibrium abundances of each species in isolation and in combinations, which is only possible in small-$N$ experiments. Abundance distributions have been exploited to fit many other models, e.g neutral theory~\cite{hubbell2001unified}.
Additional data may help estimation and model validation ~\cite{barbier2021fingerprints,hu2021emergent,fort2018quantitative}.\\
Finally, notice that although the models discussed above have a high number of parameters, their distributions have low degrees of freedom. For instance, the law of the elliptic model depends only on $\alpha$, $\mu$ and $\xi$. 



\subsection{Open mathematical and modelling problems}



\subsubsection*{Multiple equilibria and non-equilibrium attractors} 
\begin{figure}
\begin{center}\hspace*{-50pt}
\includegraphics*[width=1.1\linewidth]{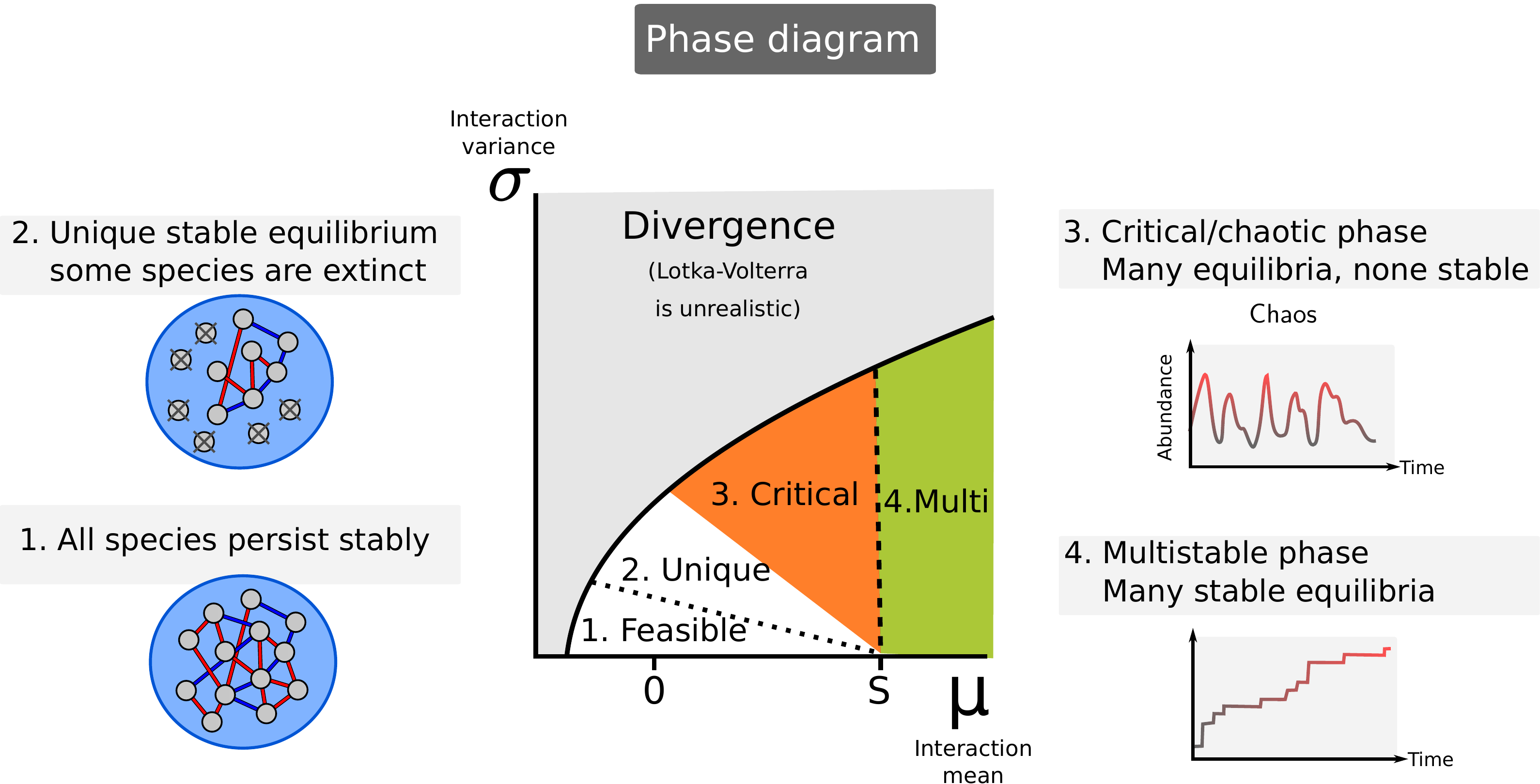}
\end{center}
\caption{Phase diagram representing the qualitative dynamical regimes of the reference random model (first obtained numerically in~\cite{kessler2015generalized}). }
\label{fig:phase}
\end{figure}

While our review focuses on conditions under which there exists a unique stable equilibrium (Sec.~\ref{sec:unique}), other regimes including multiple equilibria or non-equilibrium attractors (e.g. chaos or cycles) have been studied using physics tools~\cite{biroli2018marginally,roy2020complex}, see for instance Fig. \ref{fig:phase}. A mathematical understanding of these regimes and the associated thresholds is challenging and would represent an important step in the understanding of LV systems.

\subsubsection*{Quantitative metrics for stability} As hinted in Section \ref{sec:otherextensions} and Appendix \ref{annexe:micro1}, other dynamical properties of LV systems, such as the dynamics of fluctuations around equilibria, can also be studied using stochastic differential equation analogues to the LV ODEs. More quantitative metrics of stability are also of ecological interest, e.g. quantifying how much attractors change in response to perturbations of model parameters~\cite{barabas2014fixed}, assessing how much model parameters can change without changing the dynamical attractor (i.e. structural stability, \cite{saavedra2017structural}), or characterizing the time and trajectory of return to an attractor after a perturbation in the abundances (e.g. chap. 58 in \cite{trefethenbook}). When perturbations are small,
it is possible to linearize the dynamics around an attractor, and simply compute stability metrics for the i.i.d. or elliptic model~\cite{arnoldi2016unifying,arnoldi2018ecosystems,arnoldi2019inherent}. Beyond the linearized regime, tools such as Freidlin-Wentzell theory can help quantify the basins of dynamical attractors and transition times between them~\cite{rodriguez2020climbing}.

\subsubsection*{Counting the number of coexisting species} In Section \ref{sec:unique-equilibrium+vanishing}, we provide heuristics to compute the proportion of vanishing species for a given equilibrium and refer to articles relying on techniques from Physics (cavity method, dynamical generating functional techniques). A mathematical computation still remains out of reach and could be phrased as understanding the properties of solutions of Linear Complementarity Problems with random matrices as input. Part of the challenge lies on the fact that the properties of interest of the solution of a LCP (proportion of non-vanishing components) do not only rely on the spectral properties of the random matrices at hand. 
Table \ref{tab:LV-versus-alpha} in Section \ref{sec:unique} lists a number of open questions when studying a specific LV model.

\subsubsection*{A mathematical understanding of the cavity method} The above open question is only a part of a larger endeavour to understand the cavity method from a mathematical point of view. The cavity method is extremely versatile and in general yields to closed-form expressions accurately matching simulation results. A mathematical formalization of the cavity methods would certainly represent an important step toward the analysis of non-spectral properties of large random matrix observables.

 

\begin{thebibliography}{NHvdK{\etalchar{+}}07}

\bibitem[Abb18]{abbe}
E.~Abbe.
\newblock Community detection and stochastic block models: recent development.
\newblock {\em Journal of Machine Learning Research}, 18(177):1--86, 2018.

\bibitem[ABK{\etalchar{+}}19]{arnoldi2019fitness}
J-F. Arnoldi, M.~Barbier, R.~Kelly, G.~Barab{\'a}s, and A.~L. Jackson.
\newblock Fitness and community feedbacks: the two axes that drive long-term
  invasion impacts.
\newblock {\em bioRxiv}, page 705756, 2019.

\bibitem[ABK{\etalchar{+}}22]{ABKBJ22}
Jean-François Arnoldi, Matthieu Barbier, Ruth Kelly, György Barabás, and
  Andrew~L. Jackson.
\newblock Invasions of ecological communities: Hints of impacts in the
  invader's growth rate.
\newblock {\em Methods in Ecology and Evolution}, 13(1):167--182, 2022.

\bibitem[ABLH18]{arnoldi2018ecosystems}
J-F Arnoldi, Azenor Bideault, Michel Loreau, and Bart Haegeman.
\newblock How ecosystems recover from pulse perturbations: A theory of short-to
  long-term responses.
\newblock {\em Journal of theoretical biology}, 436:79--92, 2018.

\bibitem[ADK21]{AltDucKnow21}
Johannes Alt, Raphael Ducatez, and Antti Knowles.
\newblock Poisson statistics and localization at the spectral edge of sparse
  erd\"{o}s--r\'enyi graphs.
\newblock arxiv:2106.12519, 2021.

\bibitem[AFK21]{benarousfyodorovkhoruzhenko}
G.~Ben Arous, Y.V. Fyodorov, and B.A. Khoruzhenko.
\newblock Counting equilibria of large complex systems by instability index.
\newblock {\em Proceedings of the National Academy of Sciences},
  118(34):e2023719118, August 2021.

\bibitem[AH16]{arnoldi2016unifying}
Jean-Fran{\c{c}}ois Arnoldi and Bart Haegeman.
\newblock Unifying dynamical and structural stability of equilibria.
\newblock {\em Proceedings of the Royal Society A: Mathematical, Physical and
  Engineering Sciences}, 472(2193):20150874, 2016.

\bibitem[ALH19]{arnoldi2019inherent}
Jean-Fran{\c{c}}ois Arnoldi, Michel Loreau, and Bart Haegeman.
\newblock The inherent multidimensionality of temporal variability: how common
  and rare species shape stability patterns.
\newblock {\em Ecology letters}, 22(10):1557--1567, 2019.

\bibitem[AN21]{akjouj2021feasibility}
Imane Akjouj and Jamal Najim.
\newblock Feasibility of sparse large lotka-volterra ecosystems, 2021.

\bibitem[ARR{\etalchar{+}}15]{allhoff2015}
K.~T. Allhoff, D.~Ritterskamp, B.~C. Rall, B.~Drossel, and C.~Guill.
\newblock Evolutionary food web model based on body masses gives realistic
  networks with permanent species turnover.
\newblock {\em Scientific Reports}, 5, 2015.

\bibitem[AT12]{allesina2012stability}
S.~Allesina and S.~Tang.
\newblock Stability criteria for complex ecosystems.
\newblock {\em Nature}, 483(7388):205--208, 2012.

\bibitem[AT15]{allesina2015stability}
S.~Allesina and S.~Tang.
\newblock The stability--complexity relationship at age 40: a random matrix
  perspective.
\newblock {\em Population Ecology}, 57(1):63--75, 2015.

\bibitem[BAG95]{ben-gui-95}
G.~Ben~Arous and A.~Guionnet.
\newblock Large deviations for {L}angevin spin glass dynamics.
\newblock {\em Probability Theory and Related Fields}, 102(4):455--509, 1995.

\bibitem[Bar07]{barber2007modularity}
Michael~J. Barber.
\newblock Modularity and community detection in bipartite networks.
\newblock {\em Phys. Rev. E}, 76:066102, Dec 2007.

\bibitem[BBC18a]{bansayebilliardchazottes}
V.~Bansaye, S.~Billiard, and J.~R. Chazottes.
\newblock Rejuvenating functional responses with renewal theory.
\newblock {\em Journal of the Royal Society Interface}, 15:20180239, 2018.

\bibitem[BBC18b]{biroli2018marginally}
Giulio Biroli, Guy Bunin, and Chiara Cammarota.
\newblock Marginally stable equilibria in critical ecosystems.
\newblock {\em New Journal of Physics}, 20(8):083051, 2018.

\bibitem[BC12]{bordenave2012around}
C.~Bordenave and D.~Chafa{\"\i}.
\newblock Around the circular law.
\newblock {\em Probability surveys}, 9:1--89, 2012.

\bibitem[BCCT18]{Bordenave2018}
Charles Bordenave, Pietro Caputo, Djalil Chafaï, and Konstantin Tikhomirov.
\newblock On the spectral radius of a random matrix: An upper bound without
  fourth moment.
\newblock {\em The Annals of Probability}, 46(4), Jul 2018.

\bibitem[BCGZ21]{bordenave2021convergence}
Charles Bordenave, Djalil Chafai, and David Garc{\'i}a-Zelada.
\newblock Convergence of the spectral radius of a random matrix through its
  characteristic polynomial.
\newblock {\em {Probability Theory and Related Fields}}, 2021.

\bibitem[BDB{\etalchar{+}}11]{baskerville2011bayesian}
Edward~B. Baskerville, Andy~P. Dobson, Trevor Bedford, Stefano Allesina,
  T.~Michael Anderson, and Mercedes Pascual.
\newblock Spatial guilds in the serengeti food web revealed by a bayesian group
  model.
\newblock {\em PLOS Computational Biology}, 7(12):1--11, 12 2011.

\bibitem[BDM{\etalchar{+}}22]{botella_etal}
C.~Botella, S.~Dray, C.~Matias, V.~Miele, and W.~Thuiller.
\newblock An appraisal of graph embeddings for comparing trophic network
  architectures.
\newblock {\em Methods in Ecology and Evolution}, 13(1):203--216, 2022.

\bibitem[BDMLB21]{barbier2021fingerprints}
Matthieu Barbier, Claire De~Mazancourt, Michel Loreau, and Guy Bunin.
\newblock Fingerprints of high-dimensional coexistence in complex ecosystems.
\newblock {\em Physical Review X}, 11(1):011009, 2021.

\bibitem[BFMT16]{billiardferrieremeleardtran}
S.~Billiard, R.~Ferri\`ere, S.~M\'el\'eard, and V.C. Tran.
\newblock The effect of competition and horizontal trait inheritance on
  invasion, fixation and polymorphism.
\newblock {\em Journal of Theoretical Biology}, 411:48--58, 2016.

\bibitem[BGA{\etalchar{+}}04]{bro-etal-04}
James~H. Brown, James~F. Gillooly, Andrew~P. Allen, Van~M. Savage, and
  Geoffrey~B. West.
\newblock Toward a metabolic theory of ecology.
\newblock {\em Ecology}, 85(7):1771--1789, 2004.

\bibitem[BGBK20]{BBGK17}
Florent Benaych-Georges, Charles Bordenave, and Antti Knowles.
\newblock Spectral radii of sparse random matrices.
\newblock {\em Ann. Inst. H. Poincaré Probab. Statist.}, 56(3):2141--2161,
  2020.

\bibitem[BJRG22]{baron2022non}
Joseph~W Baron, Thomas~Jun Jewell, Christopher Ryder, and Tobias Galla.
\newblock Non-gaussian random matrices determine the stability of
  lotka-volterra communities.
\newblock {\em arXiv preprint arXiv:2202.09140}, 2022.

\bibitem[BLMV05]{bastolla2005}
U.~Bastolla, M.~Lässig, S.~C. Manrubia, and A.~Valleriani.
\newblock Biodiversity in model ecosystems, i: coexistence conditions for
  competing species.
\newblock {\em Journal of Theoretical Biology}, 235:521--530, 2005.

\bibitem[BLRT22]{billiardlemanreytran}
S.~Billiard, H.~Leman, T.~Rey, and V.C. Tran.
\newblock Continuous limits of large plant-pollinator random networks and some
  applications.
\newblock {\em MathematicS In Action}, 2022.

\bibitem[BM15]{bansayemeleard}
Vincent Bansaye and Sylvie M\'{e}l\'{e}ard.
\newblock {\em Stochastic models for structured populations}, volume~1 of {\em
  Mathematical Biosciences Institute Lecture Series. Stochastics in Biological
  Systems}.
\newblock Springer, Cham; MBI Mathematical Biosciences Institute, Ohio State
  University, Columbus, OH, 2015.
\newblock Scaling limits and long time behavior.

\bibitem[BMO14]{barabas2014fixed}
Gy{\"o}rgy Barab{\'a}s, G{\'e}za Mesz{\'e}na, and Annette Ostling.
\newblock Fixed point sensitivity analysis of interacting structured
  populations.
\newblock {\em Theoretical Population Biology}, 92:97--106, 2014.

\bibitem[BN21]{bizeul2021positive}
P.~Bizeul and J.~Najim.
\newblock Positive solutions for large random linear systems.
\newblock {\em Proceedings of the American Mathematical Society},
  149(6):2333--2348, 2021.

\bibitem[Bol01]{bollobas2001}
B.~Bollob\'{a}s.
\newblock {\em Random graphs}.
\newblock Cambridge University Press, 2 edition, 2001.

\bibitem[Bor13]{bordenave2012kernel}
Charles Bordenave.
\newblock On {E}uclidean random matrices in high dimension.
\newblock {\em Electron. Commun. Probab.}, 18:no. 25, 8, 2013.

\bibitem[BP19]{benignipeche2019kernel}
Lucas Benigni and Sandrine P\'{e}ch\'{e}.
\newblock Eigenvalue distribution of some nonlinear models of random matrices.
\newblock {\em Preprint, \url{https://arxiv.org/abs/1904.03090}}, 2019.

\bibitem[BS10]{bai2010spectral}
Z.~Bai and J.~W. Silverstein.
\newblock {\em Spectral analysis of large dimensional random matrices},
  volume~20.
\newblock Springer, 2010.

\bibitem[BSHM17]{busiello2017explorability}
Daniel~M. Busiello, Samir Suweis, Jorge Hidalgo, and Amos Maritan.
\newblock Explorability and the origin of network sparsity in living systems.
\newblock {\em Scientific Reports}, 7(1), Sep 2017.

\bibitem[Bun16]{bunin2016interaction}
Guy Bunin.
\newblock Interaction patterns and diversity in assembled ecological
  communities.
\newblock {\em arXiv preprint arXiv:1607.04734}, 2016.

\bibitem[Bun17]{bunin2017ecological}
G~Bunin.
\newblock Ecological communities with lotka-volterra dynamics.
\newblock {\em Physical Review E}, 95(4):042414, 2017.

\bibitem[BVH16]{bandeira2016sharp}
A.S. Bandeira and R.~Van~Handel.
\newblock Sharp nonasymptotic bounds on the norm of random matrices with
  independent entries.
\newblock {\em The Annals of Probability}, 44(4):2479--2506, 2016.

\bibitem[BY86]{baiyin1986rayonspectral}
Z.~D. Bai and Y.~Q. Yin.
\newblock Limiting behavior of the norm of products of random matrices and two
  problems of {G}eman-{H}wang.
\newblock {\em Probab. Theory Related Fields}, 73(4):555--569, 1986.

\bibitem[Cas90]{Case1990}
T~J Case.
\newblock Invasion resistance arises in strongly interacting species-rich model
  competition communities.
\newblock {\em Proceedings of the National Academy of Sciences},
  87(24):9610--9614, 1990.

\bibitem[CCM19]{chazottes2019time}
J-R Chazottes, P~Collet, and S~M{\'e}l{\'e}ard.
\newblock On time scales and quasi-stationary distributions for multitype
  birth-and-death processes.
\newblock {\em Annales de l'Institut Henri Poincar{\'e}, Probabilit{\'e}s et
  Statistiques}, 55(4):2249--2294, 2019.

\bibitem[CEFN22]{clenet2022equilibrium}
M.~Clenet, H.~El~Ferchichi, and J.~Najim.
\newblock Equilibrium in a large lotka-volterra system with pairwise correlated
  interactions.
\newblock {\em Stochastic Processes and their Applications, Vol. 153}, 2022.

\bibitem[CER{\etalchar{+}}19]{cirtwill}
A.R. Cirtwill, A.~Ekl\"of, T~Roslin, K.~Wootton, and D.~Gravel.
\newblock A quantitative framework for investigating the reliability of
  empirical network construction.
\newblock {\em Methods in Ecology and Evolution}, 10:902–911, 2019.

\bibitem[CFM06]{champagnatferrieremeleard}
Nicolas Champagnat, R\'egis Ferri\`{e}re, and Sylvie M\'{e}l\'{e}ard.
\newblock Unifying evolutionary dynamics: from individual stochastic processes
  to macroscopic models via timescale separation.
\newblock {\em Theoretical Population Biology}, 69:297--321, 2006.

\bibitem[Cha06]{champagnat06}
N.~Champagnat.
\newblock A microscopic interpretation for adaptative dynamics trait
  substitution sequence models.
\newblock {\em Stochastic Processes and their Applications}, 116:1127--1160,
  2006.

\bibitem[CJL{\etalchar{+}}17]{Calcagno2017}
Vincent Calcagno, Philippe Jarne, Michel Loreau, Nicolas Mouquet, and Patrice
  David.
\newblock Diversity spurs diversification in ecological communities.
\newblock {\em Nature Communications}, 8:15810, 2017.

\bibitem[CJR10]{champagnat2010convergence}
Nicolas Champagnat, Pierre-Emmanuel Jabin, and Ga{\"e}l Raoul.
\newblock Convergence to equilibrium in competitive lotka--volterra and
  chemostat systems.
\newblock {\em Comptes Rendus Mathematique}, 348(23-24):1267--1272, 2010.

\bibitem[CLV03]{ChuLuVu2003}
Fan Chung, Linyuan Lu, and Van Vu.
\newblock Spectra of random graphs with given expected degrees.
\newblock {\em Proceedings of the National Academy of Sciences - PNAS},
  100(11):6313--6318, 2003.

\bibitem[CM11]{champagnatmeleard2011}
N.~Champagnat and S.~M\'{e}l\'{e}ard.
\newblock Polymorphic evolution sequence and evolutionary branching.
\newblock {\em Probability Theory and Related Fields}, 151(1-2):45--94, 2011.

\bibitem[CMN22a]{clenet2022preprint}
M.~Cl\'enet, F.~Massol, and J.~Najim.
\newblock Equilibrium and surviving species in a large lotka-volterra system of
  differential equations.
\newblock {\em Arxiv:2205.00735}, 2022.

\bibitem[CMN22b]{clenet2022gretsi}
M.~Clenet, F.~Massol, and J.~Najim.
\newblock Surviving species in a large lotka-volterra system of differential
  equations.
\newblock In {\em 28e Colloque sur le traitement du signal et des images},
  volume 001-0257, pages p. 1029--1032, Nancy, Sep. 6--9 2022. GRETSI - Groupe
  de Recherche en Traitement du Signal et des Images.

\bibitem[CPS09]{cottle2009linear}
Richard~W Cottle, Jong-Shi Pang, and Richard~E Stone.
\newblock {\em The linear complementarity problem}.
\newblock SIAM, 2009.

\bibitem[CS13]{chengsinger2013kernel}
Xiuyuan Cheng and Amit Singer.
\newblock The spectrum of random inner-product kernel matrices.
\newblock {\em Random Matrices Theory Appl.}, 2(4):1350010, 47, 2013.

\bibitem[CT13]{cab-tou-13}
T.~Cabana and J.~Touboul.
\newblock Large deviations, dynamics and phase transitions in large stochastic
  and disordered neural networks.
\newblock {\em Journal of Statistical Physics}, 153(2):211--269, 2013.

\bibitem[dAAG20]{andreazzi}
Cecilia~Siliansky de~Andreazzi, Julia Astegiano, and Paulo~R Guimarães.
\newblock Coevolution by different functional mechanisms modulates the
  structure and dynamics of antagonistic and mutualistic networks.
\newblock {\em Oikos}, 129(2):224--237, 2020.

\bibitem[DD99]{dieckmanndoebeli}
Ulf Dieckmann and Michael Doebeli.
\newblock On the origin of species by sympatric speciation.
\newblock {\em Nature}, 400:354--357, 1999.

\bibitem[DeA18]{deangelis2018individual}
Donald~Lee DeAngelis.
\newblock {\em Individual-based models and approaches in ecology: populations,
  communities and ecosystems}.
\newblock CRC Press, 2018.

\bibitem[DG14]{deangelis2014individual}
Donald~L DeAngelis and Volker Grimm.
\newblock Individual-based models in ecology after four decades.
\newblock {\em F1000prime reports}, 6, 2014.

\bibitem[dH17]{vanderhofstad}
R.~Van der Hofstad.
\newblock {\em Random Graphs and Complex Networks}, volume~1 of {\em Cambridge
  Series in Statistical and Probabilistic Mathematics}.
\newblock Cambridge University Press, Cambridge, 2017.

\bibitem[dSSFM21]{desiqueirasantos_etal}
S.~de~Siqueira~Santos, A.~Fujita, and C.~Matias.
\newblock Spectral density of random graphs: convergence properties and
  application in model fitting.
\newblock {\em Journal of Complex Networks}, 9(6), 2021.

\bibitem[Dur07]{durrett}
R.~Durrett.
\newblock {\em Random graph dynamics}.
\newblock Cambridge University Press, New York, 2007.

\bibitem[DV13]{dovu2012kernel}
Yen Do and Van Vu.
\newblock The spectrum of random kernel matrices: universality results for
  rough and varying kernels.
\newblock {\em Random Matrices Theory Appl.}, 2(3):1350005, 29, 2013.

\bibitem[DVR{\etalchar{+}}18]{dougoud2018feasibility}
M.~Dougoud, L.~Vinckenbosch, R.P. Rohr, L-F. Bersier, and C.~Mazza.
\newblock The feasibility of equilibria in large ecosystems: A primary but
  neglected concept in the complexity-stability debate.
\newblock {\em PLoS computational biology}, 14(2):e1005988, 2018.

\bibitem[DWM02]{dunne2002connectance}
Jennifer~A. Dunne, Richard~J. Williams, and Neo~D. Martinez.
\newblock Food-web structure and network theory: The role of connectance and
  size.
\newblock {\em Proceedings of the National Academy of Sciences - PNAS},
  99(20):12917--12922, 2002.

\bibitem[EJK{\etalchar{+}}13]{eklof2013}
Anna Eklöf, Ute Jacob, Jason Kopp, Jordi Bosch, Rocío Castro-Urgal,
  Natacha~P. Chacoff, Bo~Dalsgaard, Claudio de~Sassi, Mauro Galetti, Paulo~R.
  Guimarães, Silvia~Beatriz Lomáscolo, Ana~M. Martín~González,
  Marco~Aurelio Pizo, Romina Rader, Anselm Rodrigo, Jason~M. Tylianakis,
  Diego~P. Vázquez, and Stefano Allesina.
\newblock The dimensionality of ecological networks.
\newblock {\em Ecology Letters}, 16(5):577--583, 2013.

\bibitem[EK86]{ethierkurtz}
S.N. Ethier and T.G. Kurtz.
\newblock {\em Markov Processus, Characterization and Convergence}.
\newblock John Wiley \& Sons, New York, 1986.

\bibitem[EK88]{EdelsteinKeshet1988}
L.~Edelstein-Keshet.
\newblock {\em Mathematical Models in Biology}.
\newblock SIAM Classics in Applied Mathematics. SIAM, 1988.

\bibitem[EK10]{elkaroui2010kernel}
Noureddine El~Karoui.
\newblock The spectrum of kernel random matrices.
\newblock {\em Ann. Statist.}, 38(1):1--50, 2010.

\bibitem[FK81]{Furedi1981}
Z.~F\"{u}redi and J.~Koml\'{o}s.
\newblock The eigenvalues of random symmetric matrices.
\newblock {\em Combinatorica}, 1(3):233--241, 1981.

\bibitem[FK16]{fyodorovkhoruzhenko}
Y.V. Fyodorov and B.A. Khoruzhenko.
\newblock Nonlinear analogue of the {M}ay-{W}igner instability transition.
\newblock {\em Proceedings of the National Academy of Sciences},
  113(25):6827--6832, 2016.

\bibitem[FM04]{fourniermeleard}
Nicolas Fournier and Sylvie M\'{e}l\'{e}ard.
\newblock A microscopic probabilistic description of a locally regulated
  population and macroscopic approximations.
\newblock {\em Ann. Appl. Probab.}, 14(4):1880--1919, 2004.

\bibitem[For18]{fort2018quantitative}
Hugo Fort.
\newblock Quantitative predictions from competition theory with an incomplete
  knowledge of model parameters tested against experiments across diverse taxa.
\newblock {\em Ecological Modelling}, 368:104--110, 2018.

\bibitem[FT09]{ferriere2009stochastic}
Regis Ferriere and Viet~Chi Tran.
\newblock Stochastic and deterministic models for age-structured populations
  with genetically variable traits.
\newblock In {\em ESAIM: Proceedings}, volume~27, pages 289--310. EDP Sciences,
  2009.

\bibitem[FTC09]{fau-tou-ces-09}
O.~Faugeras, J.~Touboul, and B.~Cessac.
\newblock A constructive mean-field analysis of multi population neural
  networks with random synaptic weights and stochastic inputs.
\newblock {\em Frontiers in Computational Neuroscience}, 3:1, 2009.

\bibitem[GA70]{GardnerAshby1970}
Mark~R Gardner and W~Ross Ashby.
\newblock Connectance of large dynamic (cybernetic) systems: critical values
  for stability.
\newblock {\em Nature}, 228(5273):784--784, 1970.

\bibitem[Gal18]{galla2018dynamically}
T.~Galla.
\newblock Dynamically evolved community size and stability of random
  lotka-volterra ecosystems (a).
\newblock {\em EPL (Europhysics Letters)}, 123(4):48004, 2018.

\bibitem[GAS{\etalchar{+}}17]{grilli2017feasibility}
J.~Grilli, M.~Adorisio, S.~Suweis, G.~Barab{\'a}s, J.~R Banavar, S.~Allesina,
  and A.~Maritan.
\newblock Feasibility and coexistence of large ecological communities.
\newblock {\em Nature communications}, 8:14389, 2017.

\bibitem[GBB{\etalchar{+}}06]{grimm2006standard}
Volker Grimm, Uta Berger, Finn Bastiansen, Sigrunn Eliassen, Vincent Ginot,
  Jarl Giske, John Goss-Custard, Tamara Grand, Simone~K Heinz, Geir Huse,
  et~al.
\newblock A standard protocol for describing individual-based and agent-based
  models.
\newblock {\em Ecological modelling}, 198(1-2):115--126, 2006.

\bibitem[Gem86]{geman1986spectral}
S.~Geman.
\newblock The spectral radius of large random matrices.
\newblock {\em The Annals of Probability}, pages 1318--1328, 1986.

\bibitem[GH82]{geman1982chaos}
S.~Geman and C-R. Hwang.
\newblock A chaos hypothesis for some large systems of random equations.
\newblock {\em Zeitschrift f{\"u}r Wahrscheinlichkeitstheorie und Verwandte
  Gebiete}, 60(3):291--314, 1982.

\bibitem[Gil75]{gilpin75}
M.~E. Gilpin.
\newblock Stability of feasible predator-prey systems.
\newblock {\em Nature}, 254(5496):137--139, 1975.

\bibitem[Gil77]{gillespie}
Daniel~T. Gillespie.
\newblock Exact stochastic simulation of coupled chemical reaction.
\newblock {\em J. Phys. Chem.}, 81(25):2340--2361, 1977.

\bibitem[Gin65]{ginibre1965statistical}
J.~Ginibre.
\newblock Statistical ensembles of complex, quaternion, and real matrices.
\newblock {\em Journal of Mathematical Physics}, 6(3):440--449, 1965.

\bibitem[Gir86]{girko1986elliptic}
V.L. Girko.
\newblock Elliptic law.
\newblock {\em Theory of Probability \& Its Applications}, 30(4):677--690,
  1986.

\bibitem[GKL20]{giorgikaakailemaire}
D.~Giorgi, S.~Kaakai, and V.~Lemaire.
\newblock Ibmpopsim r package, 2020.
\newblock https://cran.r-project.org/package=IBMPopSim.

\bibitem[GML16]{gravel2016}
D.~Gravel, F.~Massol, and M.~A. Leibold.
\newblock Stability and complexity in model meta-ecosystems.
\newblock {\em Nature Communications}, 7:12457, 2016.

\bibitem[GMT14]{guptametztran}
A.~Gupta, J.A.J. Metz, and V.C. Tran.
\newblock A new proof for the convergence of an individual based model to the
  trait substitution sequence.
\newblock {\em Acta Applicandae Mathematicae}, 131(1):1--27, 2014.

\bibitem[Goh77]{Goh1977AmNat}
B.~S. Goh.
\newblock Global stability in many-species systems.
\newblock {\em American Naturalist}, 111(977):135--143, 1977.

\bibitem[Gou93]{gouze1993}
Jean-Luc Gouzé.
\newblock Global behavior of n-dimensional lotka–volterra systems.
\newblock {\em Mathematical Biosciences}, 113(2):231--243, 1993.

\bibitem[HAB{\etalchar{+}}21]{hu2021emergent}
Jiliang Hu, Daniel~R Amor, Matthieu Barbier, Guy Bunin, and Jeff Gore.
\newblock Emergent phases of ecological diversity and dynamics mapped in
  microcosms.
\newblock {\em bioRxiv}, 2021.

\bibitem[HAC{\etalchar{+}}18]{hastings2018transient}
Alan Hastings, Karen~C Abbott, Kim Cuddington, Tessa Francis, Gabriel Gellner,
  Ying-Cheng Lai, Andrew Morozov, Sergei Petrovskii, Katherine Scranton, and
  Mary~Lou Zeeman.
\newblock Transient phenomena in ecology.
\newblock {\em Science}, 361(6406), 2018.

\bibitem[Hir82]{Hirsch1982}
Morris~W. Hirsch.
\newblock Systems of differential equations which are competitive or
  cooperative: I. limit sets.
\newblock {\em SIAM Journal on Mathematical Analysis}, 13(2):167--179, 1982.

\bibitem[HL11]{haegeman2011mathematical}
Bart Haegeman and Michel Loreau.
\newblock A mathematical synthesis of niche and neutral theories in community
  ecology.
\newblock {\em Journal of theoretical biology}, 269(1):150--165, 2011.

\bibitem[HLL83]{hollandlaskeyleinhardt}
P.~Holland, K.~Laskey, and S.~Leinhardt.
\newblock Stochastic blockmodels: some first steps.
\newblock {\em Social Networks}, 5:109--137, 1983.

\bibitem[HMS16]{haerter2016}
Jan~O. Haerter, Namiko Mitarai, and Kim Sneppen.
\newblock Food web assembly rules for generalized lotka-volterra equations.
\newblock {\em PloS Computational Biology}, 12(2):e1004727, 2016.

\bibitem[HMS17]{haerter2017}
Jan~O. Haerter, Namiko Mitarai, and Kim Sneppen.
\newblock Existence and construction of large stable food webs.
\newblock {\em Physical Review E}, 96(3):032406, 2017.

\bibitem[HS88]{HofbauerSigmund1988}
J.~Hofbauer and K.~Sigmund.
\newblock {\em The Theory of Evolution and Dynamical Systems: Mathematical
  Aspects of Selection}.
\newblock London Mathematical Society Stundent Texts. Cambridge University
  Press, 1988.

\bibitem[HS{\etalchar{+}}98]{hofbauer1998evolutionary}
Josef Hofbauer, Karl Sigmund, et~al.
\newblock {\em Evolutionary games and population dynamics}.
\newblock Cambridge university press, 1998.

\bibitem[Hub01]{hubbell2001unified}
S.~P. Hubbell.
\newblock {\em The unified neutral theory of biodiversity and biogeography
  (MPB-32)}, volume~32.
\newblock Princeton University Press, 2001.

\bibitem[IW89]{ikedawatanabe}
N.~Ikeda and S.~Watanabe.
\newblock {\em Stochastic Differential Equations and Diffusion Processes},
  volume~24.
\newblock North-Holland Publishing Company, 1989.
\newblock Second Edition.

\bibitem[JMM{\etalchar{+}}16]{jacquet2016}
C.~Jacquet, C.~Moritz, L.~Morissette, P.~Legagneux, F.~Massol, P.~Archambault,
  and D.~Gravel.
\newblock No complexity-stability relationship in empirical ecosystems.
\newblock {\em Nature Communications}, 7:12573, 2016.

\bibitem[JPR{\etalchar{+}}15]{James2015}
Alex James, Michael~J. Plank, Axel~G. Rossberg, Jonathan Beecham, Mark
  Emmerson, and Jonathan~W. Pitchford.
\newblock Constructing random matrices to represent real ecosystems.
\newblock {\em The American Naturalist}, 185(5):680--692, 2015.

\bibitem[Kis99]{kisdi}
Eva Kisdi.
\newblock Evolutionary branching under asymmetric competition.
\newblock {\em J. Theor. Biol.}, 197(2):149--162, 1999.

\bibitem[KJLT02]{kokkoris2002}
G.D. Kokkoris, V.A.A. Jansen, M.~Loreau, and A.Y. Troumbis.
\newblock Variability in interaction strength and implications for
  biodiversity.
\newblock {\em Journal of Animal Ecology}, 71:362--371, 2002.

\bibitem[KS15]{kessler2015generalized}
David~A Kessler and Nadav~M Shnerb.
\newblock Generalized model of island biodiversity.
\newblock {\em Physical Review E}, 91(4):042705, 2015.

\bibitem[LBP{\etalchar{+}}21]{lepersbilliardportemeleardtran}
C.~Lepers, S.~Billiard, M.~Porte, S.~M\'el\'eard, and V.C. Tran.
\newblock Inference with selection, varying population size and evolving
  population structure: Application of abc to a forward-backward.
\newblock {\em Heredity}, 126:335--350, 2021.

\bibitem[LdM13]{loreaudemaz2013}
Michel Loreau and Claire de~Mazancourt.
\newblock Biodiversity and ecosystem stability: a synthesis of underlying
  mechanisms.
\newblock {\em Ecology Letters}, 16:106--115, 2013.

\bibitem[Leg20]{legendreZEN}
S.~Legendre.
\newblock Zen, eco-evolutionary software, 2020.
\newblock https://www.biologie.ens.fr/~legendre/zen/zen.html.

\bibitem[LM96]{law1996permanence}
Richard Law and R~Daniel Morton.
\newblock Permanence and the assembly of ecological communities.
\newblock {\em Ecology}, 77(3):762--775, 1996.

\bibitem[Lot25]{lotka}
A.~J. Lotka.
\newblock {\em Elements of Physical Biology}.
\newblock Williams and Watkins, Baltimore, MD, 1925.

\bibitem[Lov12]{lovasz}
L.~Lov\'{a}sz.
\newblock {\em Large networks and graph limits}, volume~60 of {\em American
  Mathematical Society Colloquium Publications}.
\newblock American Mathematical Society, Providence, RI, 2012.

\bibitem[LT00]{LehmanTilman2000}
C.L. Lehman and D.~Tilman.
\newblock Biodiversity, stability, and productivity in competitive communities.
\newblock {\em American Naturalist}, 156(5):534--552, 2000.

\bibitem[LW19]{leewilkinson}
Clement Lee and Darren~J Wilkinson.
\newblock A review of stochastic block models and extensions for graph
  clustering.
\newblock {\em Applied Network Science}, 4, 2019.

\bibitem[M\'16]{mel-livre16}
Sylvie M\'{e}l\'{e}ard.
\newblock {\em Mod\`eles al\'{e}atoires en ecologie et evolution}, volume~77 of
  {\em Math\'{e}matiques \& Applications (Berlin) [Mathematics \&
  Applications]}.
\newblock Springer-Verlag, Berlin, 2016.

\bibitem[Mac69]{mac1969species}
Robert MacArthur.
\newblock Species packing, and what competition minimizes.
\newblock {\em Proceedings of the National Academy of Sciences},
  64(4):1369--1371, 1969.

\bibitem[May72]{may1972will}
Robert~M May.
\newblock Will a large complex system be stable?
\newblock {\em Nature}, 238(5364):413--414, 1972.

\bibitem[MGM{\etalchar{+}}96]{metzgeritzmeszenajacobsheerwaarden}
J.A.J. Metz, S.A.H. Geritz, G.~Mesz\'{e}na, F.A.J. Jacobs, and J.S.~Van
  Heerwaarden.
\newblock Adaptative dynamics, a geometrical study of the consequences of
  nearly faithful reproduction.
\newblock {\em S.J. Van Strien \& S.M. Verduyn Lunel (ed.), Stochastic and
  Spatial Structures of Dynamical Systems}, 45:183--231, 1996.

\bibitem[MGPM06]{meszena}
Géza Meszéna, Mats Gyllenberg, Liz Pásztor, and Johan~A.J Metz.
\newblock Competitive exclusion and limiting similarity: A unified theory.
\newblock {\em Theoretical population biology}, 69(1):68--87, 2006.

\bibitem[MK16]{mougi2016}
A.~Mougi and M.~Kondoh.
\newblock Food-web complexity, meta-community complexity and community
  stability.
\newblock {\em Scientific Reports}, 6:24478, 2016.

\bibitem[MLC92]{Marrow1992}
Paul Marrow, Richard Law, and C.~Cannings.
\newblock The coevolution of predator-prey interactions : Esss and red queen
  dynamics.
\newblock {\em Proceedings of the Royal Society of London. Series B: Biological
  Sciences}, 250(1328):133--141, 1992.

\bibitem[MM17a]{matias2017clustering}
Catherine Matias and Vincent Miele.
\newblock Statistical clustering of temporal networks through a dynamic
  stochastic block model.
\newblock {\em Journal of the Royal Statistical Society. Series B (Statistical
  Methodology)}, 79(4):1119--1141, 2017.

\bibitem[MM17b]{mielematias2017}
V.~Miele and C.~Matias.
\newblock Revealing the hidden structure of dynamic ecological networks.
\newblock {\em Royal Society Open Science}, 4:170251, 2017.

\bibitem[MMO{\etalchar{+}}21]{miele_etal_2021}
V.~Miele, C.~Matias, M.~Ohlmann, G.~Poggiato, S.~Dray, and W.~Thuiller.
\newblock Quantifying the overall effect of biotic interactions on species
  communities along environmental gradients.
\newblock The code for ELGRIN is part of the Econetwork R package, 2021.

\bibitem[MP67]{marcenkopastur1967math}
V.~A. Mar\v{c}enko and L.~A. Pastur.
\newblock The spectrum of random matrices.
\newblock {\em Teor. Funkci\u{\i} Funkcional. Anal. i Prilo\v{z}en. Vyp.},
  4:122--145, 1967.

\bibitem[MPV86]{mez-par-vir-(livre)86}
M.~M{\'e}zard, G.~Parisi, and M.~Virasoro.
\newblock {\em Spin Glass Theory and Beyond}.
\newblock World Scientific, 1986.

\bibitem[MR95]{molloyreed}
M.~Molloy and B.~Reed.
\newblock A critical point for random graphs with a given degree sequence.
\newblock {\em Random structures and algorithms}, 6:161--180, 1995.

\bibitem[MRA20a]{momalrobinambroise}
R.~Momal, S.~Robin, and C.~Ambroise.
\newblock Accounting for missing actors in interaction network inference from
  abundance data.
\newblock 2020.

\bibitem[MRA20b]{momalrobinambroise2}
R.~Momal, S.~Robin, and C.~Ambroise.
\newblock Tree-based inference of species interaction networks from abundance
  data.
\newblock {\em Methods in Ecology and Evolution}, 2020.

\bibitem[MRV18]{matiasrebafkavillers}
C.~Matias, T.~Rebafka, and F.~Villers.
\newblock A semiparametric extension of the stochastic block model for
  longitudinal networks.
\newblock {\em Biometrika}, 105(3):665--680, 2018.

\bibitem[MTB21]{marcus2021transition}
Stav Marcus, Ari~M. Turner, and Guy Bunin.
\newblock Local and collective transitions in sparsely-interacting ecological
  communities, 2021.

\bibitem[MY97]{murty1997}
K.G. Murty and F-T. Yu.
\newblock {\em Linear Complementarity, Linear and Nonlinear Programming}.
\newblock Internet Edition, 1997.

\bibitem[NHvdK{\etalchar{+}}07]{neutel2007}
Anje-Margriet Neutel, Johan A.~P. Heesterbeek, Johan van~de Koppel, Guido
  Hoenderboom, An~Vos, Coen Kaldeway, Frank Berendse, and Peter~C. de~Ruiter.
\newblock Reconciling complexity with stability in naturally assembling food
  webs.
\newblock {\em Nature}, 449(7162):599--602, 2007.

\bibitem[NJB13]{NuJoBa}
Scott~L Nuismer, Pedro Jordano, and Jordi Bascompte.
\newblock Coevolution and the architecture of mutualistic networks.
\newblock {\em Evolution}, 67(2):338--354, 2013.

\bibitem[OH08]{okuyamaholland}
Toshinori Okuyama and J.~Nathaniel Holland.
\newblock Network structural properties mediate the stability of mutualistic
  communities.
\newblock {\em Ecology letters}, 11(3):208--216, 2008.

\bibitem[OR14a]{o2014low}
S.~O'Rourke and D.~Renfrew.
\newblock Low rank perturbations of large elliptic random matrices.
\newblock {\em Electronic Journal of Probability}, 19, 2014.

\bibitem[OR14b]{orourkerenfrew2014}
Sean O'Rourke and David Renfrew.
\newblock {Low rank perturbations of large elliptic random matrices}.
\newblock {\em Electronic Journal of Probability}, 19(none):1 -- 65, 2014.

\bibitem[ORB07]{Otto2007}
Sonja~B. Otto, Bjorn~C. Rall, and Ulrich Brose.
\newblock Allometric degree distributions facilitate food-web stability.
\newblock {\em Nature}, 450(7173):1226--1229, 2007.

\bibitem[Pim79]{Pimm79}
Stuart~L. Pimm.
\newblock The structure of food webs.
\newblock {\em Theoretical Population Biology}, 16(2):144--158, 1979.

\bibitem[RBBB20]{roy2020complex}
Felix Roy, Matthieu Barbier, Giulio Biroli, and Guy Bunin.
\newblock Complex interactions can create persistent fluctuations in
  high-diversity ecosystems.
\newblock {\em PLoS computational biology}, 16(5):e1007827, 2020.

\bibitem[RBBC19]{roy2019numerical}
Felix Roy, Giulio Biroli, Guy Bunin, and Chiara Cammarota.
\newblock Numerical implementation of dynamical mean field theory for
  disordered systems: Application to the lotka--volterra model of ecosystems.
\newblock {\em Journal of Physics A: Mathematical and Theoretical},
  52(48):484001, 2019.

\bibitem[RBL{\etalchar{+}}19]{Romanuk2019}
Tamara~N. Romanuk, Amrei Binzer, Nicolas Loeuille, W.~Mather~A. Carscallen, and
  Neo~D. Martinez.
\newblock Simulated evolution assembles more realistic food webs with more
  functionally similar species than invasion.
\newblock {\em Scientific Reports}, 9(1):18242, 2019.

\bibitem[RD87]{robinson1987does}
James~F Robinson and Jaime~E Dickerson.
\newblock Does invasion sequence affect community structure?
\newblock {\em Ecology}, 68(3):587--595, 1987.

\bibitem[Red85]{redheffer1985volterra}
Ray Redheffer.
\newblock Volterra multipliers i.
\newblock {\em SIAM Journal on Algebraic Discrete Methods}, 6(4):592--611,
  1985.

\bibitem[RNMB16]{rohrmazza}
Rudolf~P Rohr, Russell~E Naisbit, Christian Mazza, and Louis-Felix Bersier.
\newblock Matching-centrality decomposition and the forecasting of new links in
  networks.
\newblock {\em Proceedings of the Royal Society. B, Biological sciences},
  283(1824):20152702--, 2016.

\bibitem[Rob74]{roberts74}
Alan Roberts.
\newblock The stability of a feasible random ecosystem.
\newblock {\em Nature}, 251(5476):607--608, 1974.

\bibitem[Rou79]{roughgarden}
Jonathan Roughgarden.
\newblock {\em Theory of population genetics and evolutionary ecology: an
  introduction}.
\newblock Macmillan, New York, 1979.

\bibitem[RR85]{Rummel1985}
J.~D. Rummel and J.~Roughgarden.
\newblock A theory of faunal buildup for competition communities.
\newblock {\em Evolution}, 39(5):1009--1033, 1985.

\bibitem[RSVNS20]{rodriguez2020climbing}
Pablo Rodr{\'\i}guez-S{\'a}nchez, Egbert~H Van~Nes, and Marten Scheffer.
\newblock Climbing escher’s stairs: a way to approximate stability landscapes
  in multidimensional systems.
\newblock {\em PLoS computational biology}, 16(4):e1007788, 2020.

\bibitem[SB11]{stoufferbascompte2011}
Daniel~B Stouffer and Jordi Bascompte.
\newblock Compartmentalization increases food-web persistence.
\newblock {\em Proceedings of the National Academy of Sciences - PNAS},
  108(9):3648--3652, 2011.

\bibitem[SCMA20]{servan}
Carlos~A. Serv{\'a}n, Jos{\'e}~A. Capit{\'a}n, Zachary~R. Miller, and Stefano
  Allesina.
\newblock Effects of phylogeny on coexistence in model communities.
\newblock {\em bioRxiv}, 2020.

\bibitem[SCSS88]{Sommers1988}
H.~J. Sommers, A.~Crisanti, H.~Sompolinsky, and Y.~Stein.
\newblock Spectrum of large random asymmetric matrices.
\newblock {\em Physical Review Letters}, 60(19):1895--1898, 1988.

\bibitem[SRB{\etalchar{+}}17]{saavedra2017structural}
S.~Saavedra, R.~P. Rohr, J.~Bascompte, O.~Godoy, N.J.B. Kraft, and J.~M.
  Levine.
\newblock A structural approach for understanding multispecies coexistence.
\newblock {\em Ecological Monographs}, 87(3):470--486, 2017.

\bibitem[SRG07]{santamaria}
Luis Santamaria and Miguel~A Rodriguez-Girones.
\newblock Linkage rules for plant-pollinator networks: Trait complementarity or
  exploitation barriers?
\newblock {\em PLoS biology}, 5(2):354--362, 2007.

\bibitem[Sto18]{stone2018}
L.~Stone.
\newblock The feasibility and stability of large complex biological networks: a
  random matrix approach.
\newblock {\em Scientific Reports}, 8(1):8246, 2018.

\bibitem[Szn84]{szn-84}
A.-S. Sznitman.
\newblock {\'E}quations de type de {B}oltzmann, spatialement homog{\`e}nes.
\newblock {\em Wahrscheinlichkeitstheorie und verwandte Gebiete},
  66(4):559--592, 1984.

\bibitem[TA14]{tangallesina2014}
Si~Tang and Stefano Allesina.
\newblock Reactivity and stability of large ecosystems.
\newblock {\em Frontiers in Ecology and Evolution}, 2, 2014.

\bibitem[Tak96]{takeuchi1996global}
Y.~Takeuchi.
\newblock {\em Global dynamical properties of Lotka-Volterra systems}.
\newblock World Scientific, 1996.

\bibitem[Tao12]{tao2012math}
Terence Tao.
\newblock {\em Topics in random matrix theory}, volume 132 of {\em Graduate
  Studies in Mathematics}.
\newblock American Mathematical Society, Providence, RI, 2012.

\bibitem[TC92]{TaperCase1992}
Mark~L. Taper and Ted~J. Case.
\newblock Models of character displacement and the theoretical robustness of
  taxon cycles.
\newblock {\em Evolution}, pages 317--333, 1992.

\bibitem[TE05]{trefethenbook}
Lloyd~Nicholas Trefethen and Mark Embree.
\newblock {\em Spectra and pseudospectra: the behavior of nonnormal matrices
  and operators}.
\newblock Princeton University Press, 2005.

\bibitem[TF10]{thebaultfontaine}
E.~Th\'ebault and C.~Fontaine.
\newblock Stability of ecological communities and the architecture of
  mutualistic and trophic networks.
\newblock {\em Science}, 329(5993):853--856, 2010.

\bibitem[TPA14]{tangpawarallesina2014}
Si~Tang, Samraat Pawar, and Stefano Allesina.
\newblock Correlation between interaction strengths drives stability in large
  ecological networks.
\newblock {\em Ecology Letters}, 17(9):1094--1100, 2014.

\bibitem[Ver96]{verhulst}
F.~Verhulst.
\newblock {\em Nonlinear differential equations and dynamical systems}.
\newblock Springer-Verlag, Berlin Heidelberg, 2 edition, 1996.

\bibitem[Vol31]{volterra}
V.~Volterra.
\newblock Variations and fluctuations of the number of individuals in animal
  species living together.
\newblock In R.~N. Chapman, editor, {\em Animal Ecology}, pages 409--448, New
  York, 1931. McGraw-Hill.

\bibitem[Wig67]{wigner}
Eugene~P. Wigner.
\newblock Random matrices in physics.
\newblock {\em SIAM Review}, 9(1):1--23, 1967.

\bibitem[WL14]{wangloreau2014}
Shaopeng Wang and Michel Loreau.
\newblock Ecosystem stability in space: $\alpha$, $\beta$ and $\gamma$
  variability.
\newblock {\em Ecology Letters}, 17(8):891--901, 2014.

\bibitem[Zee93]{zeeman}
M.L. Zeeman.
\newblock Hopf bifurcations in competitive three-dimensional {L}otka-{V}olterra
  systems.
\newblock {\em Dynam. Stability Systems}, 8(3):189--217, 1993.

\bibitem[ZZ02]{zeemanX2}
E.~C. Zeeman and M.~L. Zeeman.
\newblock An {$n$}-dimensional competitive {L}otka-{V}olterra system is
  generically determined by the edges of its carrying simplex.
\newblock {\em Nonlinearity}, 15(6):2019--2032, 2002.

\end{thebibliography}

\newcommand{\etalchar}[1]{$^{#1}$}


\appendix

\section{May's model} \label{ann:may_model}

\subsection{The mathematics behind May's result}

Inspired by Gardner and Ashby's numerical results \cite{GardnerAshby1970}, Robert May proposed a first mathematical model \cite{may1972will} to link the stability of an ecosystem with its complexity. 
In this model, the ecosystem is represented by a vector of $N$ functions $\boldsymbol{n}:t \mapsto (n_i(t))_{i\in [N]}$, the quantity $n_i(t)$ corresponding to the abundance of species number $i$ at time $t$.

The vector of the abundances  $\boldsymbol{n}$ satisfies a system of first order nonlinear differential equations of the form:
\beq \frac{\dd n_i}{\dd t}=F_i(\boldsymbol{n}). \label{eq:nonlinearequation_may}
\eeq
(May did not particularly consider the form of the differential
equation given by Eq.~\eqref{eq:premiere} in the introduction). 
The main interest of May lies in the study of stability of equilibria of such systems. We assume the existence of an equilibrium $\boldsymbol{n}^*=(n_i^*)_{i\in [N]}$ and write the abundance of species number $i$ as $n_i(t)=n_i^* + \varepsilon_i(t)$. Near the equilibrium, the stability of the nonlinear system \eqref{eq:nonlinearequation_may} boils down to the stability of the linear system
\beq 
\frac{\dd\bs{\varepsilon}}{\dd t}=J(\boldsymbol{n}^*)\bs{\varepsilon},\label{eq:evol_Jac_annexe}
\eeq
where $J:=J(\boldsymbol{n}^*)$ is the $N \times N$ Jacobian matrix \beq J_{ij} := \dfrac{\partial F_i }{ \partial n_j} (\boldsymbol{n}^*).\label{eq:Jac_annexe} \eeq In particular, the equilibrium is Lyapunov-stable if and only if all the eigenvalues of $J$ have negative real parts.

The main contribution of May was to model the Jacobian as a random matrix in order to use mathematical results from RMT. More precisely, May chose to replace the self-interaction coefficients by $-1$ and all the other coefficients by independent centered random variables so that:
\beq J=-I+M \label{eq:jacobian_may}\eeq  where $M_{ii}=0$ and for $i \ne j$,  $M_{ij}$ are i.i.d centered random variables with variance $V:=\var(M_{ij})$, and with a distribution independent from $N$.


May addressed the problem of determining what are the conditions on $N$ and $V$ to ensure that all the eigenvalues have negative real part. He relied on the result by Ginibre \cite{ginibre1965statistical} who proved that asymptotically in $N$, the eigenvalues of matrix $M$ are contained in a disk centered at $(-1,0)$ with radius $\sqrt{NV}$. This lead May to state the following phase transition :


\begin{prop} (May \cite{may1972will}, 1972)
If the matrix $J$ is given by \eqref{eq:jacobian_may}, the equilibrium is stable with high probability if \beq V<\frac{1}{N}\eeq
and unstable with high probability if \beq V>\frac{1}{N}.\eeq
\end{prop}

In fact, Ginibre's result is not enough to justify this phase transition but we need to understand the spectral radius of the matrix $J$. The first results on the spectral radius were obtained by Füredi and Koml\'{o}s \cite{Furedi1981} in the Hermitian case and by Bai and Yin \cite{baiyin1986rayonspectral}, Geman \cite{geman1986spectral} and Geman and Hwang \cite{geman1982chaos} in the general case. Recently, Bordenave and al \cite{bordenave2021convergence} show its convergence in probability under optimal moment assumption.


As the condition to get stability involves the number $N$ of species, what we need to deduce May's result rigorously is a concentration inequality for the spectral radius. Such an inequality has been established in \cite{Bordenave2018}.

\begin{theo}
    Let $X_N=(X_{ij})$ denote the random $N \times N$ matrix, where $X_{ij}$ are independent copies of a given symmetric complex random variable, with $\mathbb{E}[|X_{11}|^2] \le 1$. If there exists $\epsilon>0$ and $B>0$ such that $\mathbb{E}[|X_{11}|^{2+\epsilon}] \le B$, then, for any $\delta>0$, there exists a constant $K:=K(\epsilon, \delta, B)>0$, such that for any $N \in \mathbb{N}$, we have 
    \beq \mathbb{P}\left[\rho(X_N) \ge (1+\delta)\sqrt{N}\right] \le \frac{K}{(\log N)^2}.\eeq
\end{theo}

It means that, in this first model\footnote{There are slight differences between the model in \cite{Bordenave2018} and the model \eqref{eq:jacobian_may} : in \cite{Bordenave2018}, the symmetry of the law of the entries is required and all the entries of the matrix, including the diagonal entries, are i.i.d copies of the same variable while in \eqref{eq:jacobian_may}, the diagonal entries are put to zero.}, for $N$ large enough, with high probability, there is no eigenvalue of J outside the disk centered at $-1$ and of radius $\sqrt{NV}$. 

\begin{figure}[h]

  \begin{subfigure}{.46\linewidth}\label{subfig:B}
    \includegraphics[scale=0.45]{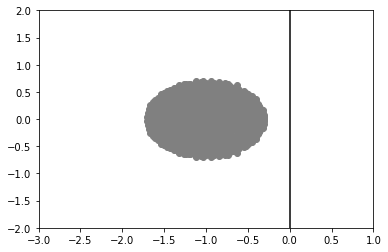}
    \caption{Stability of the spectrum.}
  \end{subfigure}%
  \hspace*{\fill}   
  \begin{subfigure}{0.46\textwidth}\label{subfig:A}
    \includegraphics[scale=0.45]{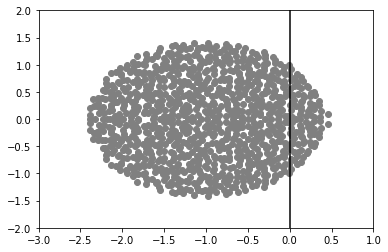}
    \caption{Some eigenvalues have positive real part.}
  \end{subfigure}%

\caption{Spectrum of the Jacobian matrix $J=M - I$, for $N=1000$ species with $C=1$, the entries $M_{ij}$, for $i \ne j$ are independent normal centered variables with variance $V$, with $V=\frac{1}{2N} < \frac{1}{N}$ in (A) and $V=\frac{2}{N} > \frac{1}{N}$ in (B).}   \label{fig:spectrum-jacobian}
\end{figure}

The spectrum of the Jacobian matrix is illustred in Figure \ref{fig:spectrum-jacobian}.

May already considered a sparse version of his initial model in \cite{may1972will}, where each possible interaction takes place with probability $C$ (connectance), independently of all the other interactions, see Section \ref{subsec:intro-sparse} for an introduction to sparsity. It means that in average, each species effectively interacts with a proportion $C$ of all the other species.
The phase transition can be now stated as :
\begin{prop}(May \cite{may1972will}, 1972)
\label{prop:may-connectance}
If $C$ is the connectance of the model, the equilibrium is stable with high probability if \beq CV<\frac{1}{N}\eeq
and unstable with high probability if \beq CV>\frac{1}{N}.\eeq
\end{prop}

Mathematically speaking, it is convenient to use the following formalism. Denote by $\Delta_{\text{ER}}$ the adjacency matrix of the Erdös-Rényi graph (see Section \ref{subsec:ER-graph} for details) meaning that each entry of $\Delta_{\text{ER}}$ has probability $C$ to be equal to $1$ and is $0$ otherwise. In other words, each species has an effect on another species with probability $C$. Then, the matrix $M$ introduced in $\eqref{eq:jacobian_may}$ is replaced by $\widetilde{M}$ equal to :
\beq \widetilde{M}:=\Delta_{\text{ER}}  \circ M = \left( [\Delta_{\text{ER}}]_{ij} M_{ij}\right)\, . 
\label{eq:erdos_renyi_may} 
\eeq
The parameter $C$ can therefore be interpreted as the average number of neighbours of a given vertex of the graph.

In this new model, the matrix $\widetilde{M}$ has the same distribution as $M$ except that $\var(\widetilde{M}_{ij})=CV$, for $i \ne j$, leading to Proposition \ref{prop:may-connectance}.

\subsection{Relation between May's model and Lotka-Volterra framework}
Consider Eq. \eqref{eq:LV} and assume that there exists an equilibrium $\bs{x}^*$. If $\bs{x}^*>0$ then one can easily compute the Jacobian at equilibrium. We provide hereafter a quick computation. Write $x_i(t)=x_i^*+\varepsilon_i(t)$ and notice that at equilibrium, since $x_i^*>0$, one has
$$
r_i =x_i^* - (\Gamma \bs{x}^*)_i\, .
$$
Now 
\begin{eqnarray*}
\frac{\dd x_i}{\dd t} &=& x_i ( r_i - x_i +(\Gamma \bs{x})_i) \,,\\
&=& (x_i^*+\varepsilon_i) ( r_i - (x_i^*+\varepsilon_i) +(\Gamma (\bs{x} +\bs{\varepsilon}))_i)\,,\\
&=& (x_i^*+\varepsilon_i) ( x_i^* - (\Gamma \bs{x}^*)_i - (x_i^*+\varepsilon_i) +(\Gamma (\bs{x} +\bs{\varepsilon}))_i)\,,\\
&=& (x_i^* +\varepsilon_i) (-\varepsilon_i +(\Gamma \bs{\varepsilon})_i) \,,\\
&=& [\mathrm{diag}(\bs{x}^*) ( -I + \Gamma)\bs{\varepsilon}]_i + o(\bs{\varepsilon})\, .
\end{eqnarray*}
Hence the Jacobian 
\begin{equation}\label{eq:jacobian-LV}
J(\bs{x^*}) =  \mathrm{diag}(\bs{x}^*)(-I +\Gamma)\,.
\end{equation}

Formally, this Jacobian resembles May's Jacobian, with important differences: 
\begin{itemize}
    \item the underlying equilibrium $\bs{x}^*$ must be feasible (i.e. $\bs{x}^*>0$). If $\Gamma$ is random, then conditions for feasibility are provided in Section \ref{subsec:feasibility}. Under these conditions, stability is granted (see for instance \cite[Corollary 1.4]{bizeul2021positive}).
    \item in the Jacobian formula $\mathrm{diag}(\bs{x}^*)(-I +\Gamma)$, there is the product of an extra matrix, namely $\mathrm{diag}(\bs{x}^*)$, with matrix $-I+\Gamma$ appears. Notice that if $\Gamma$ is random, then $\mathrm{diag}(\bs{x}^*)$ and $\Gamma$ are dependent since $\bs{x}^* = \bs{r} +\Gamma \bs{x}^*$ for a feasible equilibrium.
\end{itemize} 

Stone \cite{stone2018} considered a Jacobian matrix of this form, but with the simplifying assumption that the entries of $\textrm{diag}(\bs{x}^*)$ are independent from $\Gamma$. Then $J$ is unstable only if $\Gamma$ is unstable, and thus May's criterion would still hold. 

This independence assumption between $D$ and $\Gamma$ is strong as the subset of surviving species and their abundances are function of the interaction matrix $\Gamma$, hence not independent. 

Notice that Stone \cite{stone2018} does not consider the normalization under which the equilibrium $\bs{x}^*$ is feasible. It is thus not clear that the considered formula for the Jacobian is associated to a LV system.

\subsection{Other models}
Some authors have used more involved models for the Jacobian matrix, such as the elliptic model \ref{def:ellipticalmodel}
(see e.g. \cite{tangallesina2014}) and derived from there similar criteria for stability expressed in terms of the parameters if the model. 

We also mention other possible random models, such as studied in Ben Arous et al. \cite{benarousfyodorovkhoruzhenko,fyodorovkhoruzhenko}:
\begin{equation}
    \frac{\dd x_i}{\dd t}= - x_i(t)  + f_i(\mathbf{x})\,,
\end{equation}where $f_i(\mathbf{x})$ is a smooth random vector field which models the complexity and nonlinearity of interactions. It is assumed that for all $i\in [N]$,
\begin{equation}
    f_i(\mathbf{x})=-\frac{\partial V}{\partial x_i}(\mathbf{x})+\frac{1}{\sqrt{N}}\sum_{j=1}^N \frac{\partial A_{ij}}{\partial x_j}(\mathbf{x}),
\end{equation}
where the matrix $A$ is a random antisymmetric matrix independent of the random scalar potential $V$.

\section{Lotka-Volterra system in two dimensions}\label{section:LV-2d}

{The question of long-time behavior of Lotka-Volterra system is a very delicate one that has motivated an already large literature. Let us first discuss briefly the results for a deterministic matrix $\Gamma$ with $N$ small. The $N=2$ setting provides some basic heuristics for understanding when the dynamics admits a single equilibrium with or without species going extinct, multiple equilibria, or oscillatory behaviors (e.g. Appendix \ref{section:LV-2d}, and also \cite{hofbauer1998evolutionary,verhulst}). }\\

Recall the Lotka-Volterra system \eqref{eq:LV} for $N=2$ species and with a deterministic matrix $\Gamma$ (e.g. \cite{hofbauer1998evolutionary,verhulst}). This system generally admits four equilibria:  
\[(0,0),\qquad \left(0,\frac{r_2}{1-\Gamma_{22}}\right),\qquad \left(\frac{r_1}{1-\Gamma_{11}},0\right),\qquad \mbox{ or }\]
\begin{equation}\left(\, \dfrac{(1 -\Gamma_{22}) r_1 + \Gamma_{12} r_2 }{(1 -\Gamma_{11})(1 -\Gamma_{22}) - \Gamma_{12}\Gamma_{21}} \,,\ \dfrac{\Gamma_{21} r_1+(1 -\Gamma_{11}) r_2  }{(1 -\Gamma_{11})(1 -\Gamma_{22}) - \Gamma_{12}\Gamma_{21}}\right). \label{equilibre:LV2-2coord}\end{equation}
The stability of these equilibria can be discussed from the computation of the Jacobian matrix at these critical points. \\

\noindent \textbf{If either $x_1<0$ or $x_2<0$ at the equilibrium \eqref{equilibre:LV2-2coord},} that particular equilibrium is unfeasible and does not correspond to an admissible solution. We generally find that in these cases one species goes asymptotically to extinction, and is said to be excluded by the other species (e.g. competitive exclusion). The fact that only one species survives is a classic setting of adaptive dynamics, see Metz et al. \cite{metzgeritzmeszenajacobsheerwaarden} and Champagnat \cite{champagnat06}: this corresponds to the rule that \textit{invasion implies fixation}, meaning that the weakest species is lost when a favourable mutant arises. Provided new mutant species arrive sufficiently slowly into the system, the evolution of the population can be described by the sequence of successive dominating species or \textit{trait substitution sequence}.

\noindent \textbf{If both $x_1>0$ and $x_2>0$ at the equilibrium \eqref{equilibre:LV2-2coord}:} the fixed point $(x_1,x_2)$ is feasible. Linear stability analysis reveals that the feasible equilibrium is stable if  $\Gamma_{12} \Gamma_{21} < (1 -\Gamma_{11})(1 -\Gamma_{22})$, unstable otherwise (see e.g. chapter 6 in \cite{EdelsteinKeshet1988}). \\
When the equilibrium \eqref{equilibre:LV2-2coord} is unstable, the dynamics admits  two stable fixed points, with either species 1 or species 2 extinct (both equilibria are uninvadable, insofar as the extinct species, if introduced with small abundance, will decay exponentially). This phenomenon, known as mutual exclusion, provides a basic template for the existence of multiple stable states in Lotka-Volterra dynamics. \\
The case $\Gamma_{12}\Gamma_{21}=(1 -\Gamma_{11})(1 -\Gamma_{22})$ is singular and corresponds to a situation where one species always goes extinct~\cite{arnoldi2019fitness}. \\

Let us now discuss the possible existence of cycles. The original predator-prey model of Lotka and Volterra had antisymmetric interactions $\Gamma_{12}=-\Gamma_{21}$, and $\Gamma_{11}-1=\Gamma_{22}-1=0$ and $r$'s of opposite signs, so that the dynamics reduces to
\beq \dfrac{\dd x_1}{\dd t}=x_1(r_1+\Gamma_{12} x_2),\qquad  \dfrac{\dd x_2}{\dd t}=x_2(r_2-\Gamma_{12} x_1) \label{eq:original-prey-pred}\eeq
In that case, the dynamics admits neutral cycles around the marginally stable fixed point $(r_2/\Gamma_{12}, - r_1/\Gamma_{12})$.
Indeed, it can be checked that the function $F(t)=\Gamma_{12} (x_1(t)+x_2(t))+r_1 \log x_2(t)-r_2 \log x_1(t)$, called a first integral of \eqref{eq:LV}, remains constant over time.\\

For more general matrices $\Gamma$, the Bendixson-Dulac theorem gives a sufficient condition to show that there is no cycle (\cite{hofbauer1998evolutionary}. We also refer to \cite{Hirsch1982}, who proved that attractors of competitive or cooperative $N$-species systems could only be manifold of $N-1$ or fewer dimensions). The Bendixson-Dulax theorem in the case of the 2d-Lotka-Volterra equation is as follows:

\begin{prop}
If there exists a function $\varphi(x_1,x_2)$ such that 
\begin{multline}\frac{\partial}{\partial x_1}\Big(\varphi(x_1,x_2)\big(r_1x_1 + (\Gamma_{11}-1)x_1^2+ \Gamma_{12}x_1 x_2\big)\Big) \\ +\frac{\partial}{\partial x_2}\Big(\varphi(x_1,x_2)\big(r_2x_2 +\Gamma_{21} x_1 x_2 +(\Gamma_{22} -1) x_2^2\big)\Big)\label{dulac}\end{multline}has a constant sign in the positive quarter plane, then there is no cycle solution of the Lotka-Volterra system \eqref{eq:LV}.
\end{prop}

Let us discuss further the case where $\Gamma_{11}-1=\Gamma_{22}-1=0$, but now with symmetric interactions, $\Gamma_{12}=\Gamma_{21}$, which may arise for some competitive or mutualistic interactions. \\
If $r_1, r_2>0$ and $\Gamma_{12}>0$ (mutualistic interactions), we can choose $\varphi(x_1,x_2)= 1$ and \eqref{dulac} becomes
$r_1+r_2+\Gamma_{12}(x_1+x_2)$, which is positive on positive quarter plane, so the dynamics do not exhibit cycling. \\
If $r_1, r_2>0$ and $\Gamma_{12}<0$ (competition), there is a saddle point at $(-r_1/\Gamma_{12},-r_2/\Gamma_{12})$ and $(0,0)$ is repulsive, so there is also no possible cycle.\\
As a general heuristic, antisymmetric interactions (as in the original prey-predator model\eqref{eq:original-prey-pred}) favor cycling around fixed points, whereas symmetric interactions favor taking a shortest path toward fixed points.

Let us now consider the general case with an arbitrary matrix $\Gamma$.
Choosing $\varphi(x_1,x_2)=1/(x_1 x_2)$ for example, we obtain that \eqref{dulac} is equal to \[\frac{(\Gamma_{11}-1) x_1+(\Gamma_{22}-1)x_2}{x_1 x_2}\]
so there is no cycle if $\Gamma_{11}>1$ and $\Gamma_{22}>1$ or if $\Gamma_{11}<1$ and $\Gamma_{22}<1$.\\
In all generality, the Bendixson-Dulac theorem proves that the two-species Lotka-Volterra system admits no \textit{isolated} periodic orbit, whatever the values of the $\Gamma_{ij}$ and $r_i$ \cite[p.33]{hofbauer1998evolutionary} (the cycles exhibited above being non isolated).\\




{Let us briefly say that in higher dimensions, the complexity increases exponentially. For three competing species, Zeeman \cite{zeeman} described the compact limit sets of these systems, which are either fixed points or periodic orbits (conforming to the general result of \cite{Hirsch1982}), and found 33 different equivalence classes. For the general case of $N$ competitive species, Zeeman and Zeeman \cite{zeemanX2} have studied the carrying simplex that attracts all non-zero orbits and carries the asymptotic dynamics. }

\section{From individual-based models to Lotka-Volterra system}
\label{annexe:micro1}

As explained in the main text, the Lotka-Volterra multidimensional ordinary
differential equation~\eqref{eq:LV} or its stochastic analogue~\eqref{sde} can
both be obtained as the large population limit of a stochastic individual based
\textit{birth-death model with interactions}, see, e.g., 
\cite{bansayemeleard,mel-livre16,fourniermeleard,champagnatferrieremeleard}. 

The goal of the present appendix will be to derive \eqref{eq:LV} from individual-based models,
the other variants will be presented in Appendix \ref{annexe:micro2}.
In both appendices \ref{annexe:micro1} and \ref{annexe:micro2}, our exposition will be mainly inspired by \cite{mel-livre16}.

\subsection{One species as a birth and death process}\label{app:one_species_random}

Our starting point will be to describe the birth and death model for a single
species. This is done by means of a homogeneous
Markov jump process $(Y(t))_{t\geq 0}$ with values in $\N$. The transition probabilities of $Y$, defined by 
$P_{nm}(h) = \PP(Y({t+h}) = m \, | \, Y(t) = n) $
for $h\geq 0$ and $n,m\in\N$ are such that:
\begin{equation}
\label{geninf} 
\begin{aligned} 
P_{n,n+1}(h) &= b_n h + o(h), \quad &\text{for } & n>0\,,  \\
P_{n,n-1}(h) &= d_n h + o(h), \quad &\text{for } &  n>0\,,\\
P_{n,m}(h) \phantom{a,}& = o(h),\quad &\text{for }&|n-m|>1\,,\\
\end{aligned} 
\end{equation} 
where $(b_n)_{n\in\N}$ and $(d_n)_{n\in\N}$ are two sequences of real
non-negative numbers. For example, if individuals are exchangeable with individual birth rate $b$ and death rate $d$, then,  $b_n=b\times n$ and $d_n=d\times n,$ so that $b_n$ and $d_n$ are respectively the birth and death rates when the population is of size $n.$ We refer to \cite{mel-livre16} for more details.

We note for further use that such processes can be advantageously described by
stochastic differential equations involving random Poisson point measures
\cite{ikedawatanabe,bansayemeleard}. Denoting as $\bs N(\dd s, \dd u)$ the random
Poisson measure on $\R_+ \times \R_+$ with the intensity measure the product of
the Lebesgue measures $\dd s \otimes \dd u$, a birth and death process with
respective birth and death rate sequences $(b_n)$ and $(d_n)$ can be
written as
\begin{equation}
\label{poiss-1spec} 
Y(t) = Y(0) + \int_0^t \int_{\R_+} \left( 
 \ind_{u\leq b_{Y(s^-)}} -  
 \ind_{b_{Y(s^-)} \leq u \leq b_{Y(s^-)} + d_{Y(s^-)}} 
 \right) \bs N(\dd s, \dd u) \,. 
\end{equation} 
Such stochastic differential equations correspond to the individual-based simulation algorithms often
used by biologists (see, {\it e.g.}, Gillespie \cite{gillespie}). In these
equations, the Poisson point measure models the possible births or deaths and the indicators correspond to an
acceptation-rejection algorithm which ensures that the events occur with the
correct time-dependent and random rate $b_{Y(s^-)}$ or $d_{Y(s^-)}$. 

Using Poisson stochastic calculus, and introducing the compensated Poisson measure $\widetilde{\bs N}(\dd s, \dd u)=\bs N(\dd s, \dd u)- \dd s \otimes \dd u$, we have:
\begin{equation}\label{eq:poisson-martingale}
Y(t)=Y(0)+\int_0^t \big(b_{Y(s)}-d_{Y(s)}\big) \ \dd s + M(t)\,,
\end{equation}
where:
\[M(t) := \int_0^t \int_{\R_+} \left( 
 \ind_{u\leq b_{Y(s^-)}} -  
 \ind_{b_{Y(s^-)} \leq u \leq b_{Y(s^-)} + d_{Y(s^-)}} 
 \right) \widetilde{\bs N}(\dd s, \dd u)\]
 is a centered martingale with variance $\mathbb{E}\big(\int_0^t (b_{Y(s)}+d_{Y(s)}) \dd s\big)$. It is possible to rewrite \eqref{eq:poisson-martingale} in differential form as:
 \begin{equation}\label{eq:poisson-martingale2}
dY(t)= \big(b_{Y(t)}-d_{Y(t)}\big) \ \dd t + dM(t).
\end{equation}

\subsection{Asymptotics for one species}\label{app:asympt_one_species}

We now assume that the population size is large and introduce a
parameter $K > 0$, which is seen as a scaling parameter for the
initial population size (\textit{carrying capacity}). More precisely, we now denote our process as $Y^{K}(t)$ and the rescaled version 
\[X^K(t) :=  \frac{Y^{K}(t)}{K}\,. \] We assume that 
\[
X^K(0)=\frac{Y^{K}(0)}{K} \xrightarrow[K\to\infty]{} x_0 \quad \text{ in probability}\,,  
\]
where
$x_0 > 0$ is some deterministic or random positive real number. We also assume
that the birth and death rates depend on the scaling parameter $K$, and
denote them as $b_n^K$ and $d^K_n$ respectively. Our purpose is to study the dynamics of $X^K(t)$ in the asymptotic regime
$K\to\infty$, given different types of dependencies of the birth and death
rates on $K$. Note that since the jumps of $X^K(t)$ are of amplitude $1/K$,
the limiting process can take any value in $\R_+$. 

The first model for $b_n^K$ and $d_n^K$ will be the so-called 
\textit{logistic model}. Given three parameters $b, d,  c > 0$, this model
reads 
\begin{equation}
\label{logistic}
b_n^K = b \, n \quad \text{and} \quad 
d_n^K = d\,  n + \frac{c}{K} n^2 . 
\end{equation}According to this model, there is no interaction
between the individuals that constitute the species regarding the births,
since $b_n^K$ grows linearly with the population size. This is not the
case of the deaths, since $d^K_n$ has a quadratic component accounting for a
\textit{competition} among the individuals within the species to access the
limited amount of resources. 

From the equations~\eqref{geninf}, we easily see
that 
\begin{eqnarray*}
\EE \left(Y^K(t+h) - Y^K(t) \mid Y^K(t) = n\right) &=&
 \sum_m (m-n) P_{n,m}(h) 
 = ( b_n^K - d_n^K) h + o(h)\,, \\
\var\left(Y^K(t+h) - Y^K(t) \,  | \, Y^K(t) = n\right) &=& 
  (b_n^K + d_n^K) h + o(h)\,.
\end{eqnarray*}


On the other hand, the birth and death rates for the logistic model satisfy: for all $x\geq 0$,
\begin{align*}
    & \lim_{K\to\infty} \frac{b_{[Kx]}^K-d^K_{[Kx]}}{K} =\lim_{K\to\infty} (b-d)\frac{[Kx]}{K}-c\frac{[Kx]^2}{K^2}=
  rx - c x^2\\
&  \lim_{K\to\infty} \frac{b_{[Kx]}^K + d_{[Kx]}^K}{K^2} = \lim_{K\to\infty} (b+d)\frac{[Kx]}{K^2}+c\frac{[Kx]^2}{K^3}=0, 
\end{align*}
where $r = b-d$. We therefore get from the previous equations that
\begin{align*} 
\frac{\EE [X^K(t+h) - X^K(t) \, | \,  X^K(t)]}{h} &\simeq
 \frac{b^K_{Y^K(t)} - d^K_{Y^K(t)}}{K} \simeq 
  r X^K(t) - c X^K(t)^2, \\
\frac{\var(X^K(t+h) - X^K(t) \,  | \, X^K(t) )}{h} &\simeq 
 \frac{b^K_{Y^K(t)} + d^K_{Y^K(t)}}{K^2} \simeq 0
\end{align*} 
for large $K$.  This argument shows that the variance of the increments of
$X^K(\cdot)$ decreases faster than $h$, thus, heuristically, the stochasticity
of this process disappears for large $K$. Assuming that $x_0$ is deterministic and
using the expression of the conditional mean above, we thus infer that
$X^K(\cdot)$ converges in probability, and uniformly on every time interval $[0,T]$, to a deterministic process $x(\cdot)$
defined as the unique solution of the ODE 
\begin{equation}
\label{ode-1}
 \dot{x}(t)= r x(t) - c x(t)^2,\qquad  x(0) = x_0\ .
\end{equation}
Formally, this means
$$
\forall \varepsilon>0\,,\quad \mathbb{P} \left\{ \sup_{t\le T} \left| X^K(t) - x(t)\right| >\varepsilon\right\} \xrightarrow[K\to\infty]{} 0\, .
$$
Observe that $r = b-d$ can be seen as the population ``natural 
increase rate'' in the limit of the small population size. When $r > 0$, which
is usually the case, the competition for the available resources regulates the 
population. 

Obviously, the ODE~\eqref{ode-1} is the particular case of~\eqref{eq:LV}
obtained for $N = 1$.

The argument for showing the convergence towards the solution of~\eqref{ode-1}
can be made rigorous by establishing a compactness result over the laws of the
trajectories $X^K(\cdot)$ in the space $D([0,1],\R_+)$ of the so-called
{\it c\`adl\`ag} processes before identifying the equation satisfied by the limiting values $x(\cdot)$. The convergence finally results from the uniqueness of the limiting value. A complete proof can be found in~\cite{bansayemeleard}.

\subsection{Asymptotics for $N$ species with interactions} \label{Nspecies}

We now address the case of $N$ species with
interactions. Our (vector) process $\bs{Y}^K$ is now valued in $\NN^N$ and writes $\bs{Y}^K(t)
= ( Y_i^K(t))$ where $Y_i^K(t)$ is the population size of the
species $i$ at time $t$.  Let $\bs{b}^K=(b_i^K)$ and
$\bs{d}^K=(d_i^K)$ be two $N\times 1$ vectors with positive elements, and let $\Gamma^K = (
\Gamma_{ij}^K)$ be a $N\times N$ matrix. 

According to the model with
interactions, the individuals in the species $i$ reproduce with an individual rate
proportional to $b_i^K$, die with the individual natural death rate proportional to
$d_i^K$, and can interact with other individuals, say of the species $j$, with
a rate proportional to $\Gamma_{ij}^K-\ind_{i=j}$ resulting in an extra birth or death term. 
The indicator $\ind_{i=j}$ corresponds to the logistic competition term inside each species, as in Section \ref{app:asympt_one_species}. 
The matrix $\Gamma^K$ of the interactions may be deterministic or random. When it is random, all the computations that are presented below are made conditionally to $\Gamma^K.$ As previously, it is possible to write a stochastic differential equation involving a random Poisson point measure which generalizes~\eqref{poiss-1spec}.

Denote as $\bs{e}_i$ the $i^{\text{th}}$ canonical vector of $\RR^N$ and consider a population represented by vector $\bs{k}=(k_i)\in \NN^N$. We can now express the transition probabilities:
\begin{equation}\label{def:ratesmultid1}
\PP\left( \bs{Y}^K({t+h}) = \bs{k} + \bs{e}_i \, | \, \bs{Y}^K(t) = \bs{k}\right) =  \underbrace{\bigg(b_i^K k_i+\sum_{j:\,\Gamma^K_{ij}>0} \Gamma_{ij}^K k_i k_j\bigg)}_{\text{birth rate for species }i} h + o(h)\, . 
\end{equation}
\begin{equation}\label{def:ratesmultid2}
\PP\left( \bs{Y}^K({t+h}) = \bs{k} - \bs{e}_i \, | \, \bs{Y}^K(t) = \bs{k}\right) =  \underbrace{\bigg(d_i^K k_i+k_i^2 +\sum_{j:\,\Gamma_{ij}^K<0} |\Gamma_{ij}^K| k_i k_j \bigg)}_{\text{death rate for species }i} h +o(h)\, . 
\end{equation}
Now if $\bs{\ell}=(\ell_i)\in \NN^N$ represents another population distribution with 
$$
\|\bs{\ell}-\bs{k}\|_1 = \sum_{i} |\ell_i -k_i|\  >\  1\,,
$$ then:
\begin{equation}\label{def:ratesmultid3}
\PP\left(\bs{Y}^K({t+h}) = \bs{\ell} \, | \, \bs{Y}^K(t) = \bs{k}\right) =  o(h)\, . 
\end{equation}

Note that there is here a slight abuse of notations in \eqref{def:ratesmultid1} and \eqref{def:ratesmultid2}, where $b_i^K$ (resp. $d_i^K$) denotes to the individual birth (resp. death) rate of species $i,$ whereas in \eqref{logistic},
$b_n^K$  (resp. $d_n^K$) denotes the birth (resp. death) rate of a single species with population size $n$. In the sequel, we will use index $i \in [ N ]$ for numbering the species  and index $n \in \N$ for the size of a population.

The Lotka-Volterra equation~\eqref{eq:LV}
can be obtained along the same principle as for the single species case, when the number of species $N$ is fixed and their initial sizes are large. Consider again that the  scaling parameter $K >0$ goes to infinity and assume that each $Y_i(0)$ is of order $K$. As before, we consider
$$
\bs{X}^K(t) := \frac{\bs{Y}^K(t)}K\, .
$$
Assume that $\bs{X}^{K}(0)$ converges in probability to a deterministic vector $\bs{x}^0\in (0,\infty)^N$. The competition coefficient is rescaled as $\Gamma_{ij}^K=\Gamma_{ij}/K$, while the individual birth and death rates $b_i^K=b_i$ and $d_i^K= d_i$ are kept fixed. 
This model can be seen as a multiple species generalization of the logistic model introduced by
Equation~\eqref{logistic}. 

Write $\Gamma = ( \Gamma_{ij})$, define the vector 
$ \bs{r} = ( b_1 - d_1, \cdots, b_N - d_N)$ and recall the Hadamard product notation $\circ$ which also applies to two vectors $\bs{u}=(u_i)$ and $\bs{v}=(v_i)$ and yields $\bs{u}\circ \bs{v}=(u_i v_i)$.  
Mimicking the conditional expectation and conditional variance derivations
that follow Equation~\eqref{logistic}, one can compute that 
\begin{eqnarray*}
 \frac{\EE \left(\bs{X}^K(t+h) - \bs{X}^K(t) \, | \,  \bs{X}^K(t)\right)}{h} &\simeq&
\bs{X}^K(t) \circ \left(  \bs{r}  - \bs{X}^K(t)+  \Gamma \bs{X}^K(t) \right) \,, \\
     \frac{\Cov\left(\bs{X}^K(t+h) - \bs{X}^K(t) \, | \, \bs{X}^K(t)\right)}h  & \simeq & 0\,,
\end{eqnarray*}
for large $K$. With these heuristical derivations, we infer that the sequence of processes 
$(\bs{X}^K)$ converges in probability to the deterministic process $\bs{x}$ defined as 
the solution of the multivariate ODE 
$$
 \frac{\dd\bs{x}}{\dd t}(t) = \bs{x}(t) \circ \left(  \bs{r} - \bs{x}(t)+ \Gamma \bs{x}(t) \right) ,\qquad  \bs{x}(0) = \bs{x}^0\,.
$$
which writes componentwise
$$
\frac{\dd x_i}{\dd t}(t) = x_i(t) \left(r_i - x_i(t)+ (\Gamma \bs{x}(t))_i \right) ,\qquad  x_i(0) = x^0_i\, .
$$
This is exactly~\eqref{eq:LV}.

\section{From individual-based models to community models with noise}
\label{annexe:micro2}

The present appendix is devoted to the derivation of various models arising in Section \ref{sec:otherextensions}
from individual birth-and-death processes.



\subsection{From the stochastic individual-based process to the Feller diffusions}
\label{ann:microFeller}
We  go back to the individual model defined in Section \ref{app:asympt_one_species} but 
we now consider a different popular model for the birth and death rates. Given an
additional parameter $\sigma > 0$, we replace the rates given
by~\eqref{logistic} with 
\begin{equation}
\label{bd-feller} 
b^K_n = b n + \sigma n K 
\quad \text{and} \quad
d^K_n =d n + \frac{c}{K} n^2 + \sigma n K. 
\end{equation}  

 According to this model, the {\it individual}
birth and death rates scale with $K$, the order of the population size. 
This can be realistic when one deals with very small individuals 
such as unicellular organisms, which have small life expectancy
\cite{bro-etal-04}. 

Doing the same computation as above, we see that the conditional mean is unchanged :
\[ \frac{\EE [X^K(t+h) - X^K(t) \, | \,  X^K(t)]}{h} \simeq \left(  r X^K(t) - c X^K(t)^2\right),\]
while this time, the conditional variance becomes
\[
\frac{\var[X^K(t+h) - X^K(t) \,  | \, X^K(t) ]}{h} \simeq 2\sigma X^K(t) . 
\]
This computation shows that, here, the stochasticity does not disappear as
$K\to\infty$.  More precisely (see e.g. \cite{bansayemeleard}), $(X^K(t))_{t \ge 0}$
converges to the  solution of the so-called Feller stochastic
differential equation
\begin{equation}
\label{feller} 
\dd X(t) = (r X(t) - c X(t)^2) \dd t + \sqrt{2\sigma X(t)} \dd B(t), \quad 
 X(0) = x_0, 
\end{equation}where again $r=b-d$ and $B$ is a standard Brownian Motion. Note 
that this equation coincides with the Feller equation~\eqref{eq:Feller}  
for $N = 1.$ \\

The multivariate analogue of the Feller equation~\eqref{feller}, $N\geq 1$, can be obtained
in a similar manner. Let $\sigma > 0$ be fixed. Getting back to
Equations~\eqref{def:ratesmultid1}-\eqref{def:ratesmultid3}, let us replace the natural birth and death rates
$b_i k_i$ and $d_i k_i$ in the definition of the 
transition probabilities with $b_i k_i + \sigma k_i K$ and $d_i
k_i + \sigma k_i K$ respectively. As before, $\Gamma_{ij}^K=\Gamma_{ij}/K$. This operation is similar to the change 
from~\eqref{logistic} to~\eqref{bd-feller} above. In this situation, the 
limit in distribution of the sequence of processes $(X^K(t))_{t\ge 0}$ is given as follows :
\[
\dd X(t) = X(t) \circ \left(  r - X(t) +  \Gamma X(t) \right) \dd t 
 + \sqrt{2\sigma X(t)} \circ \dd B(t), 
\]
where $B(t)$ is a $N$-dimensional standard Brownian Motion, which is \eqref{eq:Feller} in the case when $\sigma_i = \sigma$ is the same for each species $i \in [N].$

\subsection{Mean-Field approaches}\label{app:mean-field}

In \cite{roy2019numerical}, the limit when the number of species $N$ grows to infinity is considered, starting from the SDE~\eqref{sde} that we recall here:
\begin{equation*}
dX(t) = X(t) \circ \left( \bs 1_N - X(t) + \Gamma X(t) \right) dt
  + \lambda \bs 1_N dt 
 + f(X(t)) \circ dB(t), 
\end{equation*}and that is considered on a finite time window $[0,T]$, for a given $T>0$. More precisely, we consider the limit of 
\begin{equation}\widehat{Q}_N(\dd x):=\frac{1}{N}\sum_{i=1}^N \delta_{X_i}(\dd x),
\end{equation}
which is an empirical measure on $\mathcal{C}([0,T],\R_+)$.\\

The limit $Q$ depends on the choice of a model for $\Gamma$. When the correlation $\xi = 0$ (see \eqref{def:ellipticalmodel-corr}, $\Gamma$ is then a non-centered Ginibre matrix), this limit $Q$ solves the equation
\begin{equation}
\label{meanf} 
Q = \mathbb{E} \pi(G^Q) ,
\end{equation}where, for a probability measure $\nu$ on $\mathcal{C}^+([0,T])$, $G^\nu$ is an independent Gaussian process supported by $[0,T]$ and whose law is defined by
\[\mathbb{E}(G^\nu_t)=\int z(t) \nu(dz), \qquad \Cov(G^\nu_t, G^\nu_s)=  \int z(t) z(s) \nu(dz), \]
and where, for a deterministic function
$h(\cdot) \in L^2([0,T])$, $\pi(h)$ is the probability distribution on $\mathcal
C([0,T],\R_+)$ of the diffusion process 
\[
dX(t) = X(t) \left( 1  - X(t) + h(t) \right) dt 
  + \lambda dt + f(X(t)) dB(t), 
\]
where $B(t)$ is a one-dimensional Brownian Motion. 

As mentioned in the main text, the mathematical justification for this limit uses an approach of Ben Arous and Guionnet: in
\cite{ben-gui-95} and in the subsequent contributions, this convergence is
established for related models by means of the large deviations theory, characterizing $Q$ as the
unique minimum of an adequate rate function.  This work remains to be done in
the context of the SDE~\eqref{sde}. \\


Then, a known result, due
to Sznitmann~\cite{szn-84} asserts that if the components of $X$ are exchangeable, the convergence of $\widehat{Q}_N$ to $Q$ is equivalent to the propagation of chaos: 
given an arbitrary integer $k > 0$ and a fixed
arbitrary set of integers $\{ i_1, \ldots, i_k \}$, the vector $(X_{i_1},
\ldots, X_{i_k})$ converges in distribution to a vector with probability distribution $Q^{\otimes k}$ as $N\to\infty$.  

Note that for exchangeable initial conditions and $\Gamma$ drawn from an elliptic model, the exchangeability of the components of $X$ in $\mathcal{C}([0,T],\R_+)$ holds. Beyond the elliptical model, one can
imagine more sophisticated exchangeable models, such as a model with randomized
trophic levels or a randomized space.

The last step is to rigorously study the mean-field equation \eqref{meanf} satisfied by the
law $Q$. In the context of the elliptical model, the mean-field equation was 
studied in~\cite{roy2019numerical}, where, among other conclusions, the stationary and
chaotic phases of \cite{bunin2017ecological} were recovered.

\subsection{Lotka-Volterra in models of adaptive dynamics}\label{section:AD}

A last class of Lotka-Volterra equations with noise can be obtained in the context of Adaptive Dynamics \cite{metzgeritzmeszenajacobsheerwaarden,champagnatferrieremeleard,champagnatmeleard2011,guptametztran,billiardferrieremeleardtran}. Start from the individual-based model with $N$ species of large size, so that the Proposition \ref{prop:LLN} applies and $X^K_i(t)\approx K x_i(t)$, and assume that the solution of \eqref{eq:LV_dimensions} converges to a stable equilibrium $\mathbf{x}^*$. Now, add mutations: upon birth, say with probability $\mu^K_i$, the new offspring may be a mutant who can found a new species. If the rates of appearance of mutants are sufficiently small (mutations are rare), \textit{i.e.}
\begin{equation}
    e^{-CK} \ll \mu_K \ll \frac{C}{K \log K},
\end{equation}
(see \cite{champagnat06,champagnatferrieremeleard}) the time-scales of ecology (\textit{i.e.} births and deaths) and of mutations can be separated. Then, the dynamics can be separated into three phases.

Before going into their precise description, one can mention that adaptive dynamics can be seen as an example of what is called \textit{invasion analysis}   in ecology. Other examples have been analyzed through statistical physics tools such as the cavity method (see e.g. Arnoldi et al. \cite{ABKBJ22} for  
more details) and understanding them fully remains an interesting future task (see Section \ref{sec:discussion}).\\

\noindent \textbf{First phase: invasion probability of the mutant population.} Assume that the resident species $i\in [N]$ have abundancies close to the equilibrium state $\mathbf{x}^*$ when the mutant appears. The mutant descendance constitutes the $N+1$th specie, let us denote its birth rate by $b_{N+1}$, its natural death rate by $d_{N+1}$ and by $\Gamma_{j,N+1}$ and $\Gamma_{N+1,j}$ its interaction coefficient with the specie $j\in [N]$. The intraspecific competition rate is assumed to be $-1$, as for the other species. Let us also denote by $r_{N+1}=b_{N+1}-d_{N+1}$ the natural growth rate. As long as the new mutant population remains negligible, say with a size smaller than $\lfloor K\varepsilon \rfloor$ with some small $\varepsilon>0$, its dynamics is very close to a linear birth and death process whose parameters are functions of $\mathbf{x}^*$ while the other species are unaffected by the new population. More precisely, the rates of this birth and death process are:
\[b_{N+1},\qquad \mbox{ and }\qquad d_{N+1}+\sum_{j=1}^N \Gamma_{N+1,j} x^*_j.\]Note that the non-linearities have disappeared as the state has been frozen to $(x^*,0)$.
For such a birth and death process, the probability of invasion or extinction depends on their fitness
\[\phi_{N+1}(\mathbf{x}^*)=\big(r_{N+1}-\mathbf{x}^*+\Gamma \mathbf{x}^*\big)=r_{N+1}-\sum_{j=1}^N \Gamma_{N+1,j}x_j^*.\]computed in Section \ref{app:one_species_random}. The probability that the tree associated with the above birth death process is infinite, is 
\begin{equation}\label{eq:invasion_probability}
    p_{N+1}(\x^*)=\frac{\big[\phi_{N+1}(\x^*)\big]_+}{b_{N+1}},
\end{equation}
where $[.]_+$ denotes the positive part. It is also, for the original process, the probability that the mutant population stemming from a single founder reaches size $\lfloor K\varepsilon \rfloor$ without becoming extinct, and hence it is also called the \textit{invasion probability}. With probability $1-p_{N+1}(\x^*)$, the descendance of the mutant individual gets extinct before reaching the macroscopic size $\lfloor K\varepsilon \rfloor$. Notice that $p_{N+1}(\x^*)$ belongs by definition to $[0,1]$. When the fitness $\phi_{N+1}(\x^*)$ is negative, the new population has no chance of invading the equilibrium $\x^*$. \\
By coupling the original individual-based process with linear birth and death processes, it is possible to show that the time taken by the mutant population to reach the size $\lfloor K\varepsilon \rfloor$ is roughly the same as the time taken by the birth-death process to increase from $0$ to $\lfloor K\varepsilon \rfloor$, which is 
of order \[T_1^K \propto \log \frac{K}{\phi_{N+1}(\x^*)}.\]

\noindent \textbf{Second phase: approximation by an ODE system.} Once the mutant population has reached a size $\lfloor K\varepsilon \rfloor$, following the same path as in Section \ref{Nspecies}, the evolution of the $N+1$ populations can be approximated by a system of ordinary equations as \eqref{eq:premiere} but in $\R_+^{N+1}$ and started from the initial condition $(\x^*,\varepsilon)$. Assume that the solutions converge to an equilibrium $\widetilde{\x}^*$, some components of which might be zero. This indicates that the appearance of the mutant population can result in wiping out some other species. \\
The duration of the second phase, \textit{i.e.} the time taken by the random individual-based process to enter a neighborhood of $\widetilde{\x}^*$, is of order $1$: it is the time predicted by the deterministic dynamical system.\\

\noindent \textbf{Third phase: extinction of the species corresponding to the zeros of $\widetilde{\x}^*$.} Once the stochastic individual-based proces has entered 
a neighborhood of $\widetilde{\x}^*$, say of width $\varepsilon>0$, we can show using the theory of large deviations that it stays there during an exponentially long time of order $e^{CK}$ with $C>0$ (see \cite{champagnat06}). The species of sizes less than $\lfloor K \varepsilon\rfloor$ correspond to species $i\in [N]$ can be coupled with subcritical birth-death processes such that $\phi_i(\widetilde{\x}^*)\leq 0$. These processes have birth and death rates:
\[b_i,\qquad \mbox{ and }\qquad \widetilde{d}_i=d_i+\sum_{j=1}^{N+1} \Gamma_{i,j} \widetilde{x}^*_j.\]
They get almost surely extinct, and the expected time to extinction is
\begin{equation}
\mathbb{E}_{\lfloor \varepsilon \,K \rfloor} \left[ T_i \right]=\frac{1}{b_i} \sum_{\ell\geq 1} \Big( \frac{b_i}{\widetilde{d}_i}\Big)^\ell \ \sum_{k=1}^{\lfloor \varepsilon K\rfloor -1}\frac{1}{k+\ell}
\end{equation}
(see \cite[Section 5.5.3, p.190]{mel-livre16}).\\

In Metz et al. \cite{metzgeritzmeszenajacobsheerwaarden} and Champagnat \cite{champagnat06}, the case where a system of $N=2$ species always ends with the disappearance of one of the species is considered: this corresponds to the rule that \textit{invasion implies fixation}, meaning that the weakest species is lost when a favourable mutant arises. Provided new mutant species arrive sufficiently slowly into the system, the evolution of the population can be described by the sequence of successive dominating species or \textit{trait substitution sequence}. Later, Champagnat and Méléard \cite{champagnatmeleard2011} generalized this construction to the case where species can coexist and construct the \textit{polymorphic evolution sequence} that alternates phases where the dynamics is described by the Lotka-Volterra system \eqref{eq:LV} and phases of invasion of new arriving species. Lepers et al. \cite{lepersbilliardportemeleardtran} built on such construction to provide new models of population genetics, for populations with demographies and competition. In particular, the consideration of a neutral marker (see also \cite{billiardferrieremeleardtran}) allows to describe the evolution of the genetic diversity in a Lotka-Volterra system between each occurrence of (rare) mutations and to show that the demographic variance of the marker in the species $i\in [N]$ is 
\[ \frac{2b_i}{x^*_i},\]
which also allows to recover a definition of the effective population size:
\begin{equation}
\label{def:Ne}
\mbox{Ne}_i:=x^*_i.
\end{equation}Notice that the population effective size, which partially governs the evolution of the diversity at the neutral marker, depends on the birth and death rates in the specie $i$, but also on the ones of all the species interacting with specie $i$ (which is hidden in our notation). In particular, it means that the variance in the neutral diversity within the specie $i$ depends on the competitive interactions of the latter with all the other strains.

\vspace{0.1\columnwidth}

\section{Mathematical aspects of structured models}
\label{app:structuredmodapp}

\subsection{Deterministic networks} \label{app:deterministic_model}

We gather hereafter mathematical results related to sparse networks when the adjacency matrix $\Delta$ of the network is deterministic.

In this paragraph, we propose the following model, where a single quantity $d:=d_N$ accounts for the sparsity of the network. This parameter $d$ may depend on the number of species $N$ or be finite.

Consider a $d$-regular oriented graph with $N$ vertices, each vertex having $d$ neighbours. Matrix $\Delta$ had $d$ non-null entries per row and per column and $L:= d \times N$ non-null entries overall.

One can now write the interaction matrix $\Gamma$ as 
$$
\Gamma = \frac{\Delta \circ A}{\sqrt{d}} \, ,
$$
where $A=(A_{ij})$ has i.i.d. centered entries.
A first question concerns the feasibility of the equilibrium of Lotka-Volterra system stemming from such interaction matrix. In other words, the crux of the issue is the componentwise positivity of the $N\times 1$ vector $\x_N=(x_i)_{1\le i\le N}$, solution of the linear system
\begin{equation}\label{eq:equilibrium-sparse}
\x_N = \bs{1}_N + \frac{\Gamma}{\alpha_N} \x_N\ ,
\end{equation}
where $\bs{1}_N$ is the $N\times 1$ vector with components 1 as $N$. 

Notice that since the matrix is sparse the natural normalization is $\sqrt{d}$ instead of $\sqrt{N}$ \cite{bandeira2016sharp}.

The first set involves \textit{block interaction matrix}. Let $N=d\times m$. Denote by $\mathcal{S}_m$ the group of permutations of $[m]=\{1,\cdots, m\}$ into itself, by $P_{\sigma}=(\delta_{i,\sigma(i)})_{i\in [m]}$ the permutation matrix associated to $\sigma\in \mathcal{S}_m$ with $\delta_{ik}$ the Kronecker delta function and by $J_d={\bf 1}_d {\bf 1}_d^T$ the $d\times d$ matrix of ones. We shall focus hereafter on $N \times N$ {\it block-permutation} adjacency matrices $\Delta_N$, of the form
\begin{equation}\label{eq:block-permutation}
\Delta_N = P_{\sigma}\otimes J_d = \left( \delta_{i,\sigma(i)} J_d\right)_{i\in [m]}\, ,
\end{equation}

where $N=d \times m$ and $\otimes$ designates the Kronecker product.

Such matrix $\Delta_N$ gathers species by blocks. Species of block $i$ affect species of block $\sigma(i)$ and are affected by species of block $\sigma^{-1}(i)$. 

Let's provide an example to illustrate this. Let $m=4$ and $\sigma \in \mathcal{S}_4$ defined by : 
$$
\sigma= \begin{pmatrix}
    1 & 2 & 3 & 4 \\
    2 & 4 & 3 & 1
\end{pmatrix} \, .
$$
Matrices $P_\sigma$, $\Delta_N$ and $\Delta_N \circ A$ are respectively given by :
$$
P_\sigma = \begin{pmatrix}
0&1&0&0\\
0&0&0&1\\
0&0&1&0\\
1&0&0&0
\end{pmatrix} \, , ~~~~ 
\Delta = \begin{pmatrix}
0 &J_d &0 &0\\
0 &0 &0 &J_d\\
0 &0 &J_d &0\\
J_d &0 &0 &0
\end{pmatrix} \, , ~~~~
\Delta \circ A = \begin{pmatrix}
0 &A^{(1)} &0 &0\\
0 &0 &0 &A^{(2)}\\
0 &0 &A^{(3)} &0\\
A^{(4)} &0 &0 &0
\end{pmatrix} \, ,
$$
where $A^{(i)}$, $i = 1, \dots, 4$, is a $d\times d$ random matrix.

In particular, one may assume that, for each $i$, $A^{(i)}$ has i.i.d $\mathcal{N}(0,1)$ entries. 
With a condition on $d$ depending on $N$, the same phase transition phenomenon as in the i.i.d. model \ref{theo:faisibility-iid-model} occurs, even if the normalisation parameter of the interaction matrix $\Gamma$ passes from $\sqrt{N}$ to $\sqrt{d}$.

\begin{theo}\label{theo:sparse-matrix}
Let $d\ge \log(N)$, $\alpha_N \xrightarrow[N\to\infty]{} \infty$  and denote by $\alpha_N^*=\sqrt{2\log N}$. Let $\x_N=(x_k)_{k\in [N]}$ be the solution of \eqref{eq:equilibrium-sparse} with  $\Delta_N$ given by \eqref{eq:block-permutation} and $A_N$ a random matrix with i.i.d $\mathcal{N}(0,1)$ entries, then 
\begin{enumerate}
\item If $\exists\, \varepsilon>0$ such that eventually $\alpha_N\le (1-\varepsilon) \alpha_N^*$ then 
$$
\mathbb{P}\left\{ \min_{k\in [n]} x_k>0\right\} \xrightarrow[N\to\infty]{} 0\, .
$$

\item If $\exists\, \varepsilon>0$ such that eventually $\alpha_N\ge (1+\varepsilon)\alpha_N^*$ then
$$
\mathbb{P}\left\{ \min_{k\in [N]} x_k>0\right\} \xrightarrow[n\to\infty]{} 1\, .
$$
\end{enumerate}
\end{theo}

Furthermore, under the assumptions of the second point of theorem \ref{theo:sparse-matrix}, feasibility and global stability occur simultaneously.

\begin{theo}\label{th:global_stability_sparse}
  Let $d\ge \log(N)$, $\alpha_N \xrightarrow[N\to\infty]{} \infty$ and denote by $\alpha_N^*=\sqrt{2\log N}$. Let $\x_N=(x_k)_{k\in [N]}$ be the solution of \eqref{eq:equilibrium-sparse} with $\Delta_N$ given by \eqref{eq:block-permutation} and $A_N$ a random matrix with i.i.d $\mathcal{N}(0,1)$ entries. 
  \\Then, the probability that $\x_N$ is a nonnegative and globally stable equilibrium converges to 1. 
  
  Moreover, if $\exists\, \varepsilon>0$ such that eventually $\alpha_N\ge (1+\varepsilon) \alpha_N^*$, then the probability that globally stable equilibrium $\x_N$ is feasible converges to 1.
\end{theo}

One can remark that under the first assumptions of theorem \ref{th:global_stability_sparse}, the equilibrium of the Lotka-Volterra equation is globally stable even if its feasibility is not guaranteed. In other words, some species can get abundances set to zero.

\vspace{0.2cm}
With less structure, one can get an interest in $d$-regular graphs without this block structure. Henceforth, two different approaches are possible. 

The first one is to select a $d$-regular graph which can have or not have a particular chosen structure and then study the feasibility and the stability of the equilibrium. 

The second one consists in choosing a random $d$-regular graph. A random $d$-regular graph is a graph randomly selected from $\mathcal{G}_{N,d}$, where $\mathcal{G}_{N,d}$ is the probability space of all $d$-regular graphs on $N$ vertices. In particular, if the distribution on $\mathcal{G}_{N,d}$ is the discrete uniform distribution, then all the graphs have the same probability to be taken.

Furthermore, the order of magnitude of $d$ is an important factor in the mathematics study of the feasibility and stability issues, notably if $d$ is larger or smaller than $\log(N)$.

\subsection{Kernel matrices}
\label{appendix:kernel}
The goal of this section is to gather some known results in the mathematical literature on kernel matrices. As announced in Section \ref{sec:struct}, the take-home message will be that in many situations, the limiting spectrum of a kernel matrix will be a simple deformation (in fact a linear transformation) of the Marcenko-Pastur distribution. 

The kernel matrices we will consider are of the following form : let $Y$ be a random vector in $\RR^p$
such that $\EE(Y) = 0$ and $\EE(\|Y\|^2)=1.$ Let $X_1, \ldots, X_N$ be i.i.id. copies of $Y.$ Let $g: \RR^p \times \RR^p \longrightarrow \RR$ be a symmetric matrix, denoted as the kernel, and  $f : \RR  \longrightarrow \RR$  a function called the envelope.
Typical examples will be $g(x,y) = x^Ty,$ or $\|x-y\|^2$ and $f(x)= \exp(ex)$ or $(1+x)^a$ etc. 

We will consider the (symmetric) kernel matrix $A=(A_{ij})_{1\le i,j \le N}$ where, 
\begin{equation}
\label{def:kernelmat}
A_{ij} := f(g(X_i, X_j)), 
\end{equation}

and will be interested in the asymptotics of the empirical spectral measure $\hat \mu_N := \frac{1}{N} \sum_{i=1}^N \delta_{\lambda_i(A)}$ in the regime when $p$ and $N$ grows at the same rate  
that is $\frac{p}{N} \longrightarrow \tau \in (0, \infty).$ In random matrix theory, this is called the global regime.

The first interesting and well studied case is the so-called Wishart case, when $g(x,y)= x^T y$ and $f(x)=x.$ The matrix $A$ is now just the empirical covariance matrix of the vectors $X_1, \ldots, X_N.$

If the entries of $Y$ are i.i.d., then it is well known (see \cite{marcenkopastur1967math}) that $\hat \mu_N$
converges almost surely to the Marcenko-Pastur distribution with parameter $\tau:$
\begin{equation}
\label{def:marcenkopastur}
\dd \mu_{MP, \tau}(x):= \left(1- \frac{1}{\tau}\right) \mathbf 1_{\tau>1} \delta_0 + \frac{1}{2\pi x\tau } \sqrt{(b-x)(x-a)}\mathbf{1}_{[a,b]} \dd x,
\end{equation}
where $a,b := (1 \pm \sqrt \tau)^2.$
In this model, $X_i$ represents the $p$ features of the species $i$ and $A_{ij}$ is a way to measure the similarity of species $i$ and $j.$
Note that these convergence hold for more general random vectors $Y,$ not necessarily with i.i.d. entries, for example when $Y$ is uniformly distributed on the unit sphere in $\RR^p.$

We will now study the case of a regular (differentiable) enveloppe $f$ that does not depend on the dimension $p$ of the features. There are several papers studying this problem, in particular \cite{dovu2012kernel, chengsinger2013kernel, bordenave2012kernel, elkaroui2010kernel}. We present now a general result, found in \cite{dovu2012kernel}, which can be seen as a transference principle.

\begin{theo}
\label{theo:kernel}
Let $a:= \EE(g(X_i, X_j))$ for $i \neq j$ and $b:= \EE(g(X_i, X_i)).$ Assume that $\var g(X_i, X_j) = O(1/p)$ and
\[ \forall \delta >0, \PP\left( \max_{i \neq j} |g(X_i, X_j) - \EE(g(X_i, X_j))| >\delta\right) =o(1).\]
Assume that $f$ is differentiable at $a$ and continuous at $b.$ Then, 
 if we denote by $G$ the matrix with entries
 \[ G_{ij}:=\left\{ \begin{array}{ll}
 g(X_i, X_j), & \textrm{if } i \neq j,\\
 0, & \textrm{otherwise}, 
 \end{array}
 \right. \]
and $ A$ is defined in \eqref{def:kernelmat}, then $A$ has the same limiting spectral distribution as the matrix
 \[B:= (af'(a) -f(a) +f(b)) \eye_N + f'(a) G.\]
\end{theo}

In particular, when $Y$ has i.i.d. entries or is uniform on the sphere and $g(x,y) = x^Ty,$
then the asymptotic spectral distribution of $A$  is a linear transformation of $\mu_{MP, \tau}.$

Let us now give a few ideas of the proof. The main idea is to perform a Taylor expansion of the function $f$ around $a$ (or $b$) and to use the concentration hypotheses we have made on the quantities $g(X_i, X_j) $ to justify that the two matrices have indeed the same global asymptotic regime.

More precisely, let us define the matrix
\[ C := (f(a)-af'(a)) \mathbb J_N + (af'(a)-f(a)+f(b)) \eye_N + f'(a) G,\]
with $\eye_N $ the identity matrix and $\mathbb J_N$ the $N \times N$ matrix whose entries are all one's.

Let us remark that with this description, one can also identify that possible outliers, arising from the rank one deformation $(f(a)-af'(a)) \mathbb J_N.$ We know that they can be of crucial importance for stability.\\

For $z \in \CC \setminus \RR,$ Let $m_{ A}(z):= \int \frac{1}{z-x} \dd \mu_N(x)$ the trace of the resolvant of the matrix $ A$ (also called the Stieltjes transform $ \mu_N$ ) and $m_{C}(z)$ the trace of the resolvant of the matrix $C.$ To show that the two sequences of matrices have the same limiting spectral measure, it is enough to show  (see \cite{dovu2012kernel} for the details) that 
$m_{A}(z) - m_{C}(z)$ converges to zero for all $z$ such that $\Im z >0.$ By Cauchy-Schwarz inequality and then using that $\Im z >0,$ there exists a constant $C_z$ (depending only on $z$) such that
\begin{align*}
|m_{ A}(z) - m_{C}(z)|^2 & \le \frac{1}{N} \sum_{i=1}^N  \left|\frac{1}{z-\lambda_i( A)}- \frac{1}{z-\lambda_i(C)} \right|^2 \\
& \le C_z  \frac{1}{N} \sum_{i=1}^N \left|\lambda_i( A)- \lambda_i(C)\right|^2.
\end{align*}
Now, by Hoffman-Wielandt inequality (see e.g. \cite{tao2012math}), we have that 
 \[ |m_{ A}(z) - m_{C}(z)|^2 \le C_z  \frac{1}{N} \sum_{i, j} \left|A_{ij} - C_{ij} \right|^2.\]
We now go to the comparison of $ A$ and $C$ entrywise. For $i \neq j,$ using the regularity of $f$ at point $a = \EE(f(g(X_i, X_j))),$ we have
\begin{align*}
    A_{ij}  & = f(g(X_i, X_j)) = f(a) + f'(a) (g(X_i, X_j) -a) + o(g(X_i, X_j) -a)\\
    & = C_{ij} + o(g(X_i, X_j) -a),
\end{align*}
and the rest is controlled using the concentration hypothesis on $g(X_i, X_j)$ (see \cite[lemma 1]{dovu2012kernel} for more details),
whereas 
\[ A_{ii} = f(b) + o(f(g(X_i, X_j)) -b) = C_{ii} +  o(f(g(X_i, X_j)) -b) .\]

To end this review of the mathematical literature on kernel matrices, we want to mention two interesting results. The first one can be stated in the same framework \eqref{def:kernelmat} as above. In \cite{chengsinger2013kernel}, the authors show that, considering an envelope $f$ depending on the dimension $p$ of the features, it is possible to construct an example such that the unenveloped model converges to the Marcenko-Pastur law but the enveloped model converges to a limiting measure which is not a simple linear transformation of the Marcenko-Pastur distribution.
In \cite{benignipeche2019kernel}, with motivations from neural networks, the authors considered a somehow different kernel matrix. Their starting point is two matrices $X$ of size $N_0 \times M$ and $W$ of size $N_0 \times N_1$ with i.i.d. entries with distribution $\nu_1$ and $\nu_2$ respectively, having both a second moment. The matrix $W$ can be interpreted as a matrix of weights and $X$ a matrix of features.  They form a new  $N_1 \times m$ matrix $R$ with entries  
\[R_{ij}:= f\left(\frac{WX}{\sqrt n_0}\right) \]
and finally consider the empirical covariance matrix associated to $R$ given as the $N_1 \times N_1$
matrix 
\[ M:= \frac{1}{m} RR*.\]
In both examples, the limiting measures are described through a functional equation satisfied by their respective Stieltjes transform.

It is not clear to us whether these mathematical results can be helpful
for the study of ecological systems.

\printglossary

\printglossary[type=\acronymtype]

\end{document}